\documentclass[12pt,prd, aps, showpacs, superscriptaddress,floatfix]{revtex4-1}
\usepackage{graphicx}
\usepackage{bm}
\usepackage{multirow}
\usepackage{epsfig}
\usepackage{psfrag}
\usepackage{soul}
\usepackage{multirow}
\usepackage{mathptmx}
\usepackage{mathrsfs}
\usepackage{amsmath, amssymb}
\usepackage{dcolumn}
\usepackage{epstopdf}
\usepackage{color}
\usepackage{simplewick}
\usepackage{hhline}
\usepackage{changebar}
\usepackage{subfigure}
\usepackage{placeins}
\usepackage{slashed}
\usepackage{longtable}

\usepackage{flexisym,breqn,gensymb}

\newcommand\cu{Physics Department, Columbia University, New York,
  NY 10027, USA}

\newcounter{Outline}
\newcounter{Introduction}
\newcounter{Conclusions}
\newcounter{Acknowledgments}
\newcounter{Appendix}
\newcounter{Tables}
\newcounter{Figures}
\newcommand{\ba}{\begin{eqnarray}}
\newcommand{\ea}{\end{eqnarray}}
\newcommand{\bas}{\begin{eqnarray*}}
\newcommand{\eas}{\end{eqnarray*}}
\newcommand{\be}{\begin{equation}}
\newcommand{\ee}{\end{equation}}
\newcommand{\bes}{\begin{equation*}}
\newcommand{\ees}{\end{equation*}}
\newcommand{\bi}{\begin{itemize}}
\newcommand{\ei}{\end{itemize}}

\newcommand{\bcentre}{\begin{center}}
\newcommand{\ecentre}{\end{center}}

\font\tenmsb=msbm10 scaled\magstep1
\font\sevenmsb=msbm7 scaled\magstep1
\font\fivemsb=msbm5 scaled\magstep1
\newfam\msbfam
\textfont\msbfam=\tenmsb
\scriptfont\msbfam=\sevenmsb
\scriptscriptfont\msbfam=\fivemsb

\def\rational#1#2{{\mathchoice{\textstyle{#1\over#2}}%
  {\scriptstyle{#1\over#2}}{\scriptscriptstyle{#1\over#2}}{#1/#2}}}
\def\half{\rational12}			    

\def\C{{\bb C}}				    

\def\su3{SU(3)}

\def\tD{\mbox{D}\kern-0.65em\raise0.15ex\hbox{/}\kern0.15em} 
\def\sD{\mbox{\scriptsize D}\kern-0.5em\raise0.15ex\hbox{\scriptsize/}}
\def\ssD{\mbox{\tiny D}\kern-0.42em\raise0.15ex\hbox{\tiny/}}

\def\dslash{\hbox{\(\partial\)}\kern-0.5em\raise0.15ex\hbox{/}} 

\def\su3{SU(3)}

\def\nicefrac#1#2{\leavevmode\kern.1em\raise.5ex\hbox{\the\scriptfont0 #1}\kern-
.1em/\kern-.15em\lower.25ex\hbox{\the\scriptfont0 #2}}

\newcommand{\bea}{\begin{eqnarray}}
\newcommand{\eea}{\end{eqnarray}}

\newcommand{\colvectwo}[2]{\left(\begin{array}{c}#1\\#2\end{array}\right)}
\newcommand{\rowvectwo}[2]{\left(#1\,,\,#2\right)}

\newcommand{\fmat}[4]{\left(\begin{array}{cc} #1 & #2 \\ #3 & #4 \end{array}\right)}

\newcommand{\prop}{\mathcal{G}}

\newcommand{\psibar}{\overline{\psi\vphantom{d}}}

\begin{document}
\bibliographystyle{apsrev}

\title{Lattice simulations with G-parity Boundary Conditions}

\author{N.H.~Christ}\affiliation{\cu}
\author{C.~Kelly}\affiliation{\cu}
\author{D.~Zhang}\affiliation{\cu}

\collaboration{RBC and UKQCD Collaborations}
%
%


\maketitle

\centerline{ABSTRACT}
We discuss G-parity lattice boundary conditions as a means to impose momentum on the pion ground state without breaking isospin symmetry. This technique is expected to be critical for the precision measurement of $K\rightarrow(\pi\pi)_{I=0}$ matrix elements where physical kinematics demands moving pions in the final state and the statistical noise caused by disconnected contributions will make it difficult to use multi-exponential fits to isolate this as an excited state.  We present a formalism for computing hadronic Green's functions with G-parity boundary conditions, derive the discretized action and its symmetries, discuss how the strange quark can be introduced and detail techniques for the numerical implementation of these boundary conditions. We demonstrate and test these methods using several $16^3\times 32$ dynamical domain wall ensembles with a $420$ MeV pion mass and G-parity boundary conditions in one and two spatial directions.

\refstepcounter{section}
\setcounter{section}{0}


\newpage
\section{Introduction}
\label{sec:Introduction}
Important theoretical advances~\cite{Luscher:1986pf,Lellouch:2000pv} have opened a path for the determination of physical decay amplitudes comprising multi-particle final states using lattice QCD. In particular, calculations of the magnitude of direct CP-violation in the decays of kaons into two pions are now possible~\cite{Blum:2011ng,Blum:2015ywa,Bai:2015nea}, which offers a novel and exciting test of the Standard Model. The study of such decays with physical kinematics invariably requires moving final-state particles, for which it may be difficult to isolate the signal over the typically much larger contribution involving stationary particles. This can be avoided by modifying the lattice boundary conditions to remove the state with stationary pions from the particle spectrum. For charged pions this can be achieved by imposing antiperiodic boundary conditions on just the down or up quark propagators, which results in pion momenta that are odd-integer multiples of $\pi/L$. This technique has been used successfully in the calculation of the $\Delta I=3/2$ $K\to\pi\pi$ amplitude~\cite{Blum:2011ng,Blum:2015ywa}, but is limited by the fact that it applies only to charged pions and also explicitly breaks the isospin symmetry. These issues were avoided in the aforementioned case by utilizing the Wigner-Eckart theorem to relate the desired $K^+\to\pi^+\pi^0$ decay to the unphysical decay $K^+\to\pi^+\pi^+$, for which the final state contains only charged pions and is protected from mixing with other isospin states by virtue of being the only charge-2 state with those quantum numbers. This is not the case for the calculation of the $\Delta I=1/2$ $K\to\pi\pi$ amplitude, where the final state necessarily contains neutral pions and the isospin breaking cannot be circumvented. G-parity boundary conditions (GPBC)~\cite{Wiese:1991ku,Kim:2003xt,Kim:2009fe} provide an alternative approach that results in antiperiodic charged {\it and} neutral pions, and also preserves isospin symmetry. 

The RBC \& UKQCD collaborations have successfully employed GPBC to compute the $\Delta I=1/2$ $K\to\pi\pi$ amplitude~\cite{Bai:2015nea}. This document is intended as a partner to that work in which we lay the theoretical groundwork for employing G-parity boundary conditions in zero temperature lattice simulations and perform thorough numerical investigations of a number of its key features and difficulties using three $16^3\times 32\times 16$, 2+1 flavor dynamical domain wall ensembles with GPBC in 0, 1 and 2 directions, respectively.

We begin in Section~\ref{sec:Gparity} by outlining the transformation of the pion and quark fields under G-parity, and introduce notation for later use. In Section~\ref{sec:DiscAct} we derive the appropriate discretized action in our notation that makes explicit the boundary field mixing. We also discuss the numerical implementation of GPBC. The symmetries of the lattice action are investigated in Section~\ref{sec:Symmetries}, and we detail the appropriate operators for light and heavy meson states in Sections~\ref{sec:LightHadronic} and~\ref{sec:StrangeQuark}. In the latter we also discuss how the strange quark can be introduced into this framework. In Sections~\ref{sec:Ensembles} and~\ref{sec:Results} we present numerical demonstrations of the G-parity technique on the aforementioned ensembles. Our conclusions are presented in Section~\ref{sec:Conclusions}.

A related development of charge conjugation boundary conditions, introduced for a different purpose, has been carried out by Lucini, Patella, Ramos and Tantalo~\cite{Lucini:2015hfa}.

\section{G-parity Boundary Conditions}
\label{sec:Gparity}
G-parity is conventionally defined as a product of charge conjugation and an isospin rotation by $\pi$ radians around the $y$-axis: 
\begin{equation}
\hat G = \hat C e^{-i\pi \hat I_y} = e^{-i\pi \hat I_y} \hat C
\end{equation}
where the hat-symbol is used to denote operators. Here the choice of the $y$-axis is dictated by the convention that the representation matrices for $I_x$ and $I_z$ are real while those for $I_y$ are imaginary and implies $\hat G$ commutes with isospin rotations. The charged and neutral pions are G-parity odd, therefore if imposed as a spatial boundary condition the pions become antiperiodic and have discretized momenta that are odd-integer multiples of $\pi/L$, where $L$ is the lattice spatial extent. 

For QCD we are concerned with the action of G-parity upon the quarks. In this section we derive the appropriate transformation and introduce a convenient notation for studying its implications that we will use throughout the remainder of this document.

\subsection{Action upon quark fields}
\label{sec-gponquarks}

The isospin rotation around the $y$-axis transforms the flavor-doublet of light-quark annihilation operators as follows:
\begin{dmath}{
e^{-i\pi \hat I_y}\left(\begin{array}{c}u\\d\end{array}\right) e^{i\pi \hat I_y} = e^{-i\pi\sigma_2}\left(\begin{array}{c}u\\d\end{array}\right) = \left(\begin{array}{c}-d\\u\end{array}\right)\,.\label{eq-yisospinrotq}
}\end{dmath}
To describe charge conjugation we recall the action of $\hat C$ on the fermion fields $q$ and $\bar q$~\cite{Peskin:1995ev}:
\begin{dmath}
\begin{array}{ccc}
\hat C q \hat C^{-1} = C\bar q^T & {\rm and} & \hat C \bar q  \hat C^{-1} = -q^T C^{-1} \label{eq-chargeconjdef}
\end{array}
\end{dmath}
where $C$ is the $4\times 4$ charge conjugation matrix (not to be confused with the operator $\hat C$) which obeys
\begin{dmath}
C^{-1}\gamma_\mu C = -\gamma_\mu^T\,.
\label{eqn-Cproperties}
\end{dmath}
The usual requirement that $\hat C\,^2=1$ implies that $C^T = -C$ as well as the important property that when acting on states with isospin $I$: $\hat G\,^2 = (-1)^I$.

While we will not adopt specific conventions for the Euclidean $\gamma$ matrices used in this paper, we find it convenient to choose a chiral basis where $\gamma^5$ is real and diagonal which, together with Eq.~\eqref{eq-chargeconjdef}, implies that $[C,\gamma^5] = 1$.  In addition we will assume that the matrix $C$ is unitary and real, conditions consistent with the conventions, for example, of Ref.~\cite{Peskin:1995ev}.   We summarize the properties of the matrix $C$:
\begin{equation}
C = -C^T,\quad C^{-1} = C^\dagger \quad \mbox{and} \quad  C^*=C\,.
\label{eqn-Cproperties2}
\end{equation}

Combining the two operations we find that G-parity has the following action on the quark flavor doublet:
\begin{equation}
\begin{array}{lcr}
\hat G\colvectwo{u}{d}\hat G^{-1} = \colvectwo{-C\bar d^T}{C\bar u^T} & \hspace{0.3cm}{\rm and}\hspace{0.3cm} &
\hat G\rowvectwo{\bar u}{\bar d}\hat G^{-1} = \rowvectwo{d^T C^{-1} }{-u^T C^{-1}}\,.
\end{array} \label{eqn-GPonud}
\end{equation}
Notice that this involves an explicit mixing of the flavors and also of the spinors and conjugate spinors. We will see that this leads to a number of complications.

\subsection{Notation}
\label{sec:notation}

For this document we adopt a convenient notation in which we place the field operators $u$, $d$ and their conjugates into two-component vectors,
\begin{equation}
\psi = \colvectwo{d}{C\bar u^T}\ \ {\rm and}\ \psibar = (\bar d, u^T C)\,.  \label{eq-psipsibardef}
\end{equation}
We will refer to the indices of these vectors as `flavor' indices. Note that $\psi$ and $\psibar$ transform in the same way as the quark fields under charge-conjugation (Eq.~\eqref{eq-chargeconjdef}). 
The benefit of this notation is that the action of the G-parity operator upon $\psi$ takes on a simple form:
\begin{equation}
\hat G\psi \hat G^{-1}= i\sigma_2 \psi
\quad \mbox{and} \quad
\hat G \psibar \hat G^{-1}= \psibar(-i\sigma_2) \,,\label{eqn-gppsi}
\end{equation}
where $\sigma_2$ is the second Pauli matrix.

Using Eq.~\eqref{eq-psipsibardef} we can write
\begin{align}
q &= \colvectwo{u}{d} = F_{12}C \psibar^T + F_{21}\psi\,, \label{eq-qfrompsi} \\  
\bar q &= \rowvectwo{\bar u}{\bar d} = \psi^T C F_{21} + \psibar F_{12} \label{eq-qbarfrompsi}\,,
\end{align}
and the inverse transformations,
\begin{align}
\psi &= F_{12}q + F_{21} C\bar q^T\,, \\
\psibar &= \bar q F_{21} + q^T C F_{12}\,,
\end{align}
where
\begin{align}
F_{12} \equiv \fmat{0}{1}{0}{0} = \half(\sigma_1 + i\sigma_2) & & F_{21} \equiv \fmat{0}{0}{1}{0} = \half(\sigma_1 - i\sigma_2)\,.
\end{align}
These relations allow us to transform back and forth between the two notations. For later use we also define
\begin{align}
F_{11} \equiv \fmat{1}{0}{0}{0} = \half(1 + \sigma_3) & & F_{22} \equiv \fmat{0}{0}{0}{1} = \half(1 - \sigma_3)\,.
\end{align}

\subsection{Translations of fermion fields}

In this document we define our coordinates on a torus (i.e. modulo the lattice size), such that $x_\mu = L_\mu$ is equivalent to $x_\mu = 0$; $x_\mu = -1$ is equivalent to $x_\mu= L_\mu-1$; and so on. The corresponding field variables are treated in the same way: $\psi(x_\mu=L_\mu) \equiv \psi(x_\mu=0)$; $\psi(x_\mu=-1) \equiv \psi(x_\mu=L_\mu-1)$; {\it etc}. We will assume that the spatial box dimensions are all equal to $L$ for convenience.
The action of the boundary condition upon the field is treated explicitly in the context of a spatial translation via the introduction of a coordinate-dependent coefficient given below. In the following section we will rewrite the usual covariant derivative of the fermion action in terms of the translation operator such that the symmetry of the action under these modified translations can be imposed.

The boundary condition implies that as we move from site to site along a G-parity direction from the origin we encounter the fermion fields in the following order:
\begin{equation}\begin{array}{l}
||\psi(0),\,\ldots,\, \psi(L-1)||i\sigma_2\psi(0),\,\ldots,\, i\sigma_2\psi(L-1)||-\psi(0),\,\ldots,\, -\psi(L-1)||\\-i\sigma_2\psi(0),\,\ldots,\, -i\sigma_2\psi(L-1)||\psi(0)\ldots\,,
\end{array}\end{equation}
where $||$ indicates the location of the lattice boundary, and we have suppressed the coordinates along the other three directions for clarity. This implies the action of the translation operator $\hat T_\mu$ on the fermion field in a G-parity direction $\mu$ is as follows:
\begin{dmath}
\hat T_\mu \psi(x_\mu) \hat T_\mu^{-1} = \left\{ \begin{array}{l|r}
\psi(x_\mu+1)   &  0 \leq x_\mu < L-1  \\
(i\sigma_2)\psi(0)  & x_\mu = L-1
\end{array} \right.
\end{dmath}
and
\begin{dmath}
\hat T^{-1}_\mu \psi(x_\mu) \hat T_\mu = \left\{ \begin{array}{l|r}
\psi(x_\mu-1)   &  0 < x_\mu \leq L-1  \\
(-i\sigma_2)\psi(L-1)  & x_\mu = 0
\end{array} \right.
\end{dmath}
where we have again suppressed coordinates along the other three directions. These equations can be neatly summarized by introducing unitary flavor matrices
\begin{dgroup}
\begin{dmath}
B^+_\mu(x_\mu) = \exp\left(i \delta_{x_\mu,L-1}G_\mu \pi\sigma_2/2\right)\,, \label{eq-bpdef}
\end{dmath}
\begin{dmath}
B^-_\mu(x_\mu) = \exp\left(-i \delta_{x_\mu,0}G_\mu \pi\sigma_2/2\right)\,,\label{eq-bmdef}
\end{dmath}
\end{dgroup}
for which $G_\mu$ is unity in directions with GPBC and zero otherwise. These objects are related as follows:
\begin{dgroup}
\begin{dmath}
{ [B_\mu^+(x_\mu)]^T = [B_\mu^+(x_\mu)]^\dagger = \exp\left(-i \delta_{x_\mu+1,L}G_\mu \pi\sigma_2/2\right) = B_\mu^-(x_\mu+1) }\,,  \label{eq-identbplustrans}
\end{dmath}
\begin{dmath}
{ [B_\mu^-(x_\mu)]^T = [B_\mu^-(x_\mu)]^\dagger = \exp\left(i \delta_{x_\mu-1,L-1}G_\mu \pi\sigma_2/2\right) = B_\mu^+(x_\mu-1) }\,. \label{eq-identbminustrans}
\end{dmath}
\end{dgroup}
The translations then become:
\begin{dgroup}
\begin{dmath}
\hat T_\mu\psi(x)\hat T_\mu^{-1} = B^+_\mu(x_\mu)\psi(x+\hat \mu) 
\,, \label{eq-transpsifw}
\end{dmath}
\begin{dmath}
\hat T_\mu^{-1}\psi(x)\hat T_\mu = B^-_\mu(x_\mu)\psi(x-\hat \mu)
\,. \label{eq-transpsibw}
\end{dmath}
\label{eq-transpsiboth}
\end{dgroup}
Note the four operators $\hat T_\mu$ obey the commutation relations required for elements of the group of translations, $[\hat T_\mu,\hat T_\nu] = 0$.

\section{Discretized Action}
\label{sec:DiscAct}

In this section we derive the appropriate discretized action for two light quark flavors with G-parity boundary conditions. 

\subsection{Lattice QCD action with periodic BCs in our notation}

\subsubsection{Fermion action}
We begin with the usual four-dimensional Euclidean lattice fermion action for two-flavor QCD with {\it periodic} boundary conditions in the spatial directions,
\begin{dmath}
S = \sum_x \left\{ \sum_\mu\left[ \bar q(x) U_\mu(x)\Gamma^+_\mu q(x+\hat\mu) + \bar q(x) U^\dagger_\mu(x-\hat\mu)\Gamma^-_\mu q(x-\hat\mu) \right] + m\bar q(x)q(x) \right\}\,, \label{eqn-stdaction}
\end{dmath}
where the spin matrices $\Gamma^\pm_\mu = \half(1\mp \gamma_\mu)$ for the Wilson/domain wall actions or $\Gamma^\pm_\mu = \pm\half \gamma_\mu$ for the na\"{i}ve action. Here and below the sum over each of the four components of the coordinate $x$ runs from $x_\mu = 0$ to $x_\mu = L_\mu-1$ where $x_\mu$ and $L_\mu$ are expressed in lattice units. As mentioned previously we will assume the spatial dimensions are all equal in extent: $L_x = L_y = L_z = L$ for convenience.

Note that in these expressions we interpret the quantities $q$ and $\bar q$ as Grassmann variables that would appear in a path integral. We will use these operator and Grassmann interpretations interchangeably and specify a particular choice only when it is necessary.

The above can be rewritten in terms of the quark field vectors $\psi$ and $\psibar$ defined in Eq.~\eqref{eq-psipsibardef} using Eqs.~\eqref{eq-qfrompsi} and~\eqref{eq-qbarfrompsi}:
\begin{multline}
S = \sum_x \Big\{ \sum_\mu\Big[ 
\psibar(x)\Gamma^+_\mu F_{11} U_\mu(x)\psi(x + \mu )
+\psibar(x)\Gamma^+_\mu F_{22} U^*_\mu(x)\psi(x + \mu )\\
+ \psibar(x)\Gamma^-_\mu F_{11} U^\dagger_\mu(x -\mu )\psi(x -\mu )
+\psibar(x)\Gamma^-_\mu F_{11} U^T_\mu(x -\mu )\psi(x -\mu )
\Big]\\
+ m\psibar(x)\psi(x)
\Big\}
\end{multline}
This expression can be simplified by introducing the matrix
\begin{equation}
\tilde U_\mu(x) = \fmat{U_\mu(x)}{0}{0}{U_\mu^*(x)} = F_{11}U_\mu(x) + F_{22}U^*_\mu(x)\,,
\end{equation}
which has both flavor and color indices, and with which the action becomes
\begin{dmath}
S = \sum_x \left\{ \sum_\mu\left[ \psibar(x) \tilde U_\mu(x)\Gamma^+_\mu \psi(x+\hat\mu) + \psibar(x) \tilde U^\dagger_\mu(x-\hat\mu)\Gamma^-_\mu \psi(x-\hat\mu) \right] + m\psibar(x)\psi(x) \right\}\,. \label{eqn-gpactionnobc}
\end{dmath}
For use below it is convenient to rewrite the covariant derivative in terms of the translation operators:
\begin{dmath}
\nabla_\mu \psi(x) = \half\left[ \tilde U_\mu(x)\psi(x+\hat\mu) - \tilde U^\dagger_\mu(x-\hat\mu)\psi(x-\hat\mu) \right] = \half\left[ \tilde U_\mu(x) \hat T_\mu \psi(x) \hat T_\mu^{-1} - \hat T^{-1}_\mu \tilde U_\mu^\dagger(x)\psi(x) \hat T_\mu \right]\,. \label{eq-covderivtranspbc}
\end{dmath}

\subsubsection{Gauge action}

As the fermion action is expressed in terms of the flavor-matrix gauge links, it is convenient to do the same for the gauge action. We assume the Wilson action, although it it straightforward to generalize this result. The action is 
\begin{dmath}
S_W = -\frac{\beta}{3} \sum_x \sum_{\mu, \nu>\mu}  {\rm Re}\,{\rm Tr}\,U_{\mu\nu}(x)\,, \label{eq-wilsonaction}
\end{dmath}
where
\begin{dmath}
U_{\mu\nu}(x) = U_\mu(x) U_\nu(x+\hat\mu) U_\mu^\dagger (x+\hat\nu) U_\nu^\dagger(x)
\end{dmath}
is the usual plaquette. We can construct a similar ``plaquette'' from the flavored gauge links $\tilde U_\mu$:
\begin{dmath}
\tilde U_{\mu\nu}(x) = \tilde U_\mu(x) \tilde U_\nu(x+\hat\mu) \tilde U_\mu^\dagger (x+\hat\nu) \tilde U_\nu^\dagger(x)\,,
= \fmat{U_{\mu\nu}(x)}{0}{0}{U^*_{\mu\nu}(x)}\,, \label{eq-gpplaqbcimpl}
\end{dmath}
for which
\begin{dmath}
{\rm Re}\,{\rm Tr}\,U_{\mu\nu} = \frac{1}{2}{\rm Tr}\,\tilde U_{\mu\nu} \label{eq-gpregularplaqreln}
\end{dmath}
where it is understood that the trace on the right-hand side includes the flavor indices. The Wilson action then becomes
\begin{dmath}
S_W = -\frac{\beta}{6} \sum_x \sum_{\mu, \nu>\mu}  {\rm Tr}\,\tilde U_{\mu\nu}(x)\,, \label{eq-wilsonactionflinks}
\end{dmath}

As with the covariant derivative, it is useful to express the plaquette in terms of translation operators:
\begin{dmath}
\tilde U_{\mu\nu}(x) = \tilde U_\mu(x) \hat T_\mu\tilde U_\nu(x)\hat T_\mu^{-1} \hat T_\nu\tilde U_\mu^\dagger (x)\hat T_\nu^{-1} \tilde U_\nu^\dagger(x) \label{eq-flavplaqtransop}
\end{dmath}

\subsection{Lattice QCD action with G-parity BCs}

\subsubsection{Fermion action}

The covariant derivative for G-parity BCs can be obtained by inserting Eqs.~\eqref{eq-transpsifw} and~\eqref{eq-transpsibw} into Eq.~\eqref{eq-covderivtranspbc}:
\begin{dmath}
\nabla_\mu \psi(x) = \half\left[ \tilde U_\mu(x)B^+_\mu(x_\mu)\psi(x+\hat\mu) - \left\{ \hat T^{-1}_\mu\tilde U^\dagger_\mu(x)\hat T_\mu \right\} B_\mu^-(x_\mu)\psi(x-\hat\mu) \right]\,.
\end{dmath}
The translational properties of the flavor-matrix gauge links can be found by imposing gauge invariance upon the derivative term of the action, $S_\nabla = \sum_{x,\mu} \psibar(x)\gamma^\mu\nabla_\mu \psi(x)$. 

Under a gauge transformation $V$ the field $\psi$ and its conjugate transform as
\begin{dgroup}
\begin{dmath}
{ \psi(x) = \colvectwo{d(x)}{C\bar u(x)^T} \to \tilde V(x)\psi(x)\,, }
\end{dmath}
\begin{dmath}
{ \psibar(x) = (\bar d(x), u^T(x) C) \to \psibar(x) \tilde V^\dagger(x)\,, }
\end{dmath}
\end{dgroup}
where
\begin{dmath}
{ \tilde V =  \fmat{V}{0}{0}{V^*} = F_{11}V + F_{22}V^*\,. }
\end{dmath}

The forwards component of $S_\nabla$ then transforms as,
\begin{dmath}
\sum_{x,\mu} \psibar(x)\tilde U_\mu(x)B^+_\mu(x_\mu)\gamma^\mu\psi(x+\hat\mu) \longrightarrow \sum_{x,\mu} \psibar(x)\tilde V^\dagger(x)\tilde U'_\mu(x)B^+_\mu(x_\mu)\tilde V(x+\hat\mu)\gamma^\mu\psi(x+\hat\mu)
\end{dmath}
where $\tilde U_\mu'$ is the gauge transformation of $\tilde U_\mu$. Invariance of this term under the gauge transformation then implies
\begin{dmath}
\tilde U'_\mu(x)= \tilde V(x) \tilde U_\mu(x)B^+_\mu(x_\mu)\tilde V^\dagger(x+\hat\mu) [B^+_\mu(x_\mu)]^\dagger\,.\label{eq-flavlinkgaugetrans}
\end{dmath}
For the backwards component of $S_\nabla$ the term $\hat T^{-1}_\mu\tilde U^\dagger_\mu(x)\hat T_\mu$ enters. If we assume that 
\begin{dmath}
\hat T^{-1}_\mu\tilde U^\dagger_\mu(x)\hat T_\mu = \alpha \tilde U^\dagger_\mu(x-\hat\mu) \beta
\end{dmath}
with $\alpha$ and $\beta$ flavor matrices, the backwards component of $S_\nabla$ transforms under a gauge transformation as follows:
\begin{equation}
\begin{array}{l}
\displaystyle\sum_{x,\mu} \bar\psi(x) \alpha \tilde U^\dagger_\mu(x-\hat\mu)\beta B_\mu^-(x_\mu)\gamma^\mu\psi(x-\hat\mu) \longrightarrow \\
\displaystyle\sum_{x,\mu}\bar\psi(x)\tilde V^\dagger(x) \alpha 
\left\{ B^+_\mu(x_\mu-1)\tilde V(x) [B^+_\mu(x_\mu-1)]^\dagger \tilde U^\dagger_\mu(x-\hat\mu)\tilde V^\dagger(x-\hat\mu) \right\}
\beta B_\mu^-(x_\mu) \tilde V(x-\hat\mu)\gamma^\mu\psi(x-\hat\mu)
\end{array}
\end{equation}
from which gauge invariance implies $\alpha = [B^+_\mu(x_\mu-1)]^\dagger = B^-_\mu(x_\mu)$ and $\beta = [B^-_\mu(x_\mu)]^\dagger$. Thus we find
\begin{dmath}
\hat T^{-1}_\mu\tilde U^\dagger_\mu(x)\hat T_\mu = B^-_\mu(x_\mu) \tilde U^\dagger_\mu(x-\hat\mu) [B^-_\mu(x_\mu)]^\dagger\,.~\label{eq-udagmubacktransmu}
\end{dmath}

We can now write down the covariant derivative with G-parity BCs:
\begin{dmath}
\nabla_\mu \psi(x) = \half\left[ \tilde U_\mu(x)B^+_\mu(x_\mu)\psi(x+\hat\mu) - \left\{ B_\mu^-(x_\mu)\tilde U^\dagger_\mu(x-\hat\mu)[B_\mu^-(x_\mu)]^\dagger \right\}B_\mu^-(x_\mu)\psi(x-\hat\mu) \right] 
= \half\left[ \tilde U_\mu(x)B^+_\mu(x_\mu)\psi(x+\hat\mu) - B_\mu^-(x_\mu)\tilde U^\dagger_\mu(x-\hat\mu)\psi(x-\hat\mu) \right]\,,
\end{dmath}
the corresponding (complete) fermion action
\begin{dmath}
S = \sum_x \left\{ \sum_\mu\left[ \psibar(x)\tilde U_\mu(x)\Gamma^+_\mu B^+_\mu(x_\mu)\psi(x+\hat\mu) + \psibar(x)B_\mu^-(x_\mu)\tilde U^\dagger_\mu(x-\hat\mu)\Gamma^-_\mu \psi(x-\hat\mu) \right] + m\psibar(x)\psi(x) \right\}\,, \label{eqn-lsymaction}
\end{dmath}
and the Dirac matrix,
\begin{dmath}
{\cal M}(x,y) = \sum_\mu\left[ \tilde U_\mu(x)\Gamma^+_\mu B^+_\mu(x_\mu)\delta_{x+\hat\mu,y} + B_\mu^-(x_\mu)\tilde U^\dagger_\mu(y)\Gamma^-_\mu \delta_{x-\hat\mu,y} \right] + m\delta_{x,y}\,. \label{eqn-gpdiracop}
\end{dmath}

Note that throughout this document we consistently ignore the fact that domain wall fermions have an additional discrete index associated with their coordinate in the fifth dimension, and instead treat them identically to Wilson fermions. In Appendix~\ref{appendix-dwf} we demonstrate that in our (suitably extended) $\psi$-field notation the G-parity boundary condition does not affect the fifth dimensional coordinate despite the reflection in this dimension induced by charge conjugation, further evidencing the power of this notation and justifying us dropping this index.

It is important to recognize that the introduction of G-parity boundary conditions does not alter the usual $\gamma^5$-hermiticity of the Dirac propagator,
\begin{dmath}
\prop(x,y) = \gamma^5 \prop^\dagger(y,x) \gamma^5\,. \label{eq-g5herm}
\end{dmath}
This can be seen from the expression given in Eq.~\eqref{eqn-gpdiracop} for the Dirac operator, which has the structure of the usual Euclidean lattice Dirac operator with the exception of the appearance of the $2\times2$ matrices 
$B_\mu^\pm$. We can then deduce $\gamma^5$-hermiticity following the usual steps using Eqs.~\eqref{eq-identbplustrans} and~\eqref{eq-identbminustrans} and $[\gamma^5, B_\mu^\pm]=0$.

The quark propagators obtained by inverting this Dirac matrix are $2\times 2$ matrices in flavor space as well as being matrices in spin and color space. In practise this leads to additional diagrams that must be evaluated when computing hadronic observables. In addition, we must typically compute the inverse with separate sources for each flavor, doubling the number of matrix inversions required to compute the full propagator. However in many cases this requirement can be circumvented by taking advantage of the isospin symmetry, as we demonstrate in Section.~\ref{sec-isospinact}.

\subsubsection{Gauge action}

Expressing the Wilson gauge action Eq.~\eqref{eq-wilsonactionflinks} in terms of translation operators acting on the gauge links via Eq.~\eqref{eq-flavplaqtransop}, we have
\begin{dmath}
S_W = -\frac{\beta}{6} \sum_x \sum_{\mu, \nu>\mu}  {\rm Tr}\,  \tilde U_\mu(x) \hat T_\mu\tilde U_\nu(x)\hat T_\mu^{-1} \hat T_\nu\tilde U_\mu^\dagger (x)\hat T_\nu^{-1} \tilde U_\nu^\dagger(x)\,.
\end{dmath}
The action of the translation operator upon the links has thus far been derived only for links in the same direction as the translation (Eq.~\eqref{eq-udagmubacktransmu}). The corresponding relation for links that are orthogonal to the translation can be derived by imposing gauge invariance on $S_W$. Assuming $\hat T_\mu\tilde U_\nu(x)\hat T_\mu^{-1} = \alpha \tilde U_\nu(x+\hat\mu) \beta$ and $\hat T_\nu\tilde U_\mu^\dagger (x)\hat T_\nu^{-1} = \gamma \tilde U_\mu^\dagger(x+\hat\nu) \delta$ where $\alpha-\delta$ are flavor matrices, and then applying Eq.~\eqref{eq-flavlinkgaugetrans} it is straightforward to show that the following action is gauge invariant:
\begin{dmath}
S_W = -\frac{\beta}{6} \sum_x \sum_{\mu, \nu>\mu}  {\rm Tr}\,  \tilde U_\mu(x) B^+_\mu(x_\mu) \tilde U_\nu(x+\hat\mu) [B^+_\mu(x_\mu)]^\dagger B^+_\nu(x_\nu)\tilde U_\mu^\dagger (x+\hat\nu) [B^+_\nu(x_\nu)]^\dagger \tilde U_\nu^\dagger(x)\,,
\end{dmath}
for which $\alpha = \beta^\dagger = B^+_\mu(x_\mu)$ and $\gamma = \delta^\dagger = B^+_\nu(x_\nu)$.

Combining the above results for $\alpha-\delta$ with Eq.~\eqref{eq-udagmubacktransmu} we obtain the general action of translations on the flavored gauge links:
\begin{dgroup}
\begin{dmath}
\hat T_\mu\tilde U_\nu(x) \hat T_\mu^{-1} = B_\mu^+(x_\mu)\tilde U_\nu(x+\hat\mu) [B_\mu^+(x_\mu)]^\dagger
\,,
\label{eq-transforwardU}
\end{dmath}
\begin{dmath}
\hat T^{-1}_\mu\tilde U_\nu(x) \hat T_\mu = B_\mu^-(x_\mu)\tilde U_\nu(x-\hat\mu) [B_\mu^-(x_\mu)]^\dagger
\label{eq-transbackU}
\,.
\end{dmath}
\end{dgroup}
Note that using $\hat T_\mu \tilde U_\nu(x) \tilde U^\dagger_\nu(x) \hat T_\mu^{-1} = 1$ it is easy to see that the translation operators act on $\tilde U^\dagger_\nu$ in the same way as they do on $\tilde U_\nu$.

Eq.~\eqref{eq-transforwardU} implies that as we again move from site to site along a G-parity direction $\mu$ from the origin, the gauge fields are encountered in the following order:
\begin{equation}\begin{array}{l}
||\tilde U_\mu(0),\,\ldots,\, \tilde U_\mu(L-1)||(i\sigma_2)\tilde U_\mu(0)(-i\sigma_2),\,\ldots,\, (i\sigma_2)\tilde U_\mu(L-1)(-i\sigma_2)||\tilde U_\mu(0)\ldots\,.
\end{array}\end{equation}
This implies the links transform as $\tilde U_\mu \to (i\sigma_2)\tilde U_\mu(-i\sigma_2) = \tilde U_\mu^*$ under the action of the boundary, i.e. the links must obey complex conjugate (or equivalently, charge conjugation) boundary conditions. This of course requires new gauge configurations to be generated for GPBC.

\subsection{One-flavor equivalence}
\label{sec-1fequiv}

Consider the upper component of the flavor doublet field $\psi$ for G-parity in one direction. As we move from site to site in the G-parity direction we encounter the fields in the following order:
\begin{equation}\begin{array}{l}
||d(0),\,\ldots,\, d(L-1)||C\bar u^T(0),\,\ldots,\, C\bar u^T(L-1)||-d(0),\,\ldots,\, -d(L-1)||\\\hspace{1cm}-C\bar u^T(0),\,\ldots,\, -C\bar u^T(L-1)||d(0),\,\ldots\,,
\end{array}\end{equation}
where $||$ again indicates the position of the lattice boundary and we suppress coordinates other than in the G-parity direction. This infinite series is antiperiodic in $2L$, and the subset between $0$ and $2L$ contains all of the fermionic degrees of freedom in the G-parity setup. We can therefore define a field $\Psi$ on a lattice of size $2L$ (denoting the corresponding coordinates with capital letters) with the following mapping:
\begin{dmath}
\Psi(X) = \left\{ \begin{array}{lll} d(X) & {\rm for} & 0\leq X< L \\ C\bar u^T(X-L) & {\rm for} & L\leq X< 2L\end{array}\right.\,,
\end{dmath}
which contains all the fermionic degrees of freedom and obeys antiperiodic boundary conditions in $2L$. Similarly, the gauge links ${\bf U}_\nu$ defined on this doubled lattice are mapped as follows:
\begin{dmath}
{\bf U}_\nu(X) = \left\{ \begin{array}{lll} U_\nu(X) & {\rm for} & 0\leq X< L \\ U_\nu^*(X-L) & {\rm for} & L\leq X< 2L\end{array}\right.\,.
\end{dmath}
Consider the forwards component of the action, Eq.~\eqref{eqn-lsymaction}, in the G-parity direction $\mu$, and suppress the coordinates in the other directions:
\begin{equation}\begin{array}{rl}
 \sum_{x_\mu} & \psibar(x_\mu)\tilde U_\mu(x_\mu) \Gamma^+_\mu B^+_\mu(x_\mu)\psi(x_\mu+1)  \\
 & =  \sum_{x_\mu < L-1} \left[ \bar d(x_\mu)U_\mu(x_\mu)\Gamma^+_\mu d(x_\mu+1) + u^T(x_\mu)C U^*_\mu(x_\mu)\Gamma^+_\mu C\bar u^T(x_\mu+1)\right] \\
 & \hspace{2cm}+ \bar d(L-1)U_\mu(L-1)\Gamma^+_\mu C\bar u^T(0) - u^T(L-1)C U^*_\mu(L-1)\Gamma^+_\mu d(0) \\
 & = \sum_{X_\mu < L-1} \left[ \bar\Psi(X_\mu){\bf U}_\mu(X_\mu)\Gamma^+_\mu\Psi(X_\mu+1) + \bar\Psi(X_\mu+L) {\bf U}_\mu(X_\mu+L) \Gamma^+_\mu\Psi(X_\mu+L+1)\right] \\
 & \hspace{2cm}+ \bar\Psi(L-1){\bf U}_\mu(L-1) \Gamma^+_\mu\Psi(L) - \bar\Psi(2L-1){\bf U}_\mu(2L-1)\Gamma^+_\mu\Psi(0)\\
 & = \sum_{X_\mu} \left[ \bar\Psi(X_\mu){\bf U}_\mu(X_\mu)\Gamma^+_\mu {\bf B}_\mu^+(X_\mu)\Psi(X_\mu+1)\right] \,,
 \end{array}
\end{equation}
where ${\bf B}^+_\mu(X_\mu) = \exp(i\pi \delta_{X_\mu, 2L-1})$ imposes the appropriate sign for the term crossing the boundary in $2L$. For the backwards component we likewise obtain
\begin{equation}\begin{array}{rl}
 \sum_{x_\mu} & \psibar(x)B^-_\mu(x_\mu)\tilde U^\dagger_\mu(x_\mu-1)\Gamma^-_\mu \psi(x_\mu-1)  \\
& = \sum_{X_\mu} \left[ \bar\Psi(X_\mu){\bf B}_\mu^-(X_\mu){\bf U}^\dagger_\mu(X_\mu-1)\Gamma^-_\mu \Psi(X_\mu-1)\right] \,,
 \end{array}
\end{equation}
where ${\bf B}^-_\mu(X_\mu) = \exp(i\pi \delta_{X_\mu, 0})$. 

The two terms together comprise the action for a single flavor quark field residing on a lattice of size $2L$ with antiperiodic boundary conditions in the G-parity direction $\mu$, with the only additional condition being that the gauge links on the second half of the lattice are the complex conjugates of those on the first. 

This establishes a direct equivalence between the two-flavor theory and a one-flavor theory on a doubled lattice that proves very useful when it comes to implementing these boundary conditions on a computer.

\begin{figure}[tp]
\centering
\includegraphics[width=0.3\textwidth]{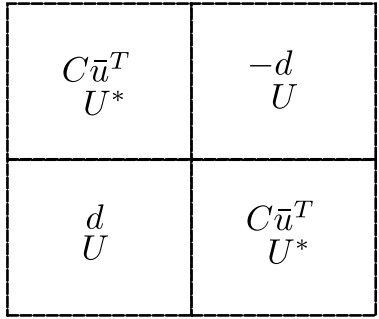}
\caption{The mapping of the quadrants of a one-flavor theory with APBC in two directions that reproduces the effect of imposing GPBC in two directions. The assignment of the quark and gauge fields are shown within the quadrants. Here the minus sign on the upper-right quadrant is related to the fact that the fields are antiperiodic under passing through a G-parity boundary twice.\label{fig-quaddlatt}}
\end{figure}

Let us consider extending this formalism to GPBC in a second direction, $y$ (with the original doubling in the $x$-direction).  This can be achieved by doubling the lattice again in the $y$-direction and imposing APBC in this second direction. However in this approach some care is required to recognize that the fermion fields on the four quadrants of the resulting lattice are not independent but are related according to Figure~\ref{fig-quaddlatt}. This has implications for example in the construction of propagator sources: a down-quark source on timeslice $\tau$ with spatial smearing function $\Theta$ centered at position $\vec x_0$ in the two flavor setup
\begin{dmath}
\eta_f(\vec x,t) = \delta_{f,0}\delta_{t,\tau}\Theta(\vec x-\vec x_0)
\end{dmath}
corresponds to the following in the twice-doubled single-flavor setup:
\begin{dmath}
\eta(\vec X,t) = \delta_{t,\tau}\left( D_{LL}(\vec X)\Theta(\vec X-\vec x_0) - D_{UR}(\vec X)\Theta(\vec X - \vec x_0 - L\hat x - L\hat y) \right)
\end{dmath}
where $D_{LL}(\vec X)$ and $D_{UR}(\vec X)$ are functions that are unity on the lower-left and upper-right quadrants, respectively, and zero elsewhere (referring to Figure~\ref{fig-quaddlatt}). Here the minus sign between the terms arises because the fermion field in the upper-right quadrant has the opposite sign to that of the lower-left quadrant. Note also that this second doubling of the lattice doubles the cost of performing the Dirac matrix inversion relative to the two-flavor approach, and therefore the twice-doubled one-flavor approach is of limited practical use for GPBCs in multiple directions. 

\begin{figure}[tp]
\centering
\includegraphics[width=0.3\textwidth]{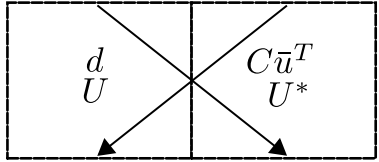}
\caption{The mapping of the upper boundary to the lower boundary in the $y$-direction of the two halves of a one-flavor theory with APBC in the $x$-direction. Here the arrows indicate the position of the next site encountered when moving through the upper boundary. Note that a minus sign must be applied when crossing (in either direction) between the lower-left and upper-right boundary segments. This setup reproduces the effect of imposing GPBC in two directions.\label{fig-orbifold}}
\end{figure}

An alternative treatment for GPBC in two directions, easily generalized to three, is to modify the boundary conditions in the $y$-direction such that passing through the boundary is accompanied by a translation by $L$ in the $x$-direction; this is illustrated in Figure~\ref{fig-orbifold}. This approach avoids further doubling for GPBC in more than one direction and therefore has the same computational cost as the two-flavor approach, but is somewhat complicated to implement numerically, and the non-nearest neighbor communications pattern in the $y$-direction is suboptimal for most parallel machines. 

We therefore conclude that this one-flavor mapping, while a useful cross-check, is not practical for high precision lattice calculations with GPBC in more than one direction. The alternative, which we employ in practise, is to implement the full two-flavor theory with the appropriate mixing of the flavors at the boundary directly.

\section{Lattice symmetries}
\label{sec:Symmetries}

The use of G-parity boundary conditions has a number of symmetry implications that we detail in this section. 

Given that the action away from the boundary is simply a rewrite of the usual lattice action, we need only consider the effects of global symmetry operations upon the boundary terms. To do so it is convenient to rewrite the matrices $B^\pm$ defined in Eqs.~\eqref{eq-bpdef} and~\eqref{eq-bmdef} as
\begin{dgroup}
\begin{dmath}
B^+_\mu(x_\mu) = 1  - G_\mu\delta_{x_\mu,L-1}(1 - i\sigma_2)\,, \label{eq-bpdefdelta}
\end{dmath}
\begin{dmath}
B^-_\mu(x_\mu) = 1 - G_\mu\delta_{x_\mu,0}(1 + i\sigma_2)\,, \label{eq-bmdefdelta}
\end{dmath}
\end{dgroup}
such that the contribution to action from the G-parity boundary conditions is contained in the following expression:
\begin{dmath}
S_{\rm GPBC} = -\sum_x \sum_\mu G_\mu \left[ \delta_{x_\mu,L-1}\psibar(x)\tilde U_\mu(x) (1 - i\sigma_2)\Gamma^+_\mu\psi(x+\hat\mu) + \delta_{x_\mu,0}\psibar(x)(1 + i\sigma_2)\tilde U^\dagger_\mu(x-\hat\mu)\Gamma^-_\mu \psi(x-\hat\mu) \right]\,.\label{eqn-lsymactionbnd}
\end{dmath}
The total action then becomes
\begin{dmath}
S = S_{\rm std} + S_{\rm GPBC}\,,\label{eq-symacttot}
\end{dmath}
where $S_{\rm std}$ is the `standard' action
\begin{dmath}
S_{\rm std} = \sum_x \left\{ \sum_\mu\left[ \psibar(x)\tilde U_\mu(x)\Gamma^+_\mu \psi(x+\hat\mu) + \psibar(x)\tilde U^\dagger_\mu(x-\hat\mu)\Gamma^-_\mu \psi(x-\hat\mu) \right] + m\psibar(x)\psi(x) \right\}
\end{dmath}
for a periodic field. Here we see that $S_{\rm GPBC}$ acts to subtract the boundary term of the standard action and add the correct G-parity flavor `twist'.

We can simplify Eq.~\eqref{eqn-lsymactionbnd} somewhat by noting that for coordinates defined on a torus (in one dimension here),
\begin{dmath}
{ \sum_{x=0}^{L-1} \psibar(x) \delta_{x,0}\psi(x-1) = \psibar(0)\psi(-1) = \psibar(L)\psi(L-1) = \sum_{x=0}^{L-1} \psibar(x+1) \delta_{x,L-1}\psi(x) }\,.
\end{dmath}
We can then write
\begin{dmath}
S_{\rm GPBC} = -\sum_x \sum_\mu G_\mu \delta_{x_\mu,L-1} \left[ \psibar(x)\tilde U_\mu(x) (1 - i\sigma_2)\Gamma^+_\mu\psi(x+\hat\mu) + \psi^T(x)C \tilde U^*_\mu(x)(1-i\sigma_2)\Gamma^+_\mu C\psibar^T(x+\hat\mu)\right]\,, \label{eqn-lsymactionbndsimp}
\end{dmath}
where for the second term we have transposed the Grassman variables and used the following identity:
\begin{dmath}
{ C [\Gamma^\pm_\mu]^T C = -\Gamma^\mp_\mu }\,. \label{eq-identCGammaC}
\end{dmath}
In terms of the usual flavor doublets we can rewrite Eq.~\eqref{eqn-lsymactionbndsimp} as
\begin{multline}
S_{\rm GPBC} = -\sum_x \sum_\mu G_\mu \delta_{x_\mu,L-1} \Big[ \bar q(x)\Gamma^+_\mu U_\mu(x) (i\sigma_2) C\bar q^T(x+\hat\mu) \\
+ \bar q(x) \Gamma^+_\mu U_\mu(x) q(x+\hat\mu)
+ q^T(x)C \Gamma^+_\mu U_\mu^*(x) C\bar q^T(x+\hat\mu) \\
+ q^T(x)C \Gamma^+_\mu U_\mu^*(x) (i\sigma_2)q(x+\hat\mu)
\Big]\,,  \label{eqn-lsymactionbndsimpqform}
\end{multline}
which proves convenient in some cases.

\subsection{Isospin}
\label{sec-isospinact}
G-parity commutes with isospin rotations around the $y$-axis by construction. This leads to a very useful identity for the quark propagators: The action of such a rotation by $\pi$ radians on the quark fields is given in Eq.~\eqref{eq-yisospinrotq}. In our G-parity notation the fields $\psi$ and $\psibar$ therefore transform as
\begin{dgroup}
\begin{dmath}
e^{-\hat I_y \pi} \psi e^{\hat I_y \pi} = (i\sigma_2)C\psibar^T
\end{dmath}
\begin{dmath}
e^{-\hat I_y \pi} \psibar e^{\hat I_y \pi} = \psi^T C(-i\sigma_2)\,.
\end{dmath}
\end{dgroup}
We write the path integral in which we have integrated over only the fermion fields on a fixed gauge background as
\begin{dmath}
\langle \cdot \rangle_\psi = \int [d\psi][d\psibar]\,(\cdot)\, \exp\left( -S[\psi,\psibar,\tilde U] \right)\,,
\end{dmath}
with which the propagator -- the inverse of the Dirac matrix on a given configuration -- can be written as
\begin{dmath}
[\prop(x,y)]_{\alpha\beta} = \langle \psi_\alpha(x)\psibar_\beta(y) \rangle_{\psi}\,,
\end{dmath}
where $\alpha$,$\beta$ are combined spin/color/flavor indices.

As the action is invariant under isospin rotation we can write
\begin{dmath}
[\prop(x,y)]_{\alpha\beta} = \langle \psi_\alpha(x)\psibar_\beta(y) \rangle_{\psi}
= \langle e^{-\hat I_y \pi} \psi_\alpha(x)\psibar_\beta(y) e^{\hat I_y \pi} \rangle_{\psi}
= \langle (i\sigma_2)C\psibar_\alpha(x) \psi_\beta(y) C(-i\sigma_2) \rangle_{\psi}
= (i\sigma_2)C [\prop(y,x)]_{\beta\alpha} C^{-1}(-i\sigma_2)\label{eq-propconjrelhilbpresc}
\end{dmath}
where we have absorbed the sign from commuting the fields using $C^{-1} = -C$.

$\gamma^5$-hermiticity then implies
\begin{equation}
\prop^*(x,y) = \gamma^5 C^{-1}(i\sigma_2) \prop(x,y) (-i\sigma_2)C \gamma^5\,\label{eqn-propconjreln}\,.
\end{equation}
Examining the flavor structure explicitly we find
\begin{equation}
\fmat{ \prop_{00}(y,z) } { \prop_{01}(y,z) }{ \prop_{10}(y,z) } { \prop_{11}(y,z) }  = \gamma^5 C^{-1}\fmat{ \prop^*_{11}(y,z) } { -\prop_{10}^*(y,z) }{ -\prop_{01}^*(y,z) } { \prop_{00}^*(y,z) } C\gamma^5\,. \label{eqn-propconjrelncpt}
\end{equation}
Notice that this implies the second column of the flavor-matrix propagator (source flavor index 1) can be obtained entirely from the first column (source flavor index 0), hence we can calculate the full propagator using only the matrix computed from sites of a single flavor. We will exploit this property in Section~\ref{sec:Results}.

\subsection{Baryon number}

Since our G-parity boundary conditions change quarks to anti-quarks, baryon number symmetry is violated.  This is dramatically illustrated by the transformation of a proton $(uud)$, with baryon number $B=1$, which becomes an anti-neutron $(\overline{d}\overline{d}\overline{u})$ with baryon number -1 at the boundary.

The baryon number violation has an additional manifestation at the quark level: The mixing of quark flavor at the G-parity boundary allows for the Wick contraction of up and down field operators:
\vspace{-0.3cm}
\begin{equation}
\contraction[0.5ex]{\prop^{(1,0)}_{y,x} = C}{\bar u}{ ^T_y }{\bar d}
\contraction[0.5ex]{\prop^{(1,0)}_{y,x} = C\bar u^T_y \bar d_x \,,\hspace{1cm}\prop^{(0,1)}_{y,x} = -}{d}{_y}{u}
\prop^{(1,0)}_{y,x} = C\bar u^T_y \bar d_x \,,\hspace{1cm}\prop^{(0,1)}_{y,x} = -d_y u^T_x C^T \,.
\vspace{-0.2cm}
\end{equation}
As a result there are typically additional diagrams involving propagators that cross the boundary. In the first of the above contractions, quark flavor flows towards the boundary on both sides. Likewise, quark flavor flows away from the boundary in the second contraction. We may interpret this as the boundary respectively destroying and creating flavor. 

Other than the mixing of the quark flavors at the boundary which we handle explicitly in our two-flavor formulation, the breaking of the baryon number symmetry is not important for calculations involving only mesonic states.

\subsection{Flavor non-singlet axial vector transformations}
\label{sec-flavsym}

The action is invariant under the flavor non-singlet vector (isospin) transformations by construction. However for an axial transformation $A = \exp(i\theta^j_A \gamma^5\sigma_j)$, we have
\begin{equation}
q \to Aq \hspace{2cm}{\rm and}\hspace{2cm} \bar q \to \bar q A\,.
\end{equation}
For Wilson/domain wall fermions, the Wilson term in Eq.~\eqref{eqn-lsymactionbndsimpqform} explicitly breaks the axial symmetry, therefore to separate the effects of G-parity we consider na\"{i}ve fermions, for which $\Gamma^\pm_\mu = \pm \gamma_\mu$. For the second and third terms of Eq.~\eqref{eqn-lsymactionbndsimpqform}, we have $A\gamma^\mu = \gamma^\mu A^\dagger$, such that the terms are invariant. For the other two terms we note
\begin{dgroup}
\begin{dmath}
{ A^T = \exp(i\theta^j_A \gamma^5\sigma_j^T) = \sigma_2 A^\dagger\sigma_2\,, } \label{eq-ATtrans}
\end{dmath}
\begin{dmath}
{ A^* = \exp(-i\theta^j_A \gamma^5\sigma_j^*) = \sigma_2 A \sigma_2\,. } \label{eq-Astartrans}
\end{dmath}
\end{dgroup}
Therefore if we write $S_{\rm GPBC} \to S_{\rm GPBC} + \Delta S_{\rm GPBC}$ (where $\Delta S_{\rm GPBC}=0$ would imply invariance of the action), we find
\begin{multline}
\Delta S_{\rm GPBC} = -\sum_x \sum_\mu G_\mu \delta_{x_\mu,L-1} \Big[ \bar q(x)\Gamma^+_\mu U_\mu(x) \big([A^\dagger]^2 - 1\big) (i\sigma_2) C\bar q^T(x+\hat\mu) \\
+ q^T(x)C \Gamma^+_\mu U_\mu^*(x) (i\sigma_2)\big( A^2 - 1 \big) q(x+\hat\mu)
\Big]\,.
\end{multline}
The action is therefore not invariant under the flavor non-singlet axial transformations. 

This explicit breaking of chiral symmetry gives rise to a non-zero chiral condensate even for zero quark mass. This may provide a useful tool when studying finite temperature QCD and was one of the original motivations for the study of G-parity boundary conditions by Wiese~\cite{Wiese:1991ku}. 

For our purposes we are interested only in the low-temperature behavior of the massive theory, where the chiral symmetry is also broken spontaneously by the dynamics and explicitly by the lattice fermion formulation. In this regime the effects of the G-parity boundary conditions enter in two ways: The first are due simply to the different sets of allowed momenta for G-parity even (integer multiples of $2\pi/L$) and odd (odd-integer multiples of $\pi/L$) states. Such effects are straightforward to take into account in our measurements. There are also more subtle effects which, in the context of the low-energy theory of interactive massive pions, enter as a change in the momentum discretization of pion loop diagrams. These effects are exponentially suppressed in $m_\pi L$ according to the Poisson summation formula and are therefore comparable in size to other, more conventional finite volume effects. As a result we need not be concerned that this boundary-induced chiral symmetry breaking will have a significant effect upon our measurements.

\subsection{Parity}

It is convenient to define the parity transformation $P$ acting on the points in our finite lattice thus: $x_i \rightarrow x^P_i = L-1-x_i$ for $i$ in spatial directions, such that the lattice coordinates are inverted: $0,1,2\ldots (L-1) \to (L-1)\ldots 2,1,0$. This is equivalent to reflecting about the midpoint $(L-1)/2$ of each spatial direction.

Under parity, the gauge links on a standard periodic lattice transform as
\begin{dmath}
\hat {\cal P} U_\mu(x) \hat {\cal P}^{-1} = U_{P(\mu)}(x^P)
\end{dmath}
where $P(i) = -i$ for spatial directions $i=1,2,3$ and $P(4) = 4$ for the time direction, and
\begin{dmath}
U_{-\mu}(x) = U_\mu^\dagger(x-\hat\mu)\,.\label{eq_Ummu}
\end{dmath}

The flavor-matrix gauge links $\tilde U_\mu = {\rm diag}(U_\mu, U^*_\mu)$ therefore transform as
\begin{dmath}
\hat {\cal P} \tilde U_\mu(x) \hat {\cal P}^{-1} = \tilde U_{P(\mu)}(x^P)\,,\label{eqn-paritylinktrans}
\end{dmath}
where the analog to the right-hand side of Eq.~\eqref{eq_Ummu} should take into account the non-trivial boundary condition on the links. To determine the appropriate form it is convenient to rewrite this equation in terms of the translation operators, then use Eq.~\eqref{eq-transbackU}:
\begin{dmath}
{ \tilde U_{-\mu}(x) = \hat T_\mu^{-1}\tilde U_\mu^\dagger(x) \hat T_\mu = B_\mu^-(x_\mu) \tilde U_\mu^\dagger(x-\hat\mu)[B_\mu^-(x_\mu)]^\dagger\,. }
\end{dmath}

We will also require the corresponding action of parity on $\tilde U_i(x-\hat i)$ for a spatial direction $i$, which can be found using $\hat {\cal P}\tilde U_\mu(x)\tilde U^\dagger_\mu(x) \hat {\cal P}^{-1}=1$ and Eq.~\eqref{eqn-paritylinktrans}, resulting in
\begin{dmath}
\hat {\cal P}\tilde U_i(x) \hat {\cal P}^{-1} = B_i^-(x^P_i) \tilde U_i(x^P-\hat i)[B_i^-(x^P_i)]^\dagger
\end{dmath}
and thus
\begin{dmath}
\hat {\cal P}\tilde U_i(x-\hat i) \hat {\cal P}^{-1} = B_i^-(x^P_i+1) \tilde U_i(x^P)[B_i^-(x^P_i+1)]^\dagger = [B_i^+(x^P_i)]^\dagger \tilde U_i(x^P)B_i^+(x^P_i)\,.
\end{dmath}

The action of parity on the down quark field $d=\psi_0$ is:
\begin{dmath}
{ \hat {\cal P} d(x) \hat {\cal P}^{-1} = \gamma^4 d(x^P)\hspace{1cm}{\rm and}\hspace{1cm}\hat {\cal P} \bar d(x) \hat {\cal P}^{-1} = \bar d(x^P)\gamma^4\,. }
\end{dmath}
For the second component, $\psi_1 = C\bar u^T$, we have
\begin{dmath}
{ \hat {\cal P} C \bar u^T(x) \hat {\cal P}^{-1} = C [\hat {\cal P}\bar u \hat {\cal P}^{-1}]^T
                                 = -\gamma^4 C \bar u^T(x^P)\,. }
\end{dmath}
The action of parity upon the full G-parity fermion doublet is therefore
\begin{dmath}
\hat {\cal P} \psi(x) \hat {\cal P}^{-1} = \gamma^4 \sigma_3 \psi(x^P)\,.
\end{dmath}

For determining the transformation properties of the action under parity it is more convenient to return to the full action, Eq.~\eqref{eqn-lsymaction}. Applying the parity transformation the first term of the action in the spatial direction $i$:
\begin{dmath}
\hat {\cal P} \psibar(x)\tilde U_i(x)\Gamma^+_i B^+_i(x_i)\psi(x+\hat i) \hat {\cal P}^{-1}
= \psibar(x^P)\gamma^4\sigma_3 \tilde U_{-i}(x^P)\Gamma^+_i B^+_i(x_i) \gamma^4\sigma_3\psi([x+\hat i]^P)
%
%
%
%
= \psibar(x^P) B^-_i(x^P)\tilde U^\dagger_i(x^P-\hat i)\Gamma^-_i \psi(x^P-\hat i)\,. \label{eqn-parproofSterm1}
\end{dmath}
%
%

Similarly, the second term transforms as
\begin{dmath}
\hat {\cal P} \psibar(x) B^-_i(x_i)\tilde U^\dagger_i(x-\hat i)\Gamma^-_i \psi(x-\hat i) \hat {\cal P}^{-1}
%
%
= \psibar(x^P) \tilde U^\dagger_i(x^P) \Gamma^+_i B^+_i(x_i^P)\psi(x^P+\hat i)\,.
\end{dmath}
These match the second and first terms of Eq.~\eqref{eqn-lsymaction} respectively, written in terms of the transformed coordinates. The temporal components are trivially invariant because $\Gamma^\pm_4$ commutes with $\gamma^4$ and $B^\pm_4 \equiv 1$. We therefore see that the action is invariant under the parity transformation.

In the above we have used
\begin{dmath}
\sigma_3 B_\mu^\pm(x_\mu) \sigma_3 = [B_\mu^\pm(x_\mu)]^\dagger
\end{dmath}
and
\begin{dmath}
B_\mu^\pm(x_\mu) = [B_\mu^\mp(x^P_\mu)]^\dagger 
\end{dmath}
which can be seen from $\delta_{x_i,L-1} = \delta_{0,L-x_i-1} = \delta_{x^P_i,0}$ and $\delta_{x_i,0} = \delta_{-x_i,0} = \delta_{L-x_i-1,L-1} = \delta_{x^P_i, L-1}$ for spatial $i$. (For $\mu=4$, $B_4^\pm$ are unit matrices hence this relation is trivially applicable.) We also recognize that for any flavor matrix $F$,
\begin{dmath}
[B_i^\pm(x_i)]^\dagger F [B_i^\pm(x_i)]^2
=
[B_i^\pm(x_i)]^\dagger [B_i^\pm(x_i)]^2 F 
= B_i^\pm(x_i)F\,,
\end{dmath}
which follows from the fact that $B^\pm_\mu$ are unitary and are either 1 or $\pm i\sigma_2$ depending on the coordinate, and that $(\pm i\sigma_2)^2 = -1$.

\subsection{Translational symmetry}
\label{sec-transsymact}

Because the gauge and fermion actions are constructed as a sum of local terms containing differences between neighboring sites which are obtained using the translation operators $\hat T_\mu$, we can write the complete
action as the sum:
\begin{equation}
S+S_W = \sum_{n_1..n_4=0}^{L-1}
      \prod_\nu\Bigl[\hat T_\nu^{n_\nu}\Bigr]
                        \Bigl[s(0)+s_W(0)\Bigr] \prod_\rho\Bigl[\hat
T_\rho^{-n_\rho}\Bigr]
\label{eq-action_rewritten}
\end{equation}
where $s(0)$ and $s_W(0)$ are the terms in the fermion and gauge actions corresponding to the point $x=0$.  Since the operator $\hat T_\mu\,^L$ is a symmetry of both the fermion and gauge actions, the summand in
Eq.~\eqref{eq-action_rewritten} depends on the integer summation variables $n_\mu$ only through $(n_\mu \mod L)$.  This implies that if we conjugate $S+S_W$ with a translation operator $\hat T_\kappa$ we will increase the summation variable $n_\kappa$ by one which simply permutes the terms in the sum and leaves the complete action unchanged.

\subsection{Translational covariance of field operators}
\label{sec-transsym}

In Eqs.~\eqref{eq-transpsifw} and~\eqref{eq-transpsibw} we observe that the na\"ive translational {\it covariance} of the quark field is broken at the boundary, where it picks up an additional matrix structure $\pm i\sigma_2$. This implies that the quantity
\begin{dmath}
\psi(\vec p,t) = \sum_{\vec x} e^{-i\vec p\cdot \vec x} \psi(\vec x,t)
\end{dmath}
is not an not an eigenstate of translation, i.e.
\begin{dmath}
\hat T_\mu\psi(\vec p,t)\hat T_\mu^{-1} \neq e^{i p_\mu}\psi(\vec p,t)
\end{dmath}
for a G-parity direction $\mu$. This is problematic as we ultimately wish to construct states of definite momentum.

We can easily form eigenstates of $i\sigma_2$ that simply pick up coefficients at the boundaries:
\begin{dmath}
\psi_\pm = \half(1\pm \sigma_2)\psi\,, \label{eq-quarkmomeigstates}
\end{dmath}
for which
\begin{dmath}
(i\sigma_2)\psi_\pm = \pm i\psi_\pm\,. 
\end{dmath}
The factor of $\half$ in Eq.~\eqref{eq-quarkmomeigstates} is arbitrary; we choose it such that $\psi$ can be interpreted as the sum of the two eigenvectors. The fields $\psi_\pm$ have the following boundary conditions: 
\begin{dgroup}
\begin{dmath}
{ \hat T_\mu \psi_\pm(x_\mu = L-1) \hat T_\mu^{-1} = \pm i \psi_\pm(x_\mu = 0) }
\end{dmath}
\begin{dmath}
{ \hat T^{-1}_\mu \psi_\pm(x_\mu = 0) \hat T_\mu = \mp i \psi_\pm(x_\mu = L-1) }
\end{dmath}
\end{dgroup}
where $\mu$ is a G-parity direction. 

The allowed discretized momenta for $\psi_\pm$ can be found by translating the momentum-space field (here we use a theory with one spatial dimension for convenience):

\begin{dmath}
\hat T \psi_\pm(p,t) \hat T^{-1} = \hat T\left(\sum_x e^{-i p\cdot x} \psi_\pm(x,t)\right)\hat T^{-1}
= { \left(\sum_{x \neq L-1} e^{-i p\cdot x}\psi_\pm(x + 1,t)\right) \pm i  e^{-i p\cdot (L-1)}\psi_\pm(0) }
= { e^{ip}\left(\sum_{x' \neq 0} e^{-i p\cdot x'}\psi_\pm(x',t)\right) \pm i e^{ip} e^{-i p\cdot L}\psi_\pm(0) }
\end{dmath}
where $x'=x+1$. These fields are therefore translationally covariant if 
\begin{equation}
\begin{array}{lclr}
ie^{-ipL} = 1 &\Rightarrow & p = \frac{\pi}{2L}(1+4n) & {\rm for}\ \psi_+ \\
-ie^{-ipL} = 1 & \Rightarrow & p = -\frac{\pi}{2L}(1+4n) &{\rm for}\ \psi_- 
\end{array}
\end{equation}
where $n$ is an arbitrary integer. The eigenvectors therefore have disjoint sets of allowed discretized momenta: For $\psi_+$ the allowed momenta are $p= \left(\ldots -7,-3,1,5,9 \ldots\right) \pi/(2L)$, whereas for $\psi_-$ they are $p= \left(\ldots -9,-5,-1,3,7 \ldots\right) \pi/(2L)$. 

For later use it is convenient to combine the eigenvectors into a single field operator whose projection is a function of momentum:
\begin{dmath}
\tilde\psi(\vec p,t) = \sum_{\vec x}  e^{-i\vec p\cdot \vec x} \prod_{j\in {\cal G}}\left[ \half \left(1 + e^{in_{p_j}\pi}\sigma_2\right)\right] \psi(\vec x,t)\,,\label{eqn-transconvfieldndir}
\end{dmath}
and
\begin{dmath}
\overline{\tilde\psi}(\vec p,t) = \sum_{\vec x} e^{-i\vec p\cdot \vec x} \psibar(\vec x,t)\prod_{j\in {\cal G}}\left[ \half \left(1 - e^{in_{p_j}\pi}\sigma_2\right)\right]\,,\label{eqn-transconvbarfieldndir}
\end{dmath}
where ${\cal G}$ is the set of directions with GPBC and $n_{p_j}$ is an integer defined via
\begin{dmath}
p_j= \frac{\pi}{2L}(1+2n_{p_j})\,. \label{eq-pinpi}
\end{dmath}
The allowed momenta for the combined field $\tilde\psi$ are now all odd-integer multiples of $\pi/(2L)$ (Below we discuss a further constraint on the momenta.). Note that the minus sign in the flavor projection of the conjugate field is required because $\psibar \to \psibar(-i\sigma_2)$ when passing through the upper boundary, and hence $\psibar_\pm = \half\psibar(1\pm\sigma_2) \to \mp i \psibar_\pm$. The conjugate field eigenvectors $\psibar_\pm$ therefore have the opposite momentum eigenvalues to $\psi_\pm$: For $\psibar_+$ the allowed momenta are $p= \left(\ldots -9,-5,-1,3,7 \ldots\right)\pi/(2L)$, whereas for $\psibar_-$ they are $p= \left(\ldots -7,-3,1,5,9 \ldots\right)\pi/(2L)$. 

\subsection{Rotational symmetry}
\label{sec-rotsymact}

Examining Eq.~\eqref{eqn-transconvfieldndir} more closely, we observe that a non-zero field operator with a definite momentum can only be created if the flavor projection operators for each momentum direction have the same sign, i.e. $e^{in_{p_i}\pi}$ is the same for all G-parity directions. With GPBC in two directions for example, this then implies
\begin{dmath}
{ n_{p_1} = n_{p_2} + 2m   \Rightarrow p_1 = p_2 + \frac{2m\pi}{L}\,, } \label{eq-npcondition}
\end{dmath}
where $m$ is an integer. In other words, the momentum components are constrained to differ only by integer multiples of $2\pi/L$. We can therefore simplify the expression for the translationally covariant field in Eq.~\eqref{eqn-transconvfieldndir} to
\begin{dmath}
\tilde\psi(\vec p,t) = \sum_{\vec x}  e^{-i\vec p\cdot \vec x} \half \left(1 + e^{in_{p}\pi}\sigma_2\right) \psi(\vec x,t)\,,\label{eqn-transconvfieldndirsimp}
\end{dmath}
where 
\begin{dmath}
{ n_p = n_{p_i} = \frac{Lp_i}{\pi} - \frac{1}{2} } \label{eqn-transcovfieldnpdef}
\end{dmath}
for each $i\in {\cal G}$.

The implications of this observation can be seen by considering the two momenta $\frac{2\pi}{L}(\frac{1}{4},\frac{1}{4},0)$ and $\frac{2\pi}{L}(\frac{1}{4},-\frac{1}{4},0)$, which are related by a cubic rotation. In the former the momentum components are identical, hence this is an allowed momentum. However, in the latter the two momentum components in the G-parity directions differ only by $\pi/L$, and therefore this is not an allowed quark momentum. This implies that the rotational symmetry has been broken at the quark level by GPBC in multiple directions. For example if we imposed GPBC in all three spatial directions in a cubic box, the na\"ive cubic symmetry of this choice is broken by this fixed relation between the three components of the allowed quark momenta.

\begin{figure}[tp]
\centering
\subfigure[$(2A,1P)\equiv(1G,1P)$]{
\includegraphics[width=0.31\textwidth]{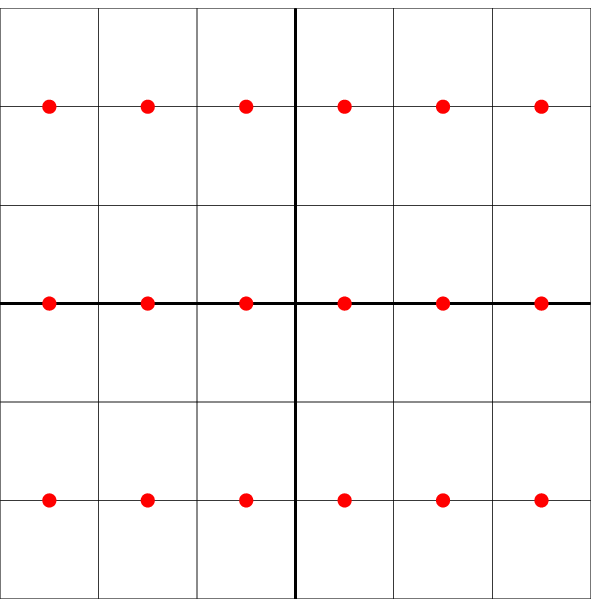}
\label{fig-mom-2A1P}
}
\hfill
\subfigure[$(2A,2A)$]{
\includegraphics[width=0.31\textwidth]{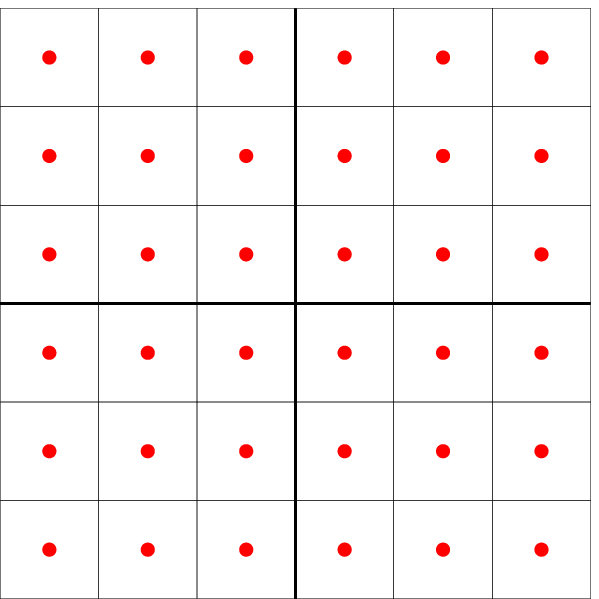}
\label{fig-mom-2A2A}
}
\hfill
\subfigure[$(1G,1G)$]{
\includegraphics[width=0.31\textwidth]{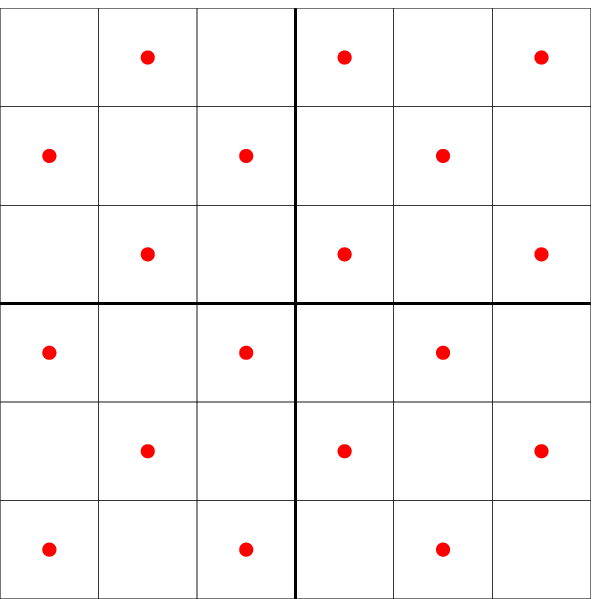}
\label{fig-mom-1G1G}
}

\caption{Allowed momenta for several two-dimensional lattice configurations. The grid spacing is $\pi/L$ and the origin is at the center of each figure. For each figure, the pair of labels in parentheses gives the lattice size in units of $L$ and the quark boundary condition for the $x$ and $y$ directions, respectively: $A$ for antiperiodic, $P$ for periodic and $G$ for G-parity.}
\end{figure}

The breaking of the rotational symmetry can be shown in a different manner by considering the structure of the Brillouin lattice: In Section~\ref{sec-1fequiv} we demonstrated that the two-flavor theory with GPBC in one direction of size $L$ is equivalent to a single-flavor theory on a lattice of size $2L$ with the quarks obeying antiperiodic boundary conditions. In Figure~\ref{fig-mom-2A1P} we plot the allowed momenta for this doubled-lattice setup. In Figure~\ref{fig-mom-2A2A} we plot those obtained if we double the lattice again and impose antiperiodic BCs in the second direction. In each case the number of points corresponds to the number of fermionic degrees of freedom, which is double in the latter compared to the former and is therefore {\it not} equivalent to GPBC in two-directions, for which the number of degrees of freedom is independent of the number of G-parity directions. (In Section~\ref{sec-1fequiv} this binding of field variables between the quadrants required us to carefully construct propagator sources in order to correctly reproduce the two-flavor theory.) We must therefore eliminate half of these points to reproduce the two-flavor theory.

Given that parity symmetry demands the Brillouin lattice be symmetric about the diagonals, and that the sites are equally spaced along both axes, there are only two possible configurations for the remaining Brillouin lattice sites: The first has all of the momentum points distributed along the positive diagonal as we plot in Figure~\ref{fig-mom-1G1G}, and matches the allowed momenta we identified above. The second is that in which the momentum sites reside in the center of the currently-unoccupied grid squares of this figure such that they lie instead along the negative diagonal. Note that the diagram is not invariant under $90^\circ$ rotations, thus showing that the cubic symmetry is broken.

It is interesting to consider why the allowed momenta lie along the positive and not the negative diagonal. The reason is due to our conventions: As mentioned in Section~\ref{sec-gponquarks}, the direction around the y-axis that we perform the isospin rotation is arbitrary. We choose to perform the rotation in the anticlockwise direction at every upper boundary of the lattice, which results in the favored direction being along the positive diagonal; we could just as easily have chosen to perform the rotation in the opposite direction at one or more boundaries, resulting in a change in the favored direction.

It is important to recognize that the choice of convention does not affect the boundary conditions on the pion wavefunction, which obeys antiperiodic boundary conditions in all cases. As a result, the pion energy remains invariant under cubic rotations, as we will demonstrate numerically in Section~\ref{sec-rotsymbreakmeas}. We do however find that the rotational symmetry breaking has a measurable impact on the amplitudes $A_{\pi}(\vec p) = |\langle \pi(\vec p)|{\cal O}_\pi(\vec p)|0\rangle|^2$ of two-point functions of pions moving in orthogonal directions, where ${\cal O}_\pi(\vec p)$ is a bilinear operator. Such an operator can be constructed as
\begin{dmath}
{\cal O}^\pm_\pi(\vec p) = \psibar_\mp(\vec p_1) \ldots \psi_\pm(\vec p_2)
\end{dmath}
where $\vec p_1 + \vec p_2 = \vec p$ and the ellipses denote spin and flavor structure that will be elaborated in the coming section. Note that both the $\psibar_-\psi_+$ and $\psibar_+\psi_-$ forms can be used providing $\vec p_1$ and $\vec p_2$ are chosen from the appropriate set of allowed momenta. In the aforementioned section we determine that the discrepancy in amplitudes can be substantially reduced by averaging the ${\cal O}^+_\pi(\vec p)$ and ${\cal O}^-_\pi(\vec p)$ forms. This observation is vital to constructing s-wave $\pi\pi$ states in our $K\to\pi\pi$ calculation.

\section{Light hadronic observables}
\label{sec:LightHadronic}

\subsection{Local light-quark bilinear operators}
\label{sec-lightmesonop}
As a result of the flavor mixing, many typical hadronic states (for example the proton) are no longer eigenstates of the Hamiltonian. For this work and in the measurement of the $K\rightarrow(\pi\pi)_{I=0}$ amplitude we are only concerned with meson states. On the lattice we form bilinear operators in the quark fields that, when applied to the vaccuum state, create a linear combination of all mesonic states with the quantum numbers specified by the operator. In this section we consider bilinear operators in which both quark fields act at the same space-time point, and in Section~\ref{sec-nonlocalops} we consider operators involving quark fields operating at different positions, for which the forms are further restricted by the requirement that the operators project onto eigenstates of the translation operator (and hence have definite momentum) in the context in which either of the quarks can independently cross the boundary, and hence change flavor.

In anticipation of including the strange quark, we henceforth refer to the fermion doublet comprising the light fields with subscript $l$.

There are three generic forms for a point bilinear operator involving only the light quark fields:
\begin{equation}\begin{array}{rl}
{\cal O}^1_{ll} &= \sum_i a_i \psibar_l \Gamma \sigma_i \psi_l \\
{\cal O}^2_{ll} &= \sum_i b_i \psibar_l \Gamma \sigma_i \psibar^T_l\\
{\cal O}^3_{ll} &= \sum_i c_i \psi^T_l \Gamma \sigma_i \psi_l\label{eqn-genbilop}
\end{array}\end{equation}
where $\Gamma$ is a generic spin-color matrix, $a$--$c$ are c-number coefficients indexed with $i\in\{0..3\}$, and we have exploited the fact that any $2\times 2$ complex matrix can be written as a linear combination of the Pauli matrices and the unit matrix (written here as $\sigma_0$).

From the color structure of the operators in Eq.~\eqref{eqn-genbilop} one can recognize that local, gauge-invariant operators of the $\mathcal{O}^1_{ll}$ form can be obtained only if the flavor matrices $\sigma_0$ or $\sigma_3$ appear, while for $\mathcal{O}^2_{ll}$ and $\mathcal{O}^3_{ll}$ the matrices $\sigma_1$ or $\sigma_2$ must be used.

Using Eq.~\eqref{eqn-gppsi} we can easily see that the G-parity odd operators are those for which the flavor structure anticommutes with $\sigma_2$, i.e. $\sigma_1$ and $\sigma_3$. Similarly, G-parity even operators are those that contain $\sigma_2$ or the unit matrix. In Table~\ref{tab-lllocop} we have compiled a list of local bilinear operators that are invariant under the above transformations and project onto states of definite G-parity eigenvalue. From the table we can easily read off the operators that create states with the quantum numbers of the pion:
\begin{equation}\begin{array}{rrr}
\pi^+ &= i\bar u \gamma^5 d &= \frac{i}{2}\psi_l^T \gamma^5 C \sigma_1 \psi_l\\
\pi^- &= -i\bar d \gamma^5 u &= -\frac{i}{2}\psibar_l \gamma^5 C \sigma_1 \psibar_l^T\\
\pi^0 &= \frac{i}{\sqrt{2}}(\bar u \gamma^5 u - \bar d \gamma^5 d) &= -\frac{i}{\sqrt{2}}\psibar_l \gamma^5 \sigma_3 \psi_l
\end{array}\label{eq-pioncreationopsgparity}\end{equation}

It is illustrative to consider here the Wick contractions of the $\pi^0$ two-point correlation function $\langle \pi^0(x)\pi^0(y) \rangle$. In addition to the usual connected diagram, this will also contain a diagram comprising two disconnected quark loops each of the form
\begin{equation}
{\rm tr}\left( \prop(x,x) \gamma^5 \sigma_3 \right)\,.\label{eq-pion-disconn-factor}
\end{equation}
In the limit of isospin symmetry such diagrams must vanish, which becomes clear if we return to standard notation but is not so obvious in this form. We can see that it vanishes in the notation of Eq.~\eqref{eq-pion-disconn-factor} as follows:
\begin{equation}\begin{array}{rl}
{\rm tr}\left( \prop(x,x) \gamma^5 \sigma_3 \right) & = {\rm tr}\left( \sigma_3^T (\gamma^5)^T \prop^T(x,x)   \right) \\
                                                    & = {\rm tr}\left( \sigma_3 \gamma^5 [\gamma^5\prop^*(x,x)\gamma^5]\right) \\
                                                    & = -{\rm tr}\left( \prop(x,x) \gamma^5 \sigma_3 \right)\,,
\end{array}\end{equation}
where on the second line we have used the $\gamma^5$ Hermiticity, Eq.~\eqref{eq-g5herm}, of the Euclidean propagator and on the third line we have used Eq.~\eqref{eqn-propconjreln}.

\begin{table}[tp]
\centering
\begin{tabular}{c|c|c|c}
\hline\hline
Operator                                  &   Std. Form                              & G-parity e-val \\
\hline
$\psibar_l \Gamma \sigma_0 \psi_l$       &  $\bar d \Gamma d - \bar u C\Gamma^T C u$  & +1            \\
$\psibar_l \Gamma \sigma_3 \psi_l$       &  $\bar d \Gamma d + \bar u C\Gamma^T C u$  & -1            \\
\hline
$\psibar_l \Gamma \sigma_1 \psibar^T_l$ &  $-\bar d (\Gamma - \Gamma^T) C u$         & -1            \\
$\psibar_l \Gamma \sigma_2 \psibar^T_l$ &  $i\bar d (\Gamma + \Gamma^T) C u$         & +1           \\
\hline
$\psi^T_l \Gamma \sigma_1 \psi_l$         &  $-\bar u C (\Gamma - \Gamma^T) d$         & -1           \\
$\psi^T_l \Gamma \sigma_2 \psi_l$         &  $-i\bar u C (\Gamma + \Gamma^T) d$        & +1            \\
\end{tabular}
\caption{The non-zero local gauge-invariant light-quark bilinear operators and their G-parity eigenvalues written is the two-flavor formalism as well as the standard form. Here $\Gamma$ is a generic spin-color matrix.\label{tab-lllocop}.}
\end{table}

\subsection{Non-local bilinear operators}
\label{sec-nonlocalops}

In this section we restrict our attention to operators of the form ${\cal O}^1_{ll}$, although the concepts can be easily generalized to the other forms. 

In practice we typically achieve better overlap with a chosen state of a particular momentum $\vec p+\vec q$ using spatially smeared operators of the form
\begin{dmath}
{\cal O}(\vec p+\vec q,t) = \sum_{\vec x,\vec y} e^{-i(\vec p\cdot\vec x + \vec q\cdot\vec y)} \phi(|\vec x-\vec y|) \psibar(\vec x,t)\Gamma\Sigma \psi(\vec y,t)\,,\label{eq-smearedbilop}
\end{dmath}
where $\phi$ is some smearing function and $\Gamma$ and $\Sigma$ are arbitrary spin and flavor matrices respectively. Here and below it is assumed that the spatial links at the time $t$ are gauge-fixed such that this operator is gauge invariant. 

To be useful, an operator $\mathcal{O}(\vec p+\vec q,t)$ specified by Eq.~\eqref{eq-smearedbilop} must add the momentum $\vec p + \vec q$ to the state to which it is applied.  That is, it must be an eigenstate of the translation operator $\hat T_j$ in the spatial $j^{th}$ direction defined in Eq.~\eqref{eq-transpsiboth}:
\begin{equation}
\hat T_j \mathcal{O}(\vec p+\vec q,t) \hat T_j^{-1} = e^{i(p_j+q_j)}\mathcal{O}(\vec p+\vec q,t).
\end{equation}
We will discuss how this can be done in the next two subsections.

\subsubsection{Point operator}
Let us first consider the special case of a local operator: $\phi(|\vec x-\vec y|) = \delta_{\vec x,\vec y}$. Under a spatial translation this becomes
\begin{dmath}
\hat T_j{\cal O}(\vec p+\vec q,t)\hat T_j^{-1} = \sum_{\vec x} e^{-i(\vec p+\vec q)\cdot\vec x}  \psibar(\vec x+\hat j,t)[B_j^+(x_j)]^\dagger\Gamma\Sigma B_j^+(x_j)\psi(\vec x+\hat j,t)
= { e^{i(p_j+q_j)}\sum_{\vec x'} e^{-i(\vec p+\vec q)\cdot\vec x'} e^{-i(p_j+q_j)L \delta_{x'_j,0}}\;\psibar(\vec x',t)[B_j^+(x'_j-1)]^\dagger\Gamma\Sigma B_j^+(x'_j-1)\psi(\vec x',t) }\,.
\end{dmath}
where $\vec x' = \vec x+\hat j$ modulo the lattice size. We see that for $\Sigma = \sigma_0$, $B_j^+$ commutes with $\Sigma$ and the above is translationally covariant providing $p_j+q_j = n\frac{2\pi}{L}$ for integer $n$; this is just the operator we identified as being a G-parity even eigenstate in the previous section. Similarly for $\Sigma =\sigma_3$, $B_\mu^+$ anticommutes and we require $p_j+q_j = (n+\frac{1}{2})\frac{2\pi}{L}$; this corresponds to the G-parity odd operator. As a result, providing the correct momenta are chosen, the local operator is an eigenstate of the translation operator.

\subsubsection{Non-local operator}

For a non-local operator, there are additional terms where only one quark crosses the boundary that render the operator non-translationally covariant: the boundary term enters for one quark but not the other. In Section~\ref{sec-transsym} we identified quark field operators, $\tilde\psi(\vec p,t)$, that are explicitly translationally covariant by virtue of their flavor-projector structure. Inserting these into Eq.~\eqref{eq-smearedbilop}, we obtain an operator that has a well defined momentum:
\begin{dmath}
\tilde{\cal O}(\vec p+\vec q,t) = \sum_{\vec x,\vec y} e^{-i(\vec p\cdot\vec x + \vec q\cdot\vec y)} \phi(|\vec x-\vec y|) \overline{\tilde\psi}(\vec x,t)\Gamma\Sigma \tilde\psi(\vec y,t)
= \sum_{\vec x,\vec y} e^{-i(\vec p\cdot\vec x + \vec q\cdot\vec y)} \phi(|\vec x-\vec y|) \psibar(\vec x,t)\half(1-e^{in_p\pi}\sigma_2)\Gamma\Sigma \half(1+e^{in_q\pi}\sigma_2)\psi(\vec y,t)
\end{dmath}
where the integers $n_p$ and $n_q$ are associated with the momenta $\vec p$ and $\vec q$ respectively, via Eq.~\eqref{eqn-transcovfieldnpdef}.

As $B_j^\pm$ commute with $\sigma_2$, the restrictions on the allowed momenta for different choices of $\Sigma$ are the same as for the local operator. The $\half(1\pm \sigma_2)$ projection operators enforce these same restrictions: For $\Sigma = \sigma_3$, on commuting the first projector through $\Sigma$ we find the projectors will cancel unless $e^{in_p\pi} = e^{in_q\pi}$ and thus $n_q = n_p + 2m$ for integer $m$. This implies $p_i = q_i + 2m\pi/L$ in G-parity directions, and thus
\begin{dmath}
p_i + q_i = 2q_i + \frac{2m\pi}{L} 
= \frac{\pi}{L} + (n_q+m)\frac{2\pi}{L}\,, 
\end{dmath}
i.e. this just restricts the total momentum to odd integer multiples of $\pi/L$, the allowed momenta of G-parity odd states. The corresponding condition for $\Sigma=\sigma_0$ requires $-e^{in_p\pi} = e^{in_q\pi}$ and so $n_q = n_p + (2m+1)$, which in turn implies $p_i + q_i = \frac{2\pi}{L}(n_p + m + 1)$, the allowed momenta of G-parity even states.

\subsubsection{Non-local neutral pion operator}

For a neutral pion the local pseudoscalar operator is $\psibar \gamma^5 \sigma_3 \psi$ and the corresponding non-local operator has the form:
\begin{dmath}
\tilde{\cal O}_\pi(\vec p+\vec q,t) = \sum_{\vec x,\vec y} e^{-i(\vec p\cdot\vec x + \vec q\cdot\vec y)} \phi(|\vec x-\vec y|) \psibar(\vec x,t)\half(1-e^{in_p\pi}\sigma_2)\gamma^5\sigma_3 \half(1+e^{in_q\pi}\sigma_2)\psi(\vec y,t)\,. \label{eq-nonlocalpi0op}
\end{dmath}
where translational covariance and the projection operators restrict the total momentum in each G-parity direction $j\in {\cal G}$ to $p_j + q_j = (2m+1)\frac{\pi}{L}$ for integer $m$ as desired.

Let us examine the quark content of this operator more carefully: Expanding the parentheses in Eq.~\eqref{eq-nonlocalpi0op} we find two independent flavor structures,
\begin{dmath}
\psibar(\vec x,t)(1+e^{in_p\pi}e^{in_q\pi})\gamma^5\sigma_3\psi(\vec y,t) = 2\psibar(\vec x,t)\gamma^5\sigma_3\psi(\vec y,t)\,,
\end{dmath}
and
\begin{dmath}
-i\psibar(\vec x,t)(e^{in_p\pi}+e^{in_q\pi})\gamma^5\sigma_1\psi(\vec y,t) = -2ie^{in_p\pi}\psibar(\vec x,t)\gamma^5\sigma_1\psi(\vec y,t)\,.
\end{dmath}
The second has unphysical flavor structure, $\psibar\gamma^5\sigma_1\psi = \bar d \gamma^5 C\bar u^T + u^T C\gamma^5 d$, and arises in cases where one of the quark fields crosses the boundary when the other does not. Na\"{i}vely this looks incorrect, but the form of the second term is, in a sense, artificial; given the translational invariance of the action, we are free to shift the boundary such that, for example, in one dimension,
\begin{dmath}
\left\langle \psibar(x)\gamma^5\sigma_1\psi(0) \right\rangle = \left\langle \psibar(x-1)\gamma^5\sigma_1 (-i\sigma_2)\psi(L-1) \right\rangle = \left\langle \psibar(x-1)\gamma^5\sigma_3\psi(L-1) \right\rangle\,,
\end{dmath}
where $\left\langle\right\rangle$ stands for the ensemble average. Thus the apparently unphysical form of the second term is merely an artifact of our arbitrary choice as to where to place the boundary.

\section{The Strange Quark}
\label{sec:StrangeQuark}

We wish to simulate with a single strange quark whose discretized action is consistent with the charge conjugation boundary conditions on the gauge fields. The most obvious choice is to also impose charge conjugation boundary conditions on the strange quark, i.e. on crossing the boundary
\begin{dmath}
{ s \rightarrow C\bar s^T\ \ {\rm and}\ \ \bar s \to s^T C\,. }
\end{dmath}
Unfortunately, with this choice of strange quark boundary condition it is impossible to form a pseudoscalar operator that projects onto the $K^0$ state and that is invariant under translations such that it will create a $K$ meson at rest. To see this, consider the two operators, $\bar s \gamma^5 d$ and $\bar u \gamma^5 s$, that transform into each other under the boundary conditions:
\begin{dgroup}
\begin{dmath}
{ \bar s \gamma^5 d \rightarrow (s^T C)\gamma^5 (C\bar u^T) = \bar u \gamma^5 s\,, }
\end{dmath}
\begin{dmath}
{ \bar u \gamma^5 s \to (-d^T C) \gamma^5 (C\bar s^T) = -\bar s \gamma^5 d\,. }
\end{dmath}
\end{dgroup}
A linear combination of these operators cannot be chosen that is invariant under translation as, under G-parity,
\begin{dmath}
\alpha \bar s \gamma^5 d + \beta \bar u \gamma^5 s \to \alpha \bar u \gamma^5 s - \beta \bar s \gamma^5 d\,,
\end{dmath}
which would simultaneously require $\alpha=\beta$ and $\alpha=-\beta$. As a result we cannot create a kaon-like state of zero momentum. This is ultimately related to the fact that the strange quark with charge conjugation BCs is periodic in $2L$, whereas the up/down quarks with GPBCs are {\it anti-}periodic in $2L$ and periodic in $4L$.

A solution is to introduce a fictional degenerate partner to the strange quark, which we refer to as the $s'$ quark, and to impose GPBC within that pair:
\begin{equation}
\hat G\left(\begin{array}{c}s'\\s\end{array}\right)\hat G^{-1} = \left(\begin{array}{c}-C\bar s^T\\C\bar s^{\prime\,T}\end{array}\right)\,.
\end{equation}
Then
\begin{dgroup}
\begin{dmath}
{ \bar s \gamma^5 d \rightarrow (s^{\prime\,T} C)\gamma^5 (C\bar u^T) = \bar u \gamma^5 s'\,, }
\end{dmath}
\begin{dmath}
{ \bar u \gamma^5 s' \rightarrow (-d^T C)\gamma^5 (-C\bar s^T) = \bar s \gamma^5 d\,, }
\end{dmath}
\end{dgroup}
and we can form eigenstates,
\begin{dmath}
\bar s \gamma^5 d \pm \bar u \gamma^5 s'\,,
\end{dmath}
with eigenvalue $\pm 1$, i.e. that obey either periodic or antiperiodic BCs. The former can be used to produce a stationary state whose physical component projects onto the neutral kaon and is therefore suitable for a $K\rightarrow\pi\pi$ calculation. 

\subsection{Local heavy-light bilinear operators}
\label{sec-hlbilinears}
\begin{table}[tp]
\centering
\begin{tabular}{c|c|c|c}
\hline\hline
Operator                                  &   Std. Form                              & G-parity e-val \\
\hline
$\psibar_h \Gamma \sigma_0 \psi_l$       &  $\bar s \Gamma d - \bar u C\Gamma^T C s'$  & +1            \\
$\psibar_h \Gamma \sigma_3 \psi_l$       &  $\bar s \Gamma d + \bar u C\Gamma^T C s'$  & -1            \\
\hline
$\psibar_h \Gamma \sigma_1 \psibar^T_l$ &  $-\bar s\Gamma C u + \bar d \Gamma^T C s'$ & -1             \\
$\psibar_h \Gamma \sigma_2 \psibar^T_l$ &  $i\bar s\Gamma C u + i\bar d \Gamma^T C s'$ & +1             \\
\hline
$\psi^T_h \Gamma \sigma_1 \psi_l$         &  $-\bar s' C \Gamma d + \bar u C \Gamma^T s$  & -1           \\
$\psi^T_h \Gamma \sigma_2 \psi_l$         &  $-i\bar s'C\Gamma d -i \bar u C \Gamma^T s$  & +1            \\
\hline
$\psibar_l \Gamma \sigma_0 \psi_h$       &  $\bar d \Gamma s - \bar s' C\Gamma^T C u$  & +1            \\
$\psibar_l \Gamma \sigma_3 \psi_h$       &  $\bar d \Gamma s + \bar s' C\Gamma^T C u$  & -1            \\
\end{tabular}
\caption{The local gauge-invariant heavy-light bilinear operators and their G-parity eigenvalues written is the two-flavor formalism as well as the standard form. Here $\Gamma$ is a generic spin-color matrix.\label{tab-hllocop}.}
\end{table}

In this section we consider local bilinear operators containing the strange quark and its fictional partner, $s'$. As with the light quarks, we write
\begin{dmath}
\psi_h = \left(\begin{array}{c}s \\ C\bar s^{\prime\,T}\end{array}\right)\,,
\end{dmath}
which is distinguished from the light-quark flavor doublet by the subscript $h$. 

Below we use the following conventions for operators that create states with the quantum numbers of the kaon:
\begin{equation}\begin{array}{rr}
K^+ &= i\bar u\gamma^5 s\\
K^- &= -i\bar s\gamma^5 u \\
K^0 &= i\bar d\gamma^5 s \\
\bar K^0 &= -i\bar s\gamma^5 d\,,
\label{eq-stdKdefs}
\end{array}\end{equation}
These transform under G-parity as follows:
\begin{equation}\begin{array}{rr}
\hat G K^+ \hat G^{-1} &= i(-d^T C)\gamma^5 (C\bar s^{\prime\,T})= -i \bar s' \gamma^5 d  \equiv K^{\prime\,+}\\
\hat G K^- \hat G^{-1} &= -i(s^{\prime\,T}C)\gamma^5 (-C\bar d^T) = +i \bar d \gamma^5 s' \equiv K^{\prime\,-} \\
\hat G K^0 \hat G^{-1} &= i(u^T C)\gamma^5 (C\bar s^{\prime\,T}) = +i \bar s' \gamma^5 u \equiv  K^{\prime\,0}\\
\hat G \bar K^0 \hat G^{-1} &= -i(s^{\prime\,T}C)\gamma^5 (C\bar u^T) = -i\bar u \gamma^5 s' \equiv \bar K^{\prime\,0}\,.
\label{eq-GPKdefs}
\end{array}\end{equation}
where we have denoted the fictional G-parity partners to the physical kaons with a prime ($'$) superscript.

For operators comprising only the heavy quark fields, the results obtained in Section~\ref{sec-lightmesonop} also apply. For heavy-light bilinears we have four operator forms:
\begin{equation}\begin{array}{rl}
{\cal O}^1_{hl} &= \sum_i d_i \psibar_h \Gamma \sigma_i \psi_l\\
{\cal O}^2_{hl} &= \sum_i e_i \psibar_h \Gamma \sigma_i \psibar^T_l \\
{\cal O}^3_{hl} &= \sum_i f_i \psi^T_h \Gamma \sigma_i \psi_l \\
{\cal O}^4_{hl} &= \sum_i g_i \psibar_l \Gamma \sigma_i \psi_h\,.
\label{eqn-genbilophl}
\end{array}\end{equation}
As before, gauge invariance restricts ${\cal O}^1_{hl}$ and ${\cal O}^4_{hl}$ to the choices $\sigma_0$ and $\sigma_3$, and likewise ${\cal O}^2_{hl}$ and ${\cal O}^3_{hl}$ are restricted to $\sigma_1$ and $\sigma_2$. However, here the lack of symmetry under interchange of the fields implies that there is no restriction on the spin-color matrices $\Gamma$. We have again compiled the operators along with their standard forms and G-parity eigenvalues in Table~\ref{tab-hllocop}.

Consider the first line of Table~\ref{tab-hllocop} with $\Gamma = \gamma^5$. We have
\begin{dmath}
\psibar_h \Gamma \sigma_0 \psi_l = (\bar s \gamma^5 d + \bar u \gamma^5 s') = i(\bar K^0 + \bar K^{\prime\,0}) \,.
\end{dmath}
This operator creates a stationary state whose physical component corresponds to the $\bar K^0$. Similarly the equivalent operator for the $K^+$ can be obtained from the sixth line of the table with $\Gamma = C\gamma^5$:
\begin{dmath}
\psi^T_h C\gamma^5 \sigma_2 \psi_l = (i\bar s'\gamma^5 d -i \bar u \gamma^5 s) = -(K^{\prime\,+} + K^+)\,.
\end{dmath}
From the table we can therefore read off the operators that project onto stationary states whose physical components correspond to the full set of charged and neutral kaons. We denote operators of definite G-parity quantum number with a tilde ($\sim$), and display the quantum number in the subscript:
\begin{equation}\begin{array}{rrr}
\tilde K^+_+ &= \frac{1}{\sqrt{2}}(K^+ + K^{\prime\,+}) = \frac{i}{\sqrt{2}}(\bar u \gamma^5 s - \bar s'\gamma^5 d) &= -\frac{1}{\sqrt{2}}\psi^T_h  C\gamma^5\sigma_2\psi_l\\
\tilde K^-_+ &= \frac{1}{\sqrt{2}}(K^- + K^{\prime\,-}) = \frac{-i}{\sqrt{2}}(\bar s \gamma^5 u - \bar d \gamma^5 s') &= \frac{1}{\sqrt{2}}\psibar_h  C\gamma^5\sigma_2\psibar_l^T\\
\tilde K^0_+ &= \frac{1}{\sqrt{2}}(K^0 + K^{\prime\,0}) = \frac{i}{\sqrt{2}}(\bar d \gamma^5 s + \bar s' \gamma^5 u) &= \frac{i}{\sqrt{2}}\psibar_l \gamma^5\psi_h\\
\bar {\tilde K}^0_+ &= \frac{1}{\sqrt{2}}(\bar K^0 + \bar K^{\prime\,0}) = \frac{-i}{\sqrt{2}}(\bar s \gamma^5 d + \bar u \gamma^5 s') &= \frac{-i}{\sqrt{2}}\psibar_h \gamma^5\psi_l
\end{array}\,.\end{equation}
We also have moving states with the opposite G-parity quantum number:
\begin{equation}\begin{array}{rrr}
\tilde K^+_- &= \frac{1}{\sqrt{2}}(K^+ - K^{\prime\,+})  &= \frac{i}{\sqrt{2}}\psi^T_h  C\gamma^5\sigma_1\psi_l\\
\tilde K^-_- &= \frac{1}{\sqrt{2}}(K^- - K^{\prime\,-})  &= -\frac{i}{\sqrt{2}}\psibar_h  C\gamma^5\sigma_1\psibar_l^T\\
\tilde K^0_- &= \frac{1}{\sqrt{2}}(K^0 - K^{\prime\,0})  &= \frac{i}{\sqrt{2}}\psibar_l \gamma^5\sigma_3\psi_h\\
\bar{\tilde K}^0_- &= \frac{1}{\sqrt{2}}(\bar K^0 - \bar K^{\prime\,0})  &= \frac{-i}{\sqrt{2}}\psibar_h \gamma^5\sigma_3\psi_l
\end{array}\,.\end{equation}

\subsection{Operators acting on the physical kaon}
\label{sec-opactphyskaon}

When measuring the $K\rightarrow\pi\pi$ amplitudes or $B_K$ in the G-parity framework we are concerned with operators that act only on the physical kaon state and not the fictional partner.  However, in order to determine matrix elements between physical states with known momenta we must work with the G-parity eigenstates that contain the fictional partner. As this transforms into the physical state when it crosses the boundary, we might expect it to make a non-trivial unphysical contribution to the measured amplitude. In order to analyze the size of this effect, consider the \textit{infinite-volume} matrix element,
\begin{dmath}
\langle \phi| {\cal O}_{\rm phys} | K^0\rangle\,,
\end{dmath}
where ${\cal O}_{\rm phys}$ is chosen to be an operator that acts on the physical kaon, does not involve the unphysical $s'$ quark operator and induces a mixing/decay to some final state $|\phi\rangle$. Now we introduce the fictional partner to the strange quark, $s'$, which introduces a new state, $|K^{\prime\,0}\rangle$, that is degenerate with $|K^0\rangle$ but has different flavor quantum numbers.  (Here the infinite-volume states $|K^0\rangle$ and $|K^{\prime\,0}\rangle$ are QCD energy and momentum eigenstates.)  This infinite-volume setup permits no mixing between these two states, hence
\begin{dmath}
\langle \phi| {\cal O}_{\rm phys} | K^{\prime\,0}\rangle = 0\,.
\end{dmath}
We can therefore combine these equations in infinite volume, giving the result
\begin{dmath}
\langle \phi| {\cal O}_{\rm phys} \underbrace{\left(| K^0\rangle + | K^{\prime\,0}\rangle\right)}_{\textstyle \sqrt{2} |\tilde K^0_+\rangle} = \langle \phi| {\cal O}_{\rm phys} | K^0\rangle\,.\label{eq-infvolKOphi}
\end{dmath}

If we restrict ourselves to momenta whose components are integer multiples of $2\pi/L$, the linear combination that we label $|\tilde K^0_+\rangle$ is a valid eigenstate of the system in infinite volume and, if viewed as a two-component wave function, ($K^0$, $K^{\prime\,0}$), will also obey G-parity boundary conditions for a finite volume of size $V=L^3$.  The infinite-volume composite particles $K^0$ and $K^{\prime\,0}$ receive only exponentially suppressed corrections when confined to move in a finite volume~\cite{Luscher:1985dn}.  Thus, up to terms $O(e^{-m_K L})$, we can obtain the infinite volume matrix element $\langle \phi| {\cal O}_{\rm phys} | K^0\rangle$ that appears on the right-hand side of Eq.~\eqref{eq-infvolKOphi} from the matrix element on the left-hand side evaluated in finite volume.  Thus,
\begin{dmath}
\langle \phi| {\cal O}_{\rm phys} | K^0\rangle = \sqrt{2}\langle \phi| {\cal O}_{\rm phys} |\tilde K^0_+\rangle_V + {\cal O}(e^{-m_K L})\,.
\label{eq-infvol1pfs}
\end{dmath}  
where the matrix element on the left-hand side is evaluated in infinite volume while that on the right in the finite volume $V$.

This is an instructive, highly simplified example of the more familiar relation between infinite-volume two-particle scattering states and finite-volume energies and matrix elements discovered by L\"uscher~\cite{Luscher:1990ux} and by Lellouch and  L\"uscher~\cite{Lellouch:2000pv}.  If we consider the case where only the $s$-wave two-particle scattering phase shift is non-zero, the analysis in Ref.~\cite{Luscher:1990ux} can be understood as adjusting the relative momentum of the two-pions until the $s$-wave two-particle scattering state combines with the non-$s$-wave component of L\"uscher's Helmholtz function to produce a function obeying the finite-volume boundary conditions.  In our simpler case the momenta of the $|K^0\rangle$ and $|K^{\prime\,0}\rangle$ states are adjusted so that the combined state $(|K^0\rangle + |K^{\prime\,0}\rangle)/\sqrt{2}$ obeys the required G-parity boundary conditions.  

The Lellouch-Luscher factor $\propto \partial(\delta_0+\phi)/\partial k$ of Ref.~\cite{Lellouch:2000pv} corrects for the higher partial waves that are present in the finite volume $\pi\pi$ state and affect its normalization but do not contribute directly to the $s$-wave decay matrix element.  The factor $\sqrt{2}$ in Eq.~\ref{eq-infvol1pfs} plays a similar role, correcting for the shift in the normalization of the state caused by the otherwise irrelevant $|K^{\prime\,0}\rangle$ component of the finite-volume eigenstate, $|\tilde K^0_+\rangle$.

In the discussion above, we have focused on the treatment of the kaon state that appears on the right-hand side of the matrix element in Eq.~\eqref{eq-infvolKOphi} when G-parity boundary conditions are present.  Of course, the left-hand state $\langle\phi|$ in that matrix element is also affected by these boundary conditions.  In the case of the mixing parameter $B_K$ the $\langle \overline{K}\,^0|$ state appears on the left-hand side of that matrix element and can be treated in the same way as proposed for the $|K^0\rangle$ state on the right-hand side.  In the case that the left-hand state $\langle\phi|$ is an interacting multi-particle final state there are additional contributions arising from the scattering, the form of which depends on the particles in question. In practice we are primarily interested in $K\rightarrow\pi\pi$ decays, where the pions interact in the finite volume and individually obey antiperiodic boundary conditions. Here the contribution is completely described by the Lellouch-L\"{u}scher formula~\cite{Lellouch:2000pv,Lin:2001ek} generalized to the antiperiodic case~\cite{Blum:2012uk}.

\subsection{Strange quark determinant}
\label{sec-ssdeterminant}

With the introduction of the fictional $s'$ quark, the theory now contains four flavors: two degenerate pairs, one light and one heavy.  In addition to the effect of the fictional $s'$ as a valence quark that must appear in states which are eigenstates of spatial momenta, that was discussed in the previous section, the $s'$ quark will also appear as a sea quark through the fermion determinant of the heavy-quark Dirac operator which by definition acts on the $s/s'$ quark doublet.  However, in practice we wish to simulate a 2+1 flavor theory with a single strange quark species so the contribution of the $s'$ quark must be removed.

We can represent the Dirac matrix in this heavy flavor space schematically as
\begin{equation}
{\cal M}_{s/s'} = \left(\begin{array}{cc}{\cal M}_{s\leftarrow s} & {\cal M}_{s\leftarrow s'} \\ {\cal M}_{s'\leftarrow s} & {\cal M}_{s'\leftarrow s'}\end{array}\right)\,.
\end{equation}
The determinant of this block matrix is
\begin{dmath}
{\rm det}({\cal M}_{s/s'}) = {\rm det}\left( {\cal M}_{s\leftarrow s} - {\cal M}_{s\leftarrow s'}{\cal M}^{-1}_{s'\leftarrow s'}{\cal M}_{s'\leftarrow s} \right){\rm det}\left({\cal M}_{s'\leftarrow s'}\right)\,.
\end{dmath} 

In infinite volume the flavor mixing components vanish and the determinant reduces to the product of the determinants of the elements connecting individual quark flavors:
\begin{equation}
{\rm det}({\cal M}_{s/s'}) \stackrel{L\rightarrow\infty}{=}  {\rm det}({\cal M}_{s\leftarrow s}){\rm det}({\cal M}_{s'\leftarrow s'}) = \left[{\rm det}({\cal M}_{s\leftarrow s})\right]^2\,,
\end{equation}
where in the last equality we have used the degeneracy of the $s$ and $s'$ quarks. In this limit the contribution of a single flavor can be obtained simply by taking the square root of the two-flavor determinant. 

At finite volume the Dirac matrix cannot be factored and the rooting prescription does not result in the determinant of a local operator. The non-locality of the resulting effective action then leaves no guarantee that the continuum limit of the rooted determinant defines a theory that lies in the correct universality class. This issue has received much attention in the context of staggered fermions, where each quark flavor corresponds to four `tastes' which are coupled at high momenta and where a rooting prescription is used to reduce the number of quark tastes from four to one.  Fortunately our situation is easier to analyze.

We will consider two quite similar determinants.  The first is the case of direct interest: an $s/s'$ doublet with the $s$ and $s'$ species connected at a boundary by G-parity boundary conditions.  In this case the determinant does not factorize and taking the square root of that determinant produces an effective action which does not precisely correspond to that of a local field theory of fermions.  The second case also involves a degenerate $q_1/q_2$ doublet but each obeys independent charge-conjugation boundary conditions and the resulting determinant is a simple square whose square root corresponds to the Pfaffian of a local field theory of a single fermion species obeying charge-conjugation boundary conditions~\cite{Kim:2009fe}.

We will demonstrate that these two determinants differ by terms which fall exponentially as the system size grows so that the square root of the G-parity determinant differs from that of a proper local theory by terms which can be safely neglected.  These two theories are identical if the boundary terms are neglected: both are theories of two flavors of fermion obeying open boundary conditions.  The theories differ only because of the boundary terms.  For the $s/s'$ theory the boundary terms of the Wilson/domain wall theory can be written explicitly as
\begin{eqnarray}
-\overline{s}'(L-1)\Gamma_x^+C U_x(L-1)\overline{s}^T(0)
&+&\overline{s}(L-1)\Gamma_x^+C U_x(L-1){\overline{s}'}^T(0)  
\nonumber \\
-{s'}^T(L-1)C\Gamma_x^- U_x^\dagger(L-1) s(0)
&+&{s}^T(L-1)C\Gamma_x^- U_x^\dagger(L-1) s'(0)
\label{eq:BD-G}
\end{eqnarray}
For simplicity, we have assumed that G-parity boundary conditions are imposed only in the $x$ direction.

\begin{figure}
\centering
\includegraphics[width=0.4\textwidth]{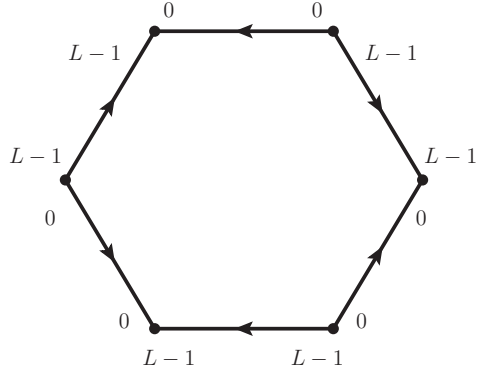}
\caption{A term in the graphical expansion of the quark determinant in powers of the boundary terms of the Dirac operator.  Here the directed lines represent lattice quark propagators for a Dirac operator obeying open boundary conditions and the vertices the boundary terms appearing in Eqs.~\eqref{eq:BD-G} or \eqref{eq:BD-C}.  In this figure we show only a leading term when $L$ is large where the fermion propagators join operators located at the same side of the volume.}
\label{fig:SqrtDet}
\end{figure}

For the case of the doublet of fermions obeying charge-conjugation boundary conditions, the corresponding boundary terms are similar:
\begin{eqnarray}
\sum_{i=1,2} \left\{
       \overline{q}_i(L-1)\Gamma_x^+C U_x(L-1)\overline{q}_i^T(0)
               +q_i^T(L-1)C\Gamma_x^+ U_x^\dagger (L-1)q_i(0) \right\}
\label{eq:BD-C}
\end{eqnarray}
We can now compare these two theories by expanding their respective determinants in powers of these boundary terms.  These expansions will take the form of a sum of products of closed loops of the sort shown in Fig.~\ref{fig:SqrtDet}, where the vertices joining two quark lines correspond to the bilinear terms in Eqs.~\eqref{eq:BD-G} and \eqref{eq:BD-C} and the directed lines in that figure correspond to fermion propagators with open boundary conditions.  For our explicit examples in which G-parity or charge-conjugation boundary conditions are imposed in only the $x$-direction,  the vertices joining two outgoing quark lines will have the gamma and link matrix structure $\Gamma_x^+C U_x(L-1)$ while those at which two fermion lines are absorbed will contain $C\Gamma_x^+U_x(L-1)^*$.

As suggested in Fig.~\ref{fig:SqrtDet} the leading term when $L$ is large will involve fermion propagators which connect operators in two boundary terms at the same side of the volume.  That is either $x=0$ is connected with $x=0$ or $x=L-1$ is connected with $x=L-1$.  A propagator joining operators located at $x=0$ and $x=L-1$ will be exponentially suppressed.  The connections that are shown in Fig.~\ref{fig:SqrtDet} correspond to a term which is not exponentially suppressed.

Inspecting Eqs.~\eqref{eq:BD-G} and \eqref{eq:BD-C}, we can recognize that the graphs describing terms of the form shown in Fig.~\ref{fig:SqrtDet}, which are not exponentially suppressed, are identical for these two theories.  For the theory with charge-conjugation boundary conditions a graph such as that in Fig.~\ref{fig:SqrtDet} will correspond to a specific Feynman amplitude with a prefactor of 1/3 because of the symmetry of the graph under cyclic shifts of the vertices by two positions along the ring and an additional factor of 2 to account for the two independent flavors.

Exactly the same Feynman amplitude will appear in the G-parity theory with the same factor of 1/3 arising from the cyclic symmetry.  The factor of 2 also appears because for the G-parity theory the propagators must alternate between that of an $s$ and that of an $s'$ quark as one moves around the ring.  The exchange of $s$ and $s'$ everywhere yields the factor of two.  The two minus signs which appear in Eq.~\eqref{eq:BD-G} do not appear in Eq.~\eqref{eq:BD-C} .  However, as can be seen from the pattern of contractions which appears in Fig.~\ref{fig:SqrtDet} these minus signs always appear in pairs for the G-parity case and hence have no effect.

We conclude that if we neglect terms which are exponentially suppressed, our rooted $s/s'$ determinant equals the Pfaffian corresponding to a local theory of fermions obeying charge conjugation boundary conditions.  We therefore expect no subtle difficulties to arise from our use of the square root of this determinant.

\section{$2+1$f Domain Wall Fermion Ensembles}
\label{sec:Ensembles}
In this and the following section we present results for full QCD simulations on a $16^3\times 32$ lattice volume with G-parity boundary conditions in zero, one and two directions. We begin by discussing the generation of the G-parity gauge 
configurations and then present and interpret the results of a variety of measurements made on these ensembles.

\subsection{Simulation Parameters and Generation}

In Section~\ref{sec-1fequiv} we discussed the `one-flavor' implementation of G-parity boundary conditions in a single direction whereby one simulates with a lattice of doubled extent and antiperiodic quark boundary conditions. This technique is very easy to implement in any lattice code library, but suffers from an additional factor of $2^{n-1}$ increase in computational cost when applied in $n>1$ G-parity directions. A much cleaner approach is simply to simulate with two separate quark fields that mix at the boundary; we refer to this as the `two-flavor' approach. In practise this requires significant code modifications to handle the increased memory size of the two-flavor field and its unusual boundary condition, as well as careful implementation of the complex conjugate boundary conditions on the gauge fields. Nevertheless, using the one-flavor method as a cross-check, we implemented the two-flavor technique in the CPS++ library using the Bagel/BFM library~\cite{Boyle:2009vp} for optimized fermion inversions on IBM Blue Gene/Q machines~\cite{Boyle:2012iy}. An independent implementation in the Grid framework~\cite{Boyle:2015tjk} has since been developed and cross-verified.

For our first fully dynamical simulations, we generated two 2+1 flavor domain wall ensembles with the Iwasaki gauge action at $\beta=2.13$ ($a^{-1}= 1.73(3)$ GeV~\cite{Allton:2008pn}) and lattice size $16^3\times 32\times 16$ with G-parity boundary conditions in one and two directions. We henceforth refer to these as the GP1 and GP2 ensembles respectively. For both ensembles we used an input light quark mass of $m_{u/d}=0.01$ and a strange quark mass of $m_s=0.032$. These parameters were chosen to match those of a previously generated ensemble with periodic boundary conditions, details of which were originally published in Ref.~\cite{Allton:2007hx}, although there a heavier strange mass of 0.04 was used; the ensembles were later extended with the value of $m_s=0.032$ used for this study, which is closer to the physical strange mass. For comparison with the G-parity ensembles we make use of the latter configurations and refer to this ensemble by the label GP0.

\subsection{Ensemble generation}

In most domain wall simulations, the matrix inversions performed to evaluate the fermion determinant in the hybrid Monte Carlo evolution (and of course for the measurements) are obtained using the conjugate gradient algorithm. This algorithm requires the fermion matrix to be Hermitian and positive-definite, and to ensure this one typically inverts instead the product ${\cal M}^\dagger {\cal M}$, where ${\cal M}$ is the Dirac matrix. The determinant thus obtained represents the contribution of two degenerate flavors. The contribution of a single strange quark is obtained using the rational hybrid Monte Carlo (RHMC) algorithm, whereby a rational approximation to the square root of the degenerate two-flavor determinant is obtained. 

With G-parity boundary conditions the Dirac matrix is intrinsically two-flavor and therefore the square of the determinant describes four flavors. As a result we must also use RHMC in the light-quark sector in order to obtain the two-flavor determinant; as this is an exact square there are no systematics introduced by this procedure. For the strange quark we must use the fourth root to obtain the contribution of a single flavor, and the implications of this rooting were discussed in Section~\ref{sec-ssdeterminant}. In practise we found the use of RHMC in the light sector required more precise rational approximations than is is typically required for the strange quark in order to obtain good Metropolis acceptance. This was not a severe hindrance for these cheap ensembles, but for our physical-point ensembles that we generated to measure the $K\rightarrow(\pi\pi)_{I=0}$ amplitudes~\cite{Bai:2015nea} we found the associated linear algebra overheads were significant and limited the value of introducing multiple Hasenbusch masses to reduce the inversion cost as well as making it difficult to take advantage of mixed-precision techniques.

Note that, although it was not used for this analysis, the RHMC algorithm for the light quarks can be replaced by the ``exact one-flavor action'' developed by TWQCD~\cite{Ogawa:2009ex,Chen:2014hyy}, whereby a Hermitian and positive-definite operator is constructed describing a single flavor (or two flavors with GPBC) of Wilson/domain wall fermions. In Ref.~\cite{Jung:2017xef} we describe how to implement this technique in a highly efficient manner and demonstrate a factor of 4.2$\times$ improvement in the speed of generating the G-parity ensembles for our $K\to\pi\pi$ analysis.

The ensembles used for the analysis in this document were generated using three layers of nested Omelyan integrators. The lowest level integrator (with the shortest time-step) comprises the gauge field and conjugate momentum; the mid-level integrator contains the strange-quark action ; and the top-level integrator contains the light-quark action. Each integrator uses an Omelyan parameter of $\lambda =0.2$ and we update the lowest-level integrator 8 times for every update of the strange-quark force in order to improve the acceptance; the other two Omelyan integrators both update their child integrators with the usual 1:1 cadence, each with 1 step. Each trajectory on the G-parity ensembles comprised 5 steps with a step size of 0.2, and the GP0 ensemble used 4 steps of step size 0.25. With this tuning we obtained 65\% Metropolis acceptance for the GP1 ensemble and 63\% on the GP2 ensemble. A 79\% Metropolis acceptance was reached for the GP0 ensemble.

\subsection{Ensemble properties}
\label{sec-ensproperties}

\begin{figure}[tp]
\centering
\includegraphics*[width=0.32\textwidth]{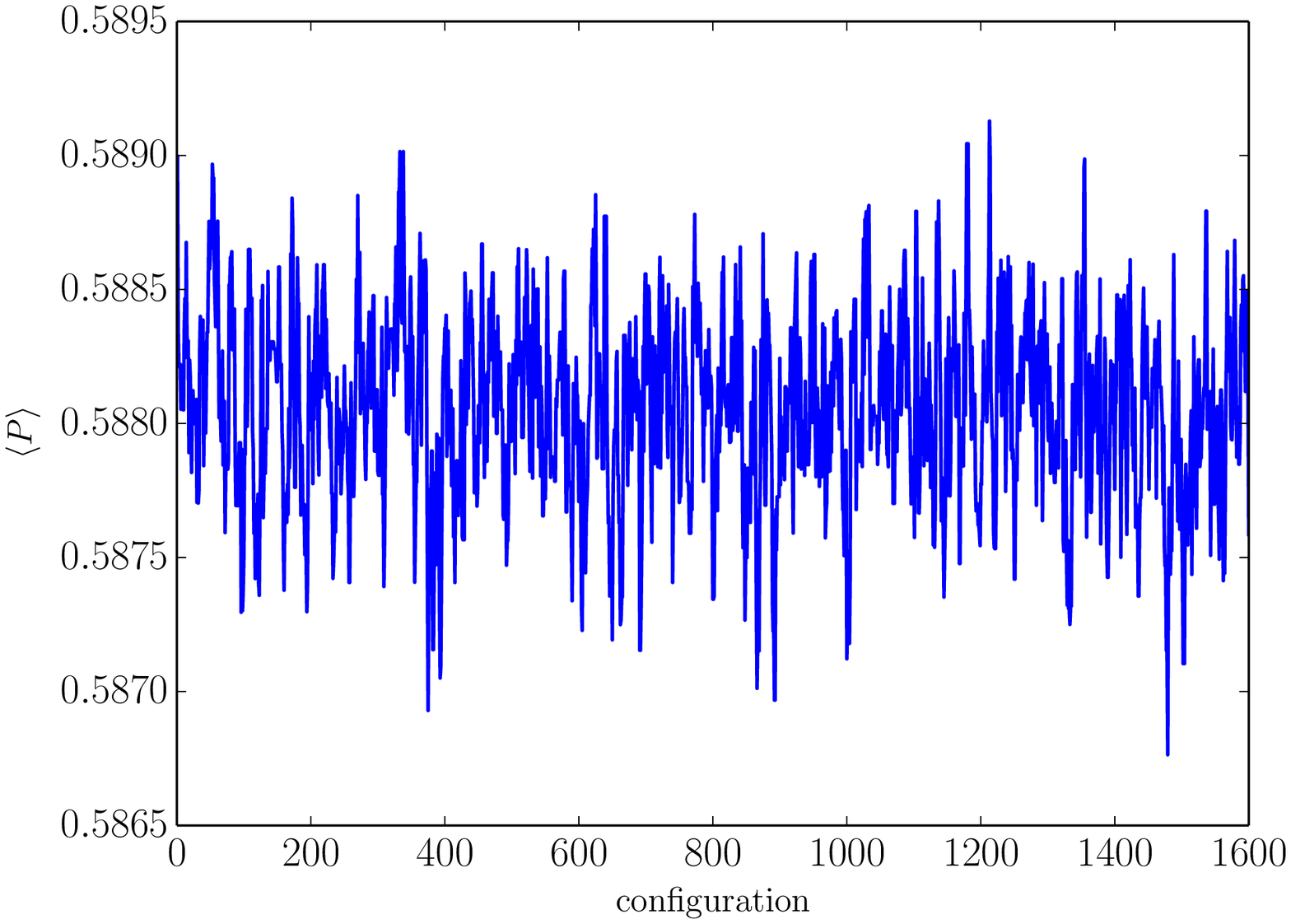}
\includegraphics*[width=0.32\textwidth]{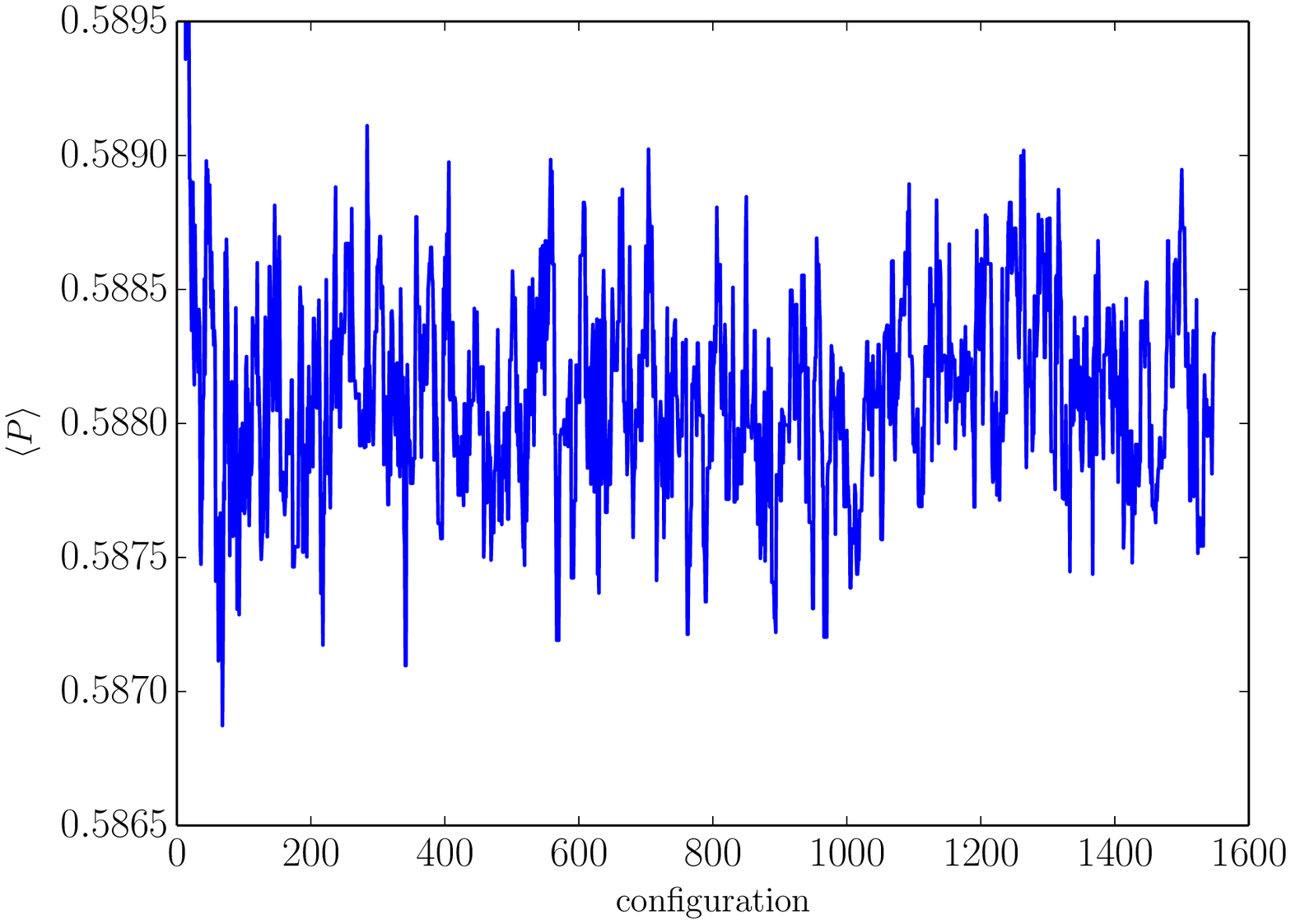}
\includegraphics*[width=0.32\textwidth]{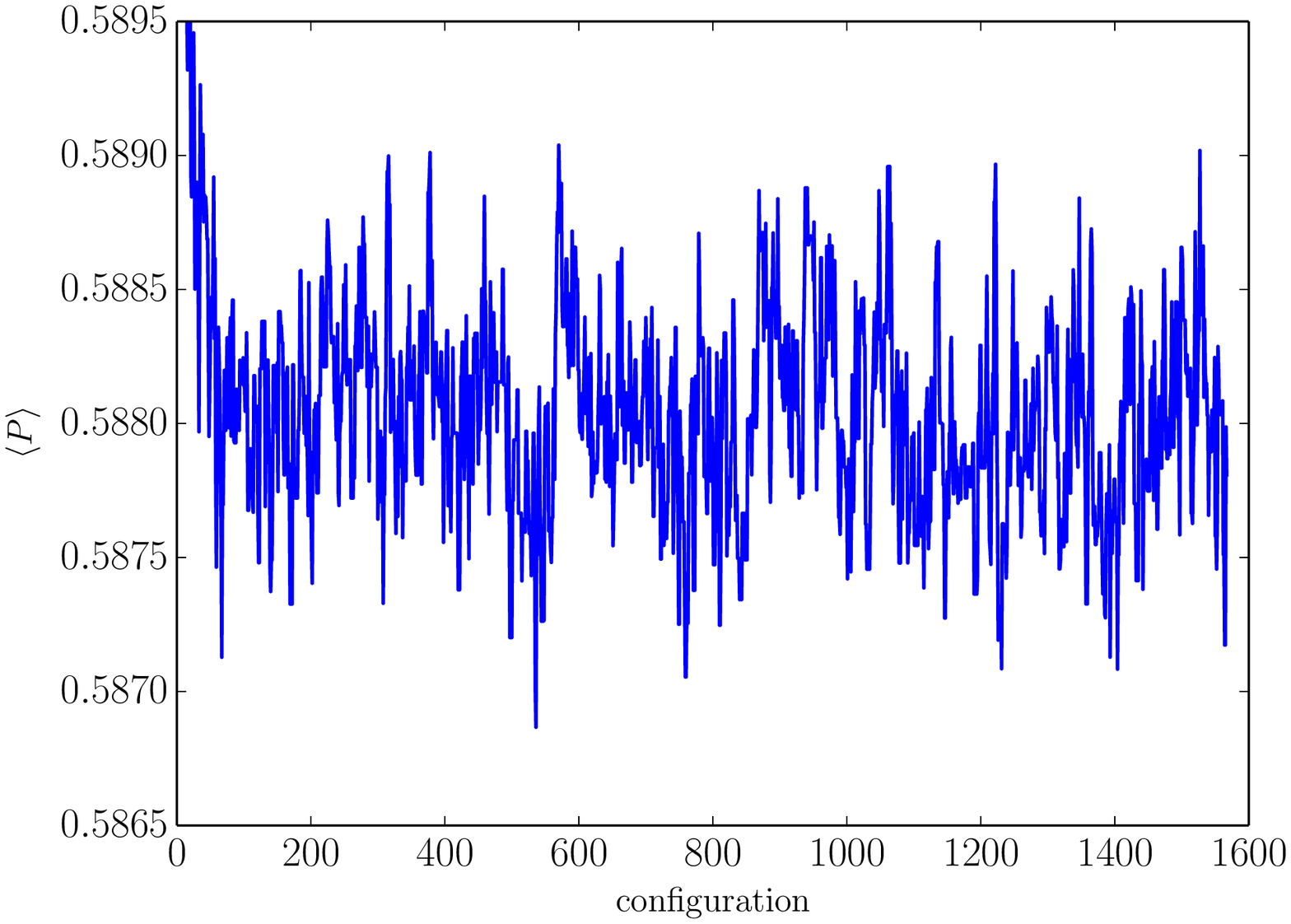}\\

\includegraphics*[width=0.32\textwidth]{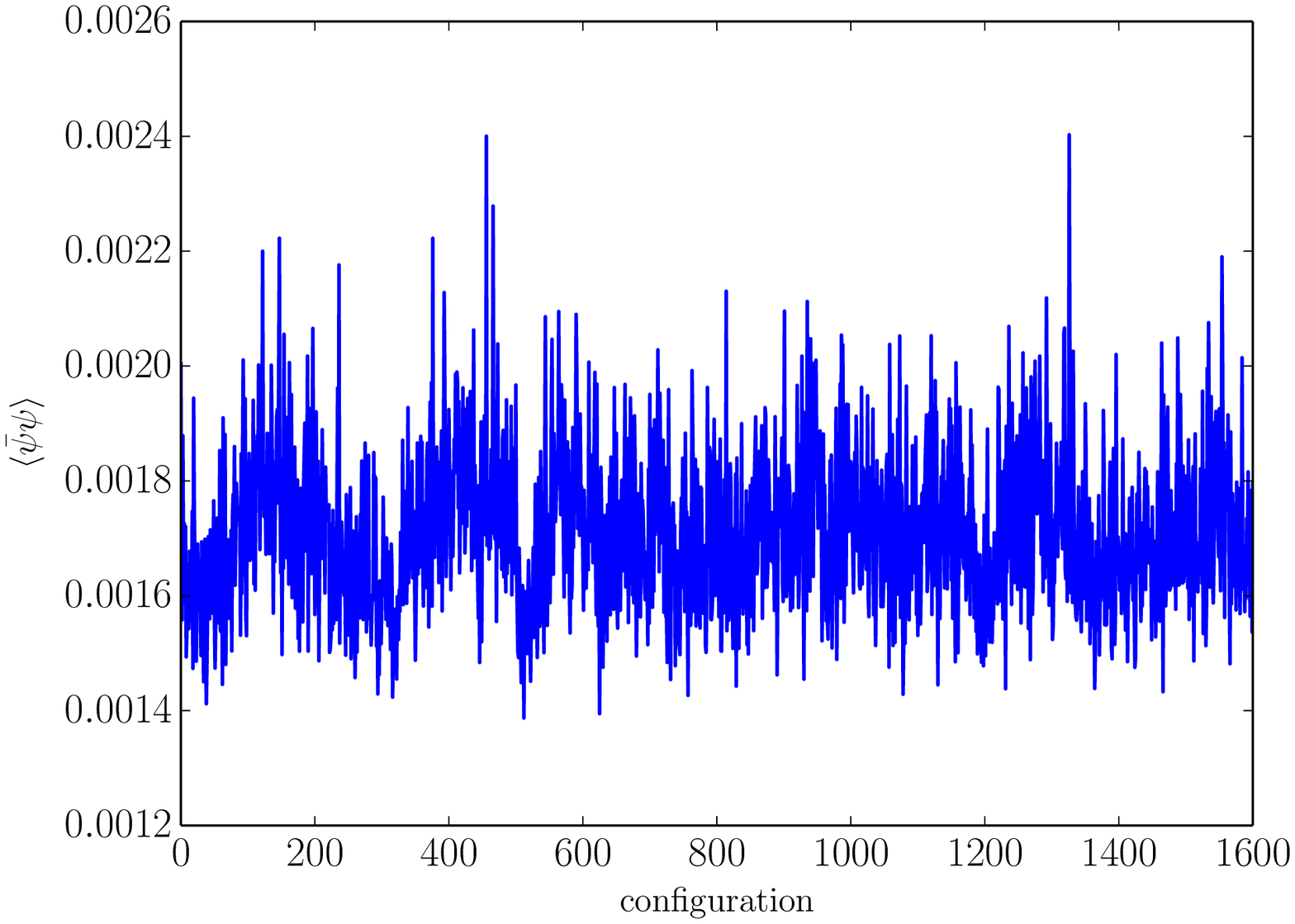}
\includegraphics*[width=0.32\textwidth]{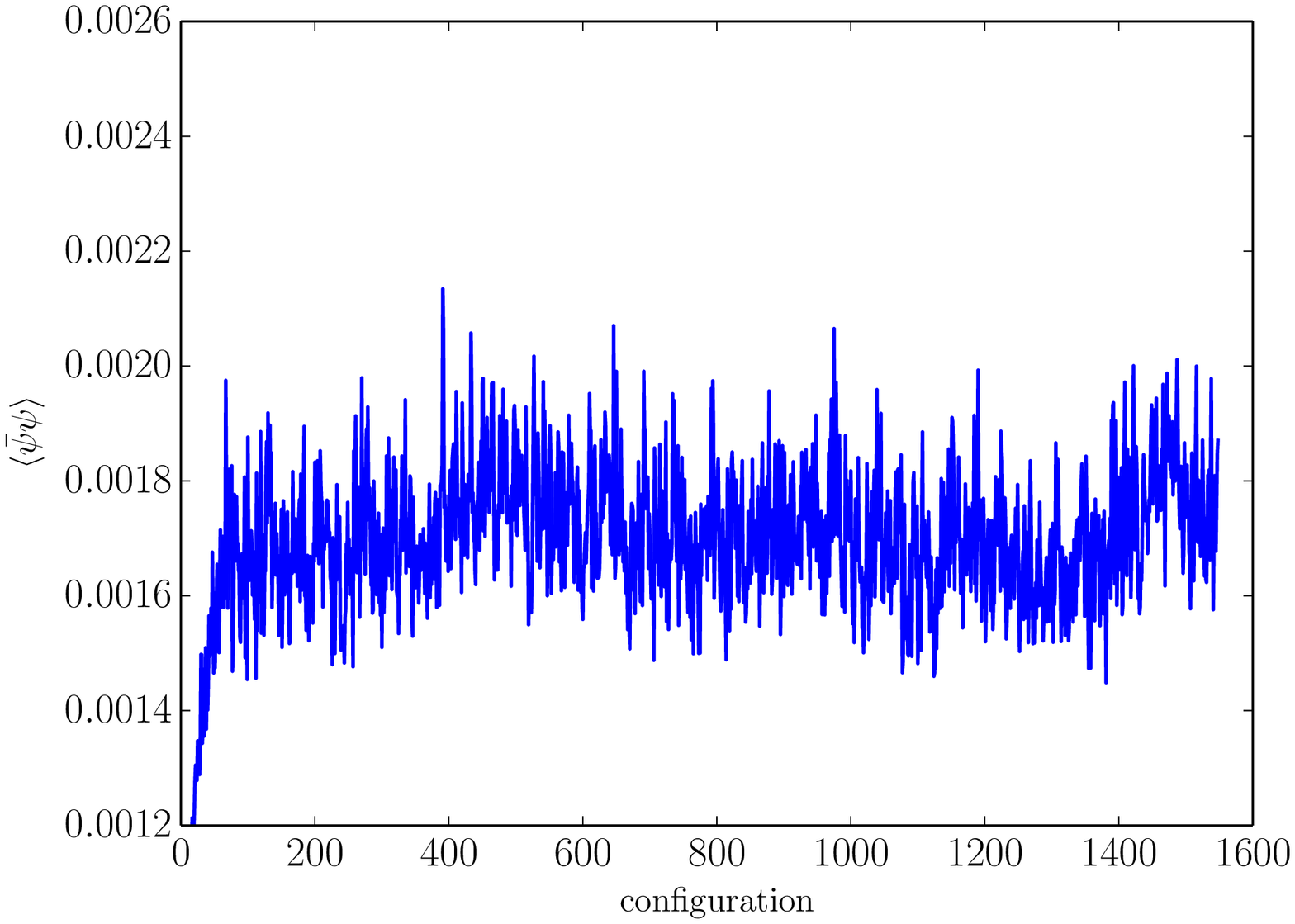}
\includegraphics*[width=0.32\textwidth]{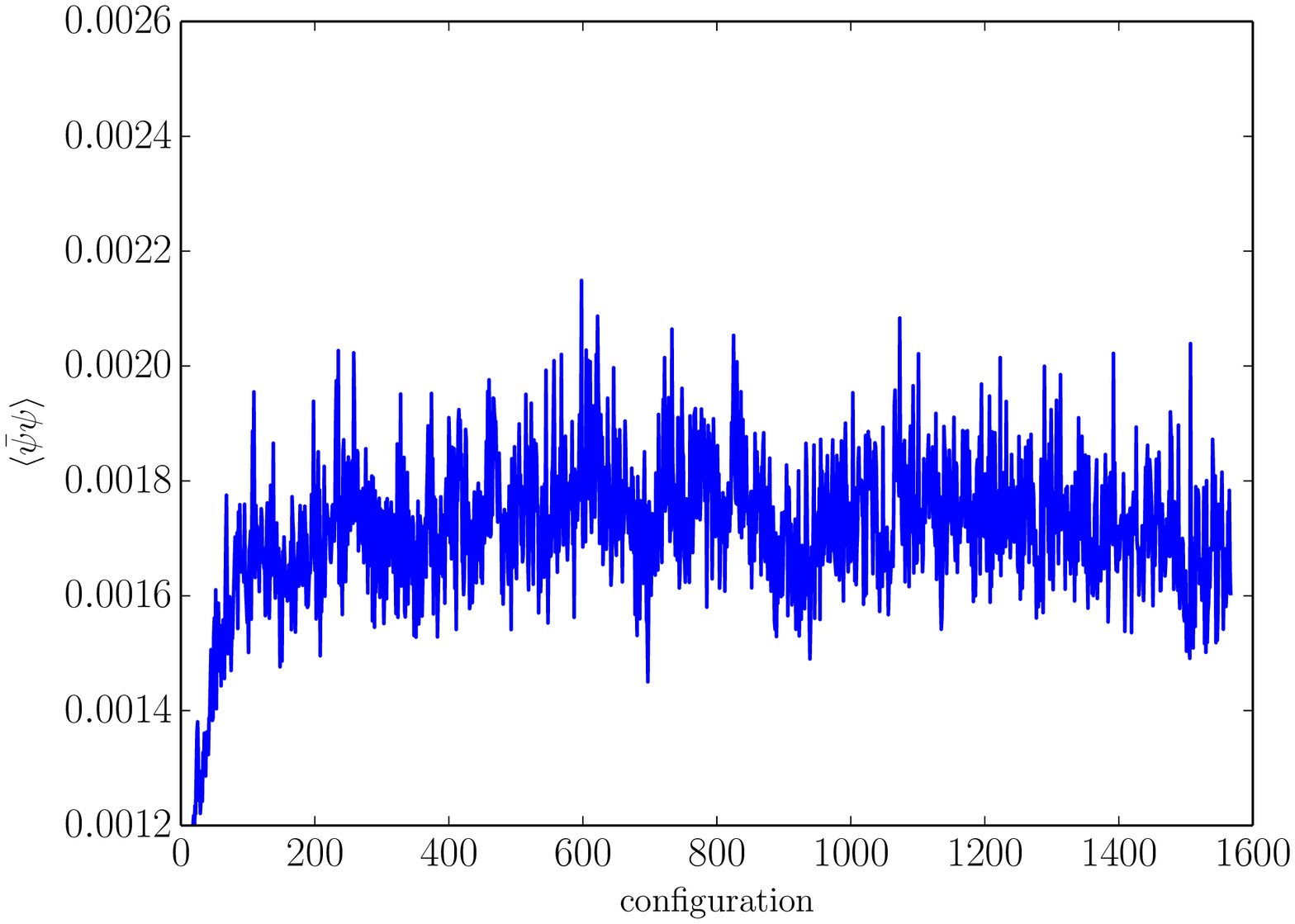}\\

\includegraphics*[width=0.32\textwidth]{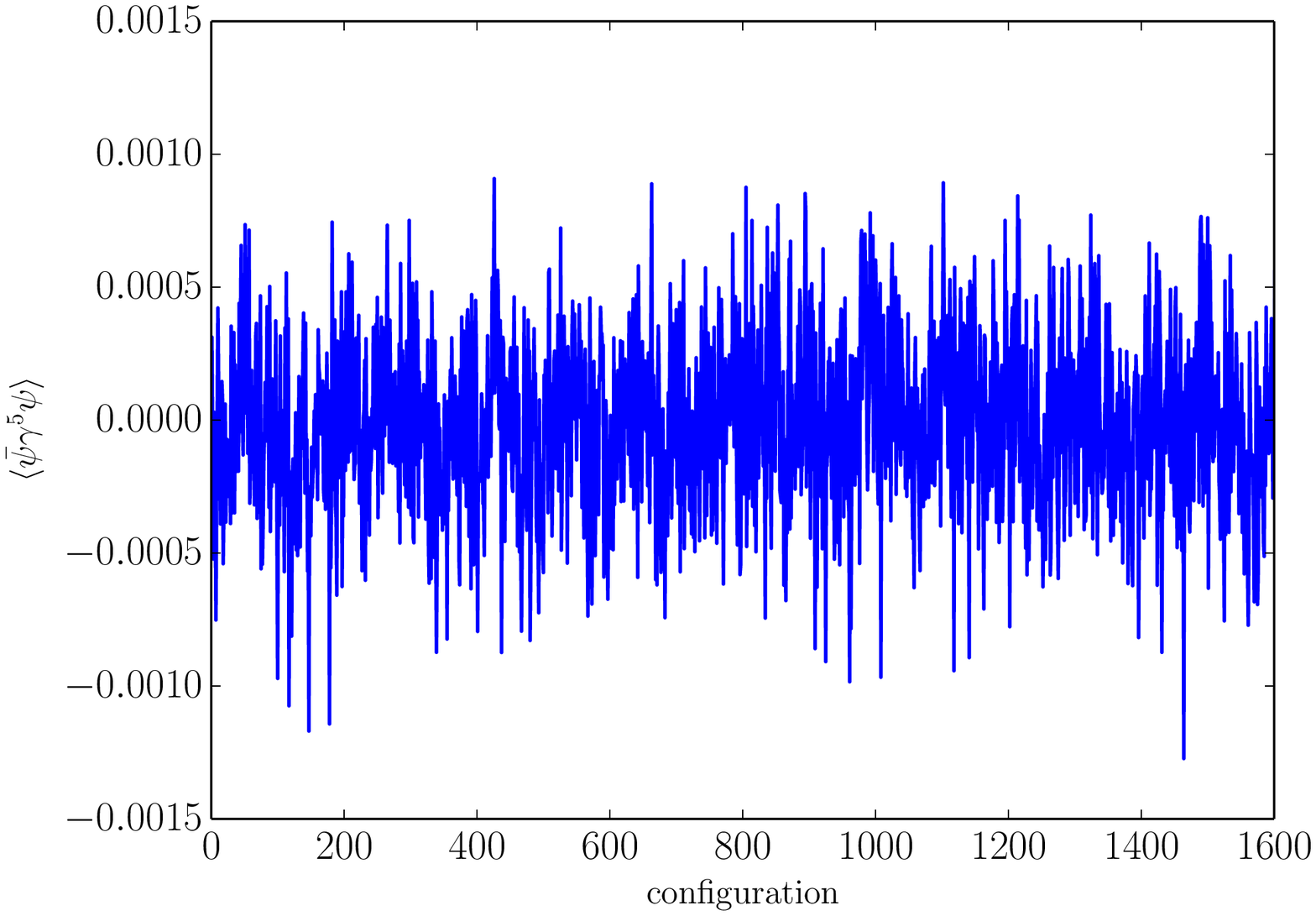}
\includegraphics*[width=0.32\textwidth]{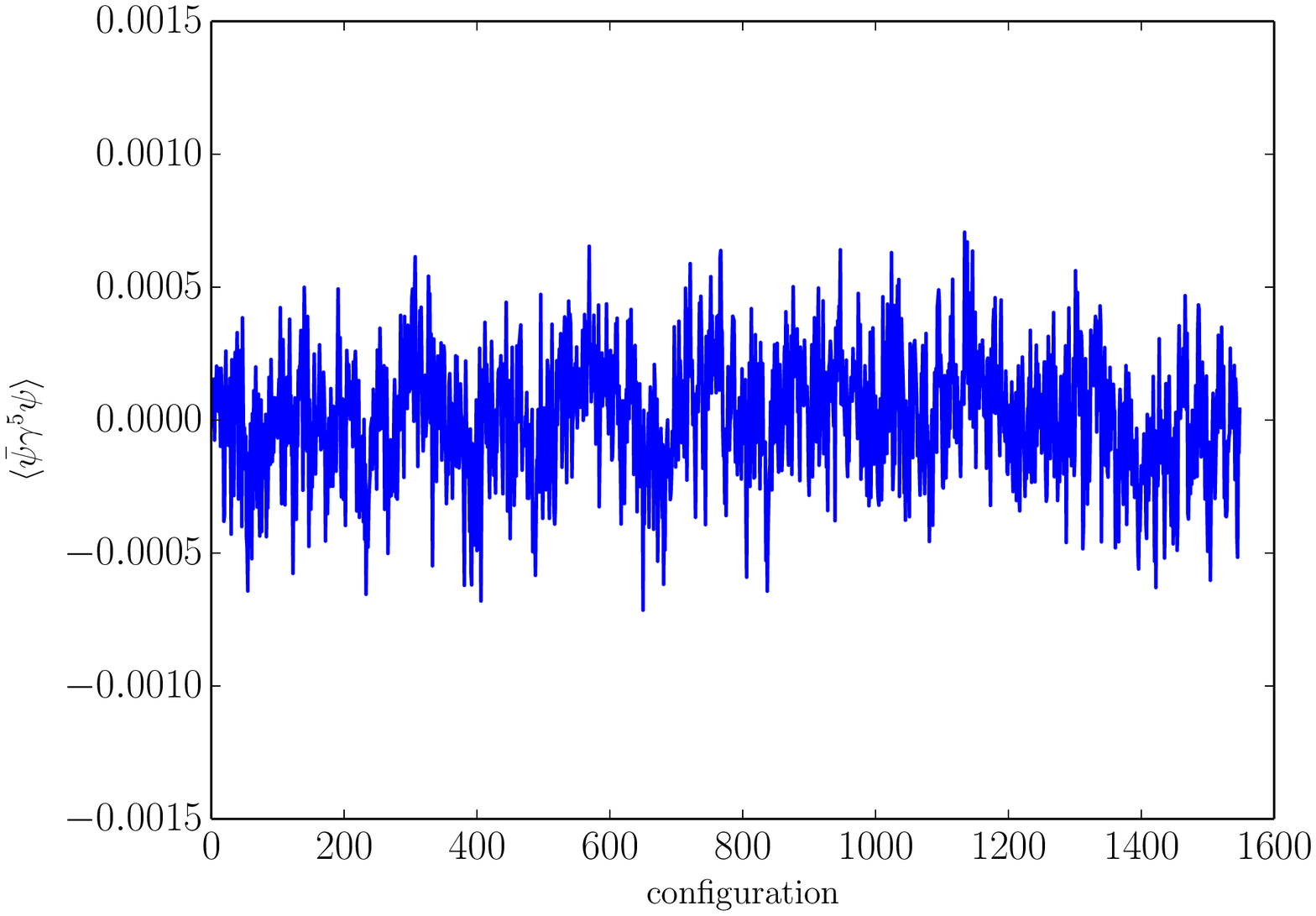}
\includegraphics*[width=0.32\textwidth]{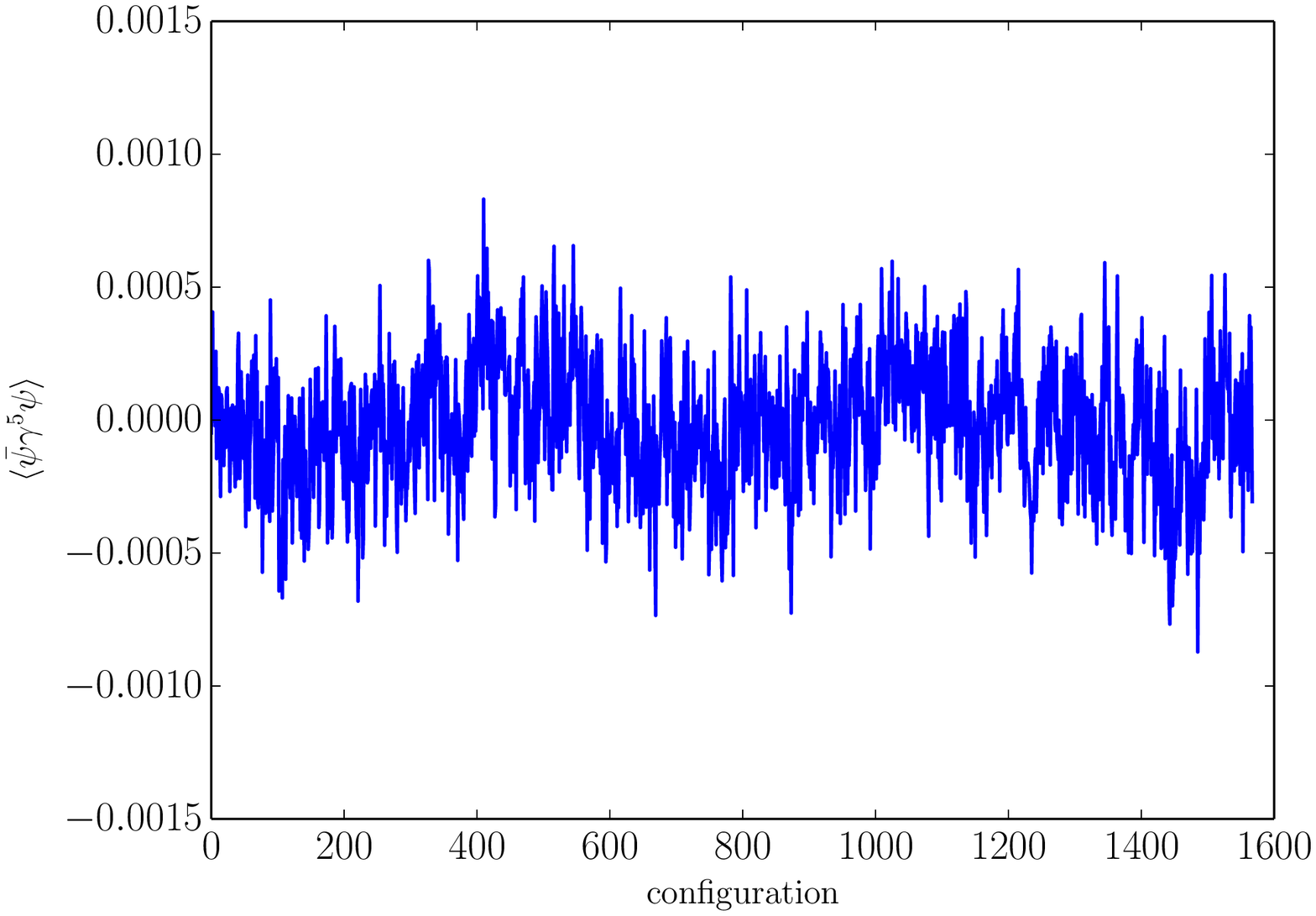}\\

\includegraphics*[width=0.32\textwidth]{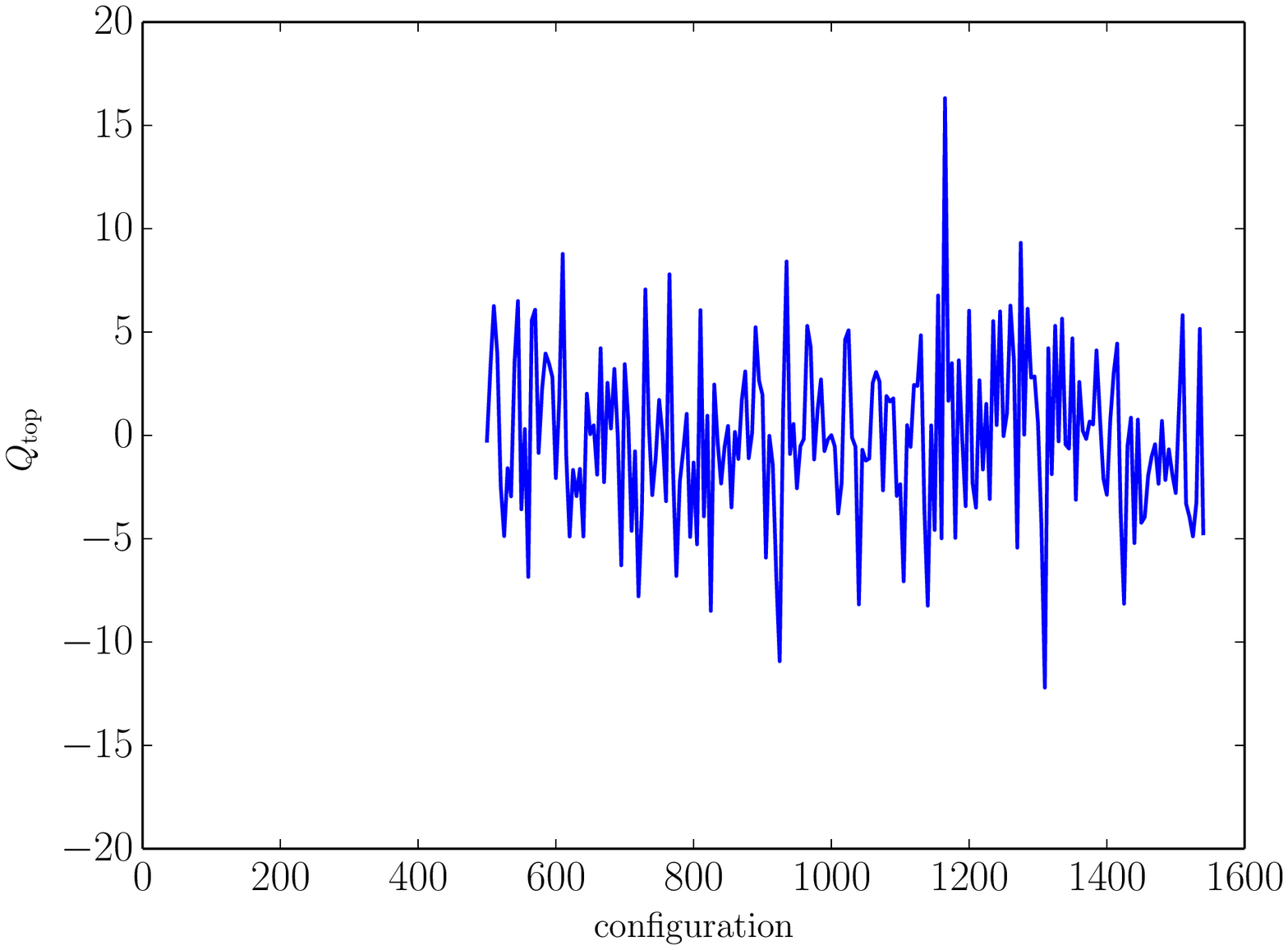}
\includegraphics*[width=0.32\textwidth]{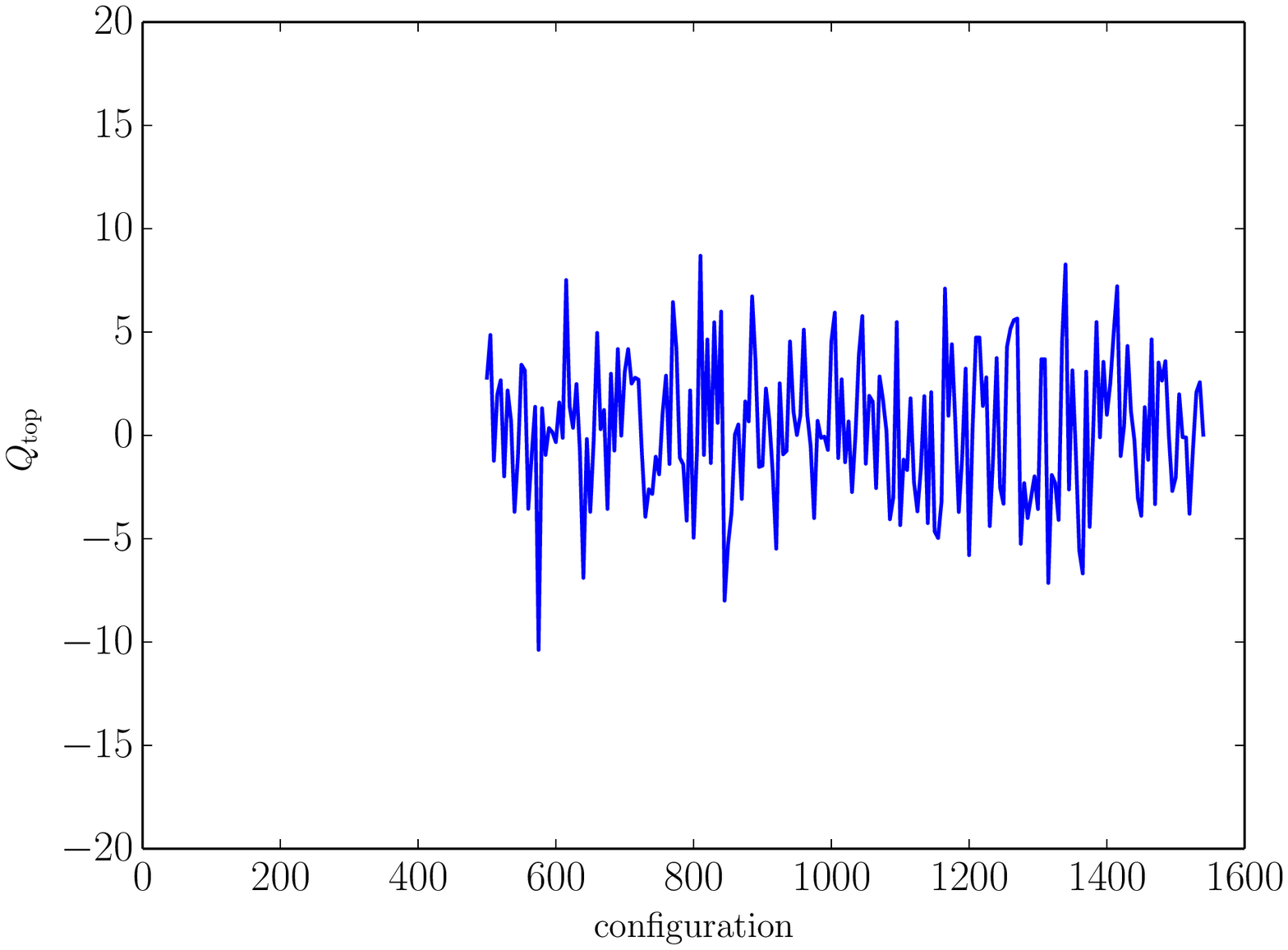}
\includegraphics*[width=0.32\textwidth]{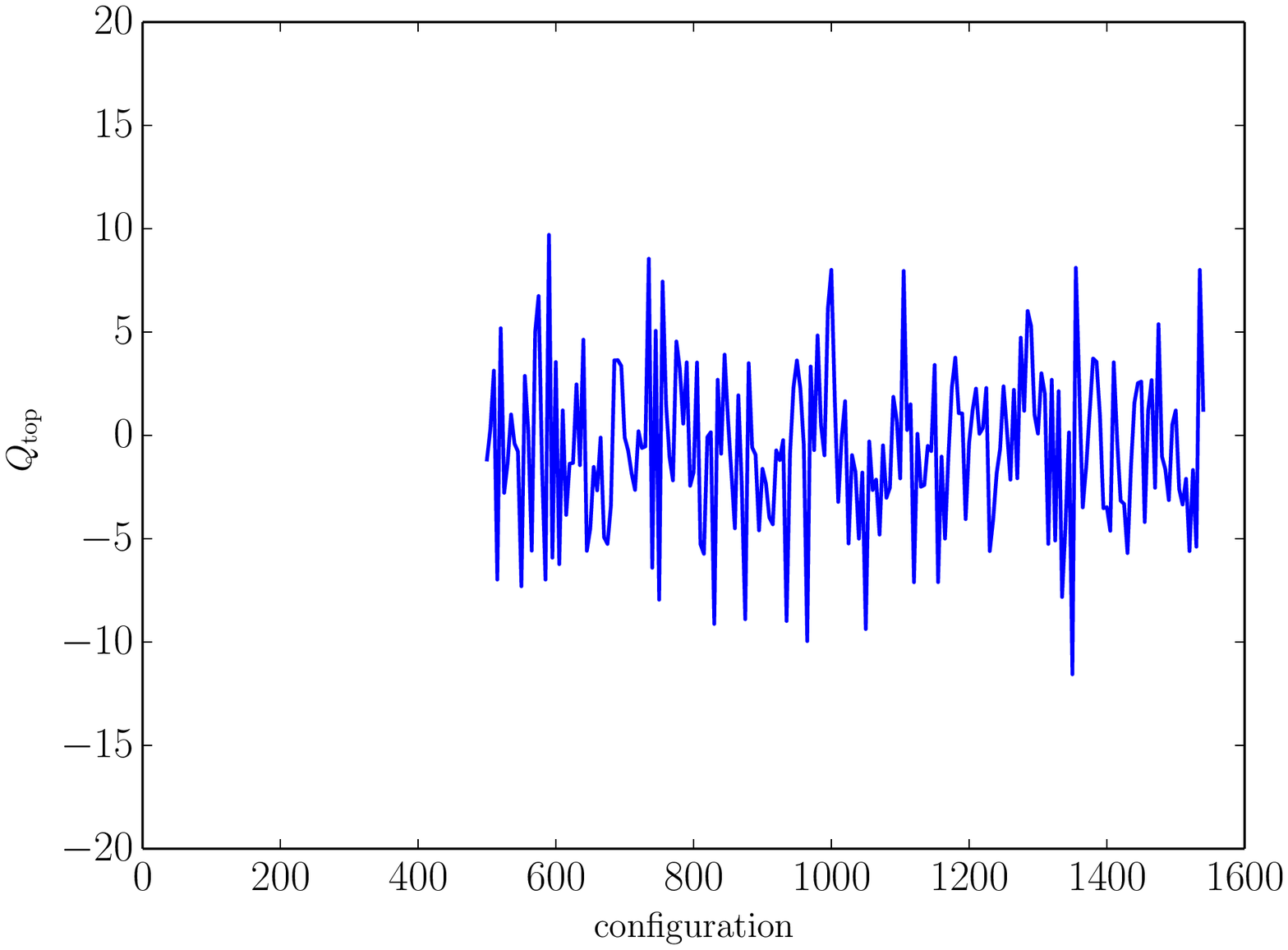}\\
\caption{
The evolution of the average plaquette (first line), chiral condensate (second line), pseudoscalar condensate (third line), and the topological charge (final line) for the 16GP0, 16GP1 and 16GP2 ensembles from left to right respectively.
\label{fig-evolplots}
}
\end{figure}

\begin{figure}[tp]
\centering
\includegraphics*[width=0.32\textwidth]{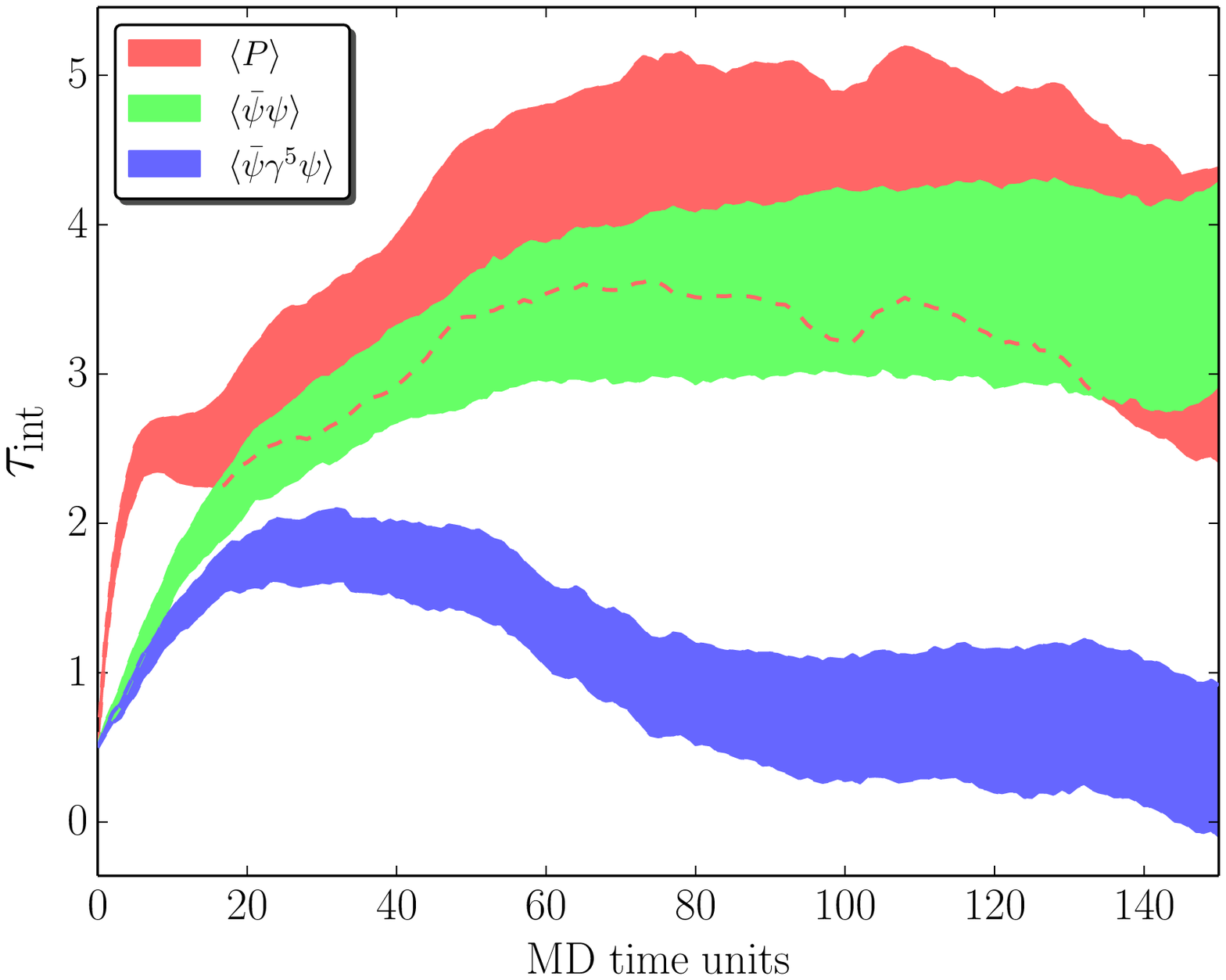}
\includegraphics*[width=0.32\textwidth]{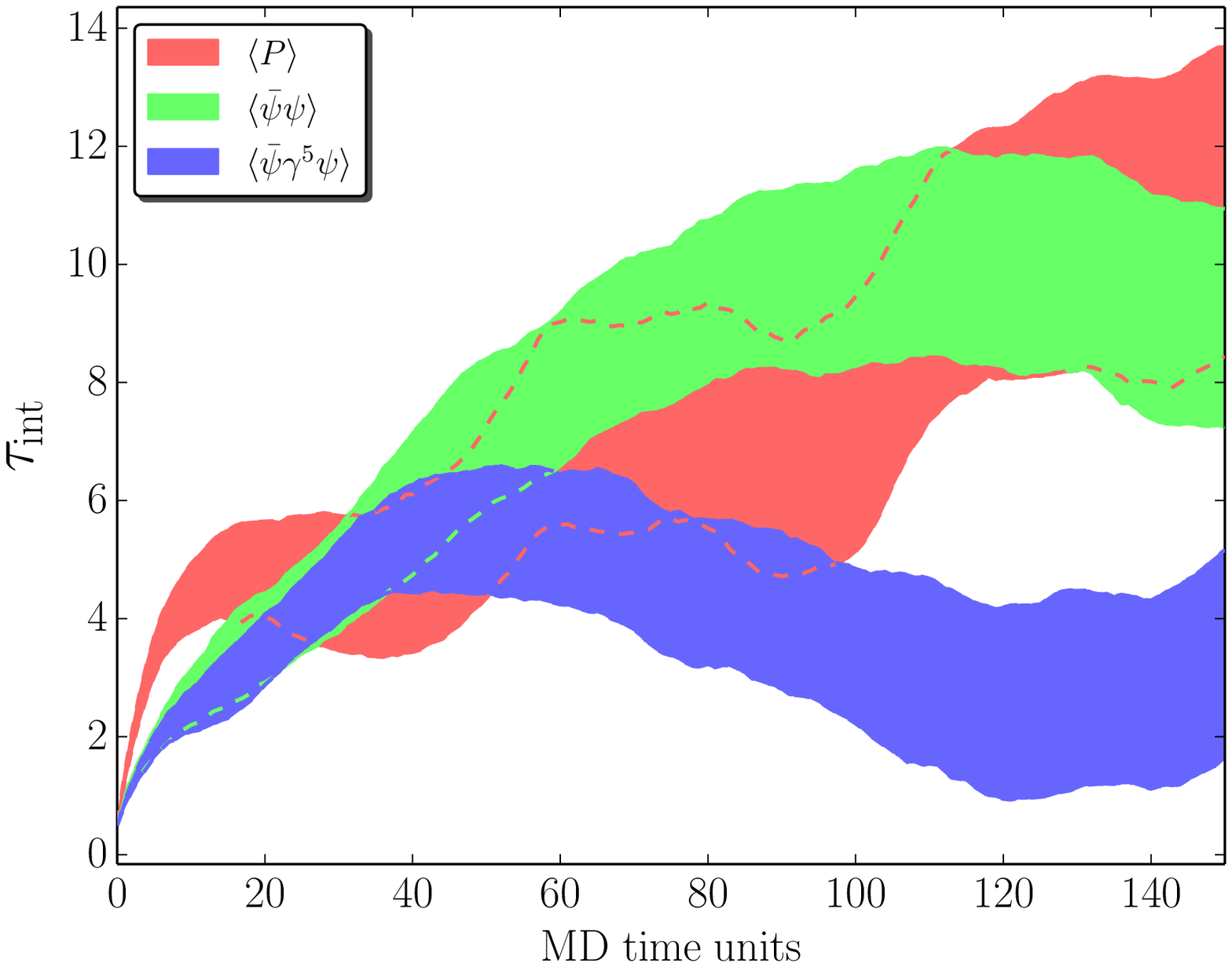}
\includegraphics*[width=0.32\textwidth]{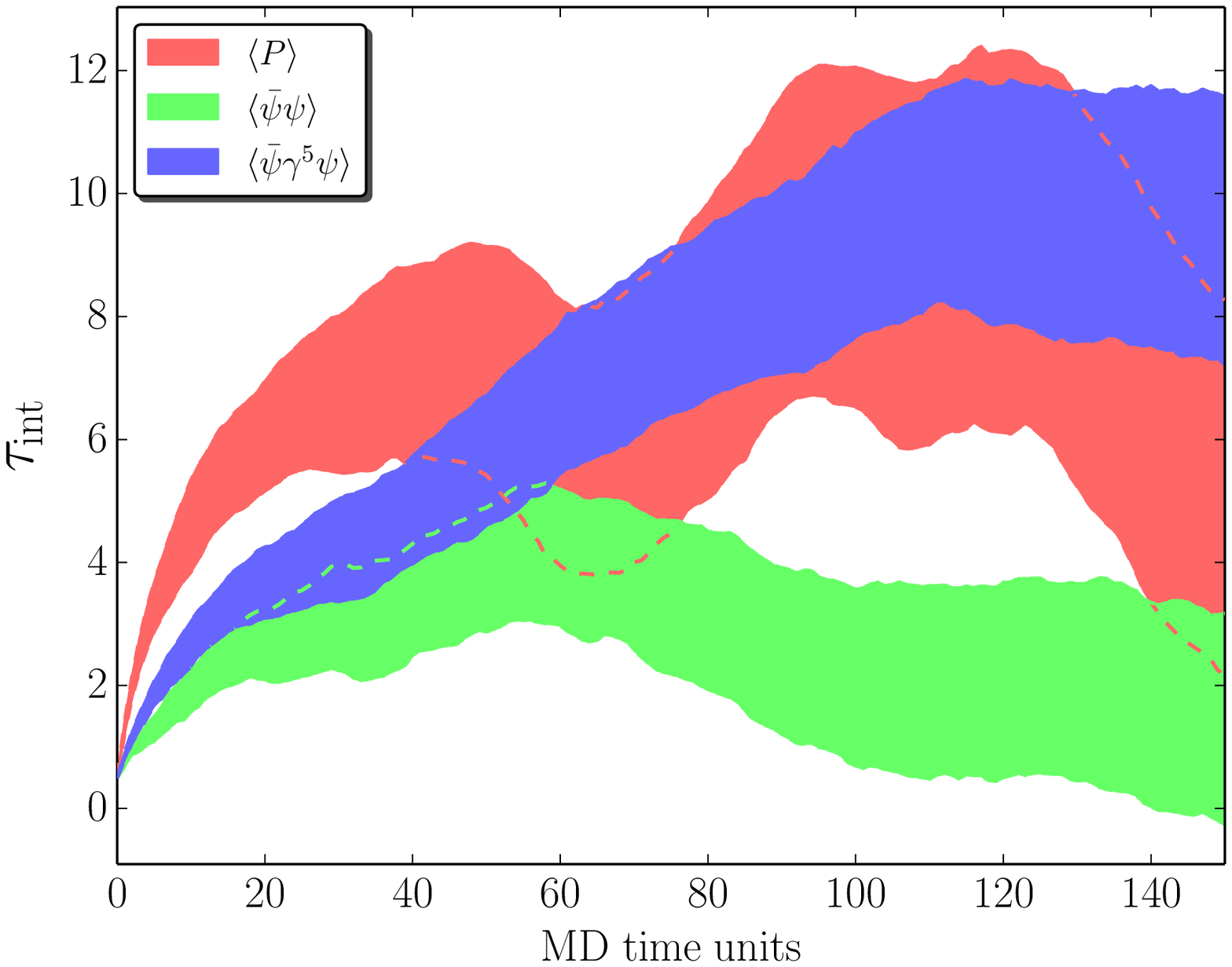}

\caption{
The integrated autocorrelation time as a function of the Molecular Dynamics time separation.
\label{fig-autocorr}
}
\end{figure}

For each of the three ensembles we measured the plaquette, chiral and pseudoscalar condensates after every trajectory. We also measured the topological charge after every fifth trajectory starting from 500 MD time units of evolution. We plot the Monte Carlo time histories of these quantities in Figure~\ref{fig-evolplots}. From the plaquette, chiral and pseudoscalar condensates we measure the accumulated integrated autocorrelation times as a function of the MD time separation, the results of which are shown in Figure~\ref{fig-autocorr}. Here the errors were obtained by binning the correlations $C(t,t+\Delta)$ with a fixed separation $\Delta$ over neighboring configurations, using the first technique described in Section II.D of Ref.~\cite{Arthur:2012opa}. We also measured the autocorrelation using the topological charge measurements, but found that these measurements, performed every 5 MD time units, were sufficiently separated that no observable correlations were found within the statistical errors; we therefore excluded these data from the figures. The figures suggest integrated autocorrelation times of ${\sim}15$ MD time units for the G-parity ensembles, and ${\sim}10$ for the GP0 ensemble, such that there are 30 and 20 MD time units between independent measurements, respectively.

The expectation values of the plaquette, chiral condensate and pseudoscalar density are given in Table~\ref{ens-expvals}. Here the errors are determined after binning uniformly over 40 MD time units. We observe excellent consistency between the plaquettes, and the pseudoscalar density is consistent with zero indicating good topological sampling. The chiral condensate agrees well between the GP0 and GP1 ensembles but differs between the GP0 and GP2 ensembles by 1.9(7)\% level. While this may be simply statistics, as supported by the agreement between the GP0 and GP1 ensembles, this difference may also be attributed to the explicit breaking of the flavor non-singlet axial symmetry by the boundary conditions that we identified in Section~\ref{sec-flavsym}. We would expect that any change in the chiral condensate would also manifest directly in the pion energy, which is related to the chiral condensate at leading order in chiral perturbation theory ($\chi$PT). In Section~\ref{sec-measpion} below, we find that the pion energy on this ensemble is indeed different than expected at the 1.5\% level, but the value obtained is {\it lower} than expected, in the opposite direction to that suggested by $\chi$PT for an upwards shift in the chiral condensate. We therefore conclude that the observed effect is likely statistical in nature or that a more elaborate analysis using finite-volume $\chi$PT is needed.

\begin{table}[tp]
\begin{tabular}{c|ccc}
\hline\hline
Ens. & $\langle P \rangle$ & $\langle \bar\psi\psi \rangle$ & $\langle \bar\psi\gamma^5\psi \rangle$  \\
\hline
GP0 &  0.588048(23) & 0.0017101(98) & 0.000005(14)\\
GP1 &  0.588113(31) & 0.0017073(94) & 0.000032(19)\\
GP2 &  0.587987(38) & 0.0017434(67) & -0.000029(23)
\end{tabular}
\caption{The expectation values of the plaquette (second column), chiral condensate (third column) and pseudoscalar density (fourth column) for each ensemble.\label{ens-expvals} }
\end{table}

\section{Results}
\label{sec:Results}
In this section we present measurements of the residual mass, the axial-current renormalization factor (via the Ward-Takahashi identity), the pion and kaon energies and decay constants, and the neutral kaon mixing parameter, $B_K$. These quantities are computed on each of the three ensembles discussed above, and compared to examine the effects of the boundary.

We perform our measurements for all three ensembles using 104 configurations in the range 500-1530, measuring every 10 MD time units. In order to reduce the effect of autocorrelations in the data we consistently bin (average) over $4$ successive configurations ($40$ MD time units) for simplicity.

We use Coulomb gauge-fixed wall source propagators to compute our correlation functions. Light and heavy quark propagators are generated with unitary masses, $m_l=0.01$ and $m_h=0.032$ respectively. On the GP0 ensemble we project onto zero momentum on the source timeplane, whereas for the G-parity ensembles we project onto non-zero source momenta: on the GP1 ensemble we use momenta $\vec p = \pm\frac{\pi}{2L}(1,0,0)$ and on the GP2 ensemble we use $\vec p = \pm\frac{\pi}{2L}(1,1,0)$. The sources are placed on time-slice $0$ on all configurations, and we generate propagators with both periodic ($p$) and antiperiodic ($a$) temporal boundary conditions. From these we take linear combinations $F=p+a$ and $B=p-a$ that project out the forwards and backwards moving components of the propagator respectively, effectively doubling the temporal periodicity and significantly reducing round-the-world finite-time-extent effects.

For the G-parity ensembles we must construct the $2\times 2$ flavor matrix propagators, with elements
\begin{dmath}
\Theta_{fi}(\vec x,t; \vec p) = \sum_{\vec y} \prop_{fg}(\vec x,t ; \vec y,0) \tilde V^\dagger_{gh}(\vec y)\eta_{hi}(\vec y) e^{-i\vec p\cdot y}\,,
\end{dmath}
where $f$--$i$ are flavor indices, $\tilde V$ is the gauge-fixing matrix, and $\eta$ is a generic, flavor-matrix-valued source. For a wall source, $\eta$ is just the unit matrix for all sites. In order to obtain $\Theta$ one can simply invert propagators for both source flavor indices separately, i.e. construct $\Theta_{f1}$ and $\Theta_{f2}$ (for all $f$) with two inversions of the two-flavor Dirac matrix. With this approach, one must perform four inversions in order to both compute $\Theta(\vec p)$ and its parity partner $\Theta(-\vec p)$. 

In most cases we can use the exact relation between the propagator and its complex conjugate, Eq.~\eqref{eqn-propconjreln}, to obtain both $\Theta(\vec p)$ and $\Theta(-\vec p)$ with only two inversions of the Dirac matrix as follows: Taking the complex conjugate of the solution vector,
\begin{dmath}
\Theta^*(\vec x,t; \vec p) = \sum_{\vec y} \prop^*(\vec x,t ; \vec y,0) \tilde V^T(\vec y)\eta^*(\vec y) e^{i\vec p\cdot y}\,,
                         = \sum_{\vec y} \left[\gamma^5 C^{-1} \sigma_2\prop(\vec x,t ; \vec y,0)\sigma_2 C \gamma^5\right]\left[ \sigma_2 \tilde V^\dagger(\vec y)\sigma_2 \right]\eta^*(\vec y) e^{i\vec p\cdot y}\,.
\end{dmath}
For any source that has the property $C \gamma^5 \eta^*(\vec y) = \eta(\vec y) C \gamma^5$ (which trivially applies to any source proportional to the unit matrix, such as the wall source in question), the above reduces to
\begin{dmath}
\Theta^*(\vec x,t; \vec p) = \gamma^5 C^{-1} \sigma_2\Theta(\vec x,t; -\vec p)\sigma_2 C\gamma^5\,.
\end{dmath}
Multiplying from both sides by $\sigma_2$ and examining the equation in component form, we find
\begin{dmath}
 \left(\begin{array}{cc}\Theta^*_{22}(\vec x,t; \vec p)  & -\Theta^*_{21}(\vec x,t; \vec p)  \\ -\Theta^*_{12}(\vec x,t; \vec p)  & \Theta^*_{11}(\vec x,t; \vec p) \end{array}\right) = \gamma^5 C^{-1}\left(\begin{array}{cc}\Theta_{11}(\vec x,t; -\vec p)  & \Theta_{12}(\vec x,t; -\vec p)  \\ \Theta_{21}(\vec x,t; -\vec p)  & \Theta_{22}(\vec x,t; -\vec p) \end{array}\right)C\gamma^5\,.
\end{dmath}
Notice that we can obtain the first column of the left-hand side, i.e. those elements with source index 2, from the first column of the right-hand side, which contains only elements with source index 1. Using this relation we can obtain both $\Theta(\vec p)$ and $\Theta(-\vec p)$ by computing only $\Theta_{f1}(\vec p)$ and $\Theta_{f1}(-\vec p)$, from which we obtain $\Theta_{f2}(-\vec p)$ and $\Theta_{f2}(\vec p)$, respectively.

\subsection{Pion two-point function}
\label{sec-measpion}
Using the momentum-eigenstate field operators obtained in Section~\ref{sec:Symmetries} we derived the appropriate operator for a neutral pion state in Eq.~\eqref{eq-nonlocalpi0op}. Taking $p_i = \pi/2L$ for each G-parity direction (i.e. $n_p = 0$ in Eq.~\eqref{eqn-transconvfieldndirsimp}), and utilizing a wall source and point sink, we obtain the following form for the pion two-point function: 
\begin{dmath}
C^{LW}_{\pi_0\pi_0}(t) 
= \sum_{\vec x,\vec y, \vec z} \left\langle \left[e^{2i\vec p\cdot \vec x}\frac{-i}{\sqrt{2}}\bar\psi_l(\vec x,t)\sigma_3\gamma^5\psi_l(\vec x,t)\right]\left[ e^{-i\vec p\cdot (\vec y+\vec z)}\frac{-i}{\sqrt{2}}\bar\psi_l(\vec y,0)\sigma_3\half(1+\sigma_2)\gamma^5\psi_l(\vec z,0) \right]\right\rangle
= \frac{1}{4}\sum_{\vec x,\vec y, \vec z} e^{2i\vec p\cdot \vec x}e^{-i\vec p\cdot (\vec y+\vec z)}{\rm tr}\left\{ \prop_l^\dagger(\vec x,t;\vec z,0)\sigma_3\prop_l(\vec x,t;\vec y,0)\sigma_3(1+\sigma_2)  \right\}
=
\frac{1}{4}\sum_{\vec x}e^{2i\vec p\cdot \vec x}{\rm tr}\left\{  \Theta_l^\dagger(\vec x,t; -\vec p,0) \sigma_3 \Theta_l(\vec x,t; \vec p,0) \sigma_3 (1+\sigma_2) \right\}\,, \label{eq-pi0corr}
\end{dmath}
where $\Theta_l$ are the light-quark solution vectors defined above. The $LW$ superscript indicates that the correlation function has a local (point) sink and wall source. At the sink location we have not included the $(1-\sigma_2)$ projector as it is not necessary for a local operator. We remind the reader that Wick contractions resulting in disconnected loops at the sink location vanish because of the $\gamma^5$-hermiticity of the propagator and its relation to its complex conjugate (cf. Section~\ref{sec-lightmesonop}).

It is straightforward to show that the last line of Eq.~\eqref{eq-pi0corr} can also be obtained by considering the charged pion operators given in Eq.~\eqref{eq-pioncreationopsgparity}, which is necessary due to the exact isospin symmetry of the formalism. We therefore generically assign the source/sink operator with the label $P$ for pseudoscalar.

The corresponding pion two-point function for the GP0 ensemble is,
\begin{dmath}
C^{LW}_{PP}(t) = \sum_{\vec x}{\rm tr}\left\{ \Theta_l^\dagger(\vec x,t;\vec 0,0)\Theta_l(\vec x,t; \vec 0,0) \right\}\,. \label{eq-picorrnogp}
\end{dmath}

For all ensembles we compute the correlation functions with both the forwards ($F$) and backwards ($B$) moving propagators and average the results (after the appropriate temporal reflection) to improve statistics. As the amount of data is large, we were unable to accurately resolve a correlation matrix and therefore performed uncorrelated fits (i.e. with a diagonal correlation matrix) here and for the other results presented in this document. 

For use below we also fit the correlation functions with axial-vector sink operators appropriate for neutral pions, 
\begin{dmath}
A_\mu = \frac{-i}{\sqrt{2}}(\bar u\gamma^\mu\gamma^5 u - \bar d\gamma^\mu\gamma^5 d) = \frac{i}{\sqrt{2}}\bar\psi_l \gamma^\mu\gamma^5\sigma_3\psi
%
%
\end{dmath}
for $\mu$ in the temporal direction and each of the spatial directions in which GPBC are applied. Here we have chosen the phase convention such that the two-point function describing a neutral pion annihilated by the time-component axial-vector operator,
\begin{dmath}
C^{LW}_{A_4P}(t) 
= \sum_{\vec x,\vec y, \vec z} \left\langle \left[e^{2i\vec p\cdot \vec x}\frac{i}{\sqrt{2}}\bar\psi_l(\vec x,t)\sigma_3\gamma^4\gamma^5\psi_l(\vec x,t)\right]\left[ e^{-i\vec p\cdot (\vec y+\vec z)}\frac{-i}{\sqrt{2}}\bar\psi_l(\vec y,0)\sigma_3\half(1+\sigma_2)\gamma^5\psi_l(\vec z,0) \right]\right\rangle
= \frac{1}{4}\sum_{\vec x,\vec y, \vec z} e^{2i\vec p\cdot \vec x}e^{-i\vec p\cdot (\vec y+\vec z)}{\rm tr}\left\{ \prop_l^\dagger(\vec x,t;\vec z,0)\sigma_3\gamma^4\prop_l(\vec x,t;\vec y,0)\sigma_3(1+\sigma_2)  \right\}\,,\label{eq-APcorr}
\end{dmath}
is real and positive. With this convention the correlators with spatial axial-vector sinks are pure-imaginary, which we will show is necessary to fulfill the axial Ward-Takahashi identity. 

We also compute the wall-source, wall-sink $PP$ correlation function, which takes the following form on the G-parity ensembles:
\begin{dmath}
C^{WW}_{PP}(t) 
= \sum_{\vec r,\vec s,\vec y, \vec z} \left\langle\left[e^{i\vec p\cdot (\vec r+\vec s)}\frac{-i}{\sqrt{2}}\bar\psi_l(\vec r,t)\sigma_3\half(1-\sigma_2)\gamma^5\psi_l(\vec s,t) \right]\left[ e^{-i\vec p\cdot (\vec y+\vec z)}\frac{-i}{\sqrt{2}}\bar\psi_l(\vec y,0)\sigma_3\half(1+\sigma_2)\gamma^5\psi_l(\vec z,0) \right]\right\rangle
=
\frac{1}{8}\sum_{\vec r,\vec s}e^{i\vec p\cdot(\vec r + \vec s )}{\rm tr}\left\{\Theta_l^\dagger(\vec r,t; -\vec p,0) \sigma_3(1-\sigma_2) \Theta_l(\vec s,t; \vec p,0) \sigma_3 (1+\sigma_2)  \right\}\,, \label{eq-PPWWcorr}
\end{dmath}
where again we have suppressed the gauge fixing matrices at source and sink.

On each ensemble we simultaneously fit the various correlation functions with a common exponent. We also include the point-sink two-point function with the $A_4$ operator at the sink and the Hermitian conjugate of this operator (ensuring a positive sign) at the source, in order to improve the signal for the exponent. The data were fit to the following forms,
\begin{equation}
C^{LW}_{O_1 O_2}(t) = {\cal N}^{LW}_{O_1 O_2}\left[ e^{-Et} \pm e^{-E(2T-t)} \right] \label{eq-twoptfitform}
\end{equation}
where $O_1$ and $O_2$ are the sink and source operators respectively, $T=32$ is the lattice temporal length, and ${\cal N}$ are the amplitudes. We use the notation $P$ for the pseudoscalar operator and $A_\mu$ for the axial operators. The sign of the backwards propagating component (the second term in the brackets) depends on the how the operators transform under time reflection: The $PP$ and $A_i P$, for $i$ in a spatial direction, require the plus (cosh-like) form and the $A_4 P$ the minus (sinh-like) form. Note the temporal length is doubled due to our use of the $F$ and $B$ combinations of temporal boundary condition.

\begin{figure}[tp]
\includegraphics[width=0.48\textwidth]{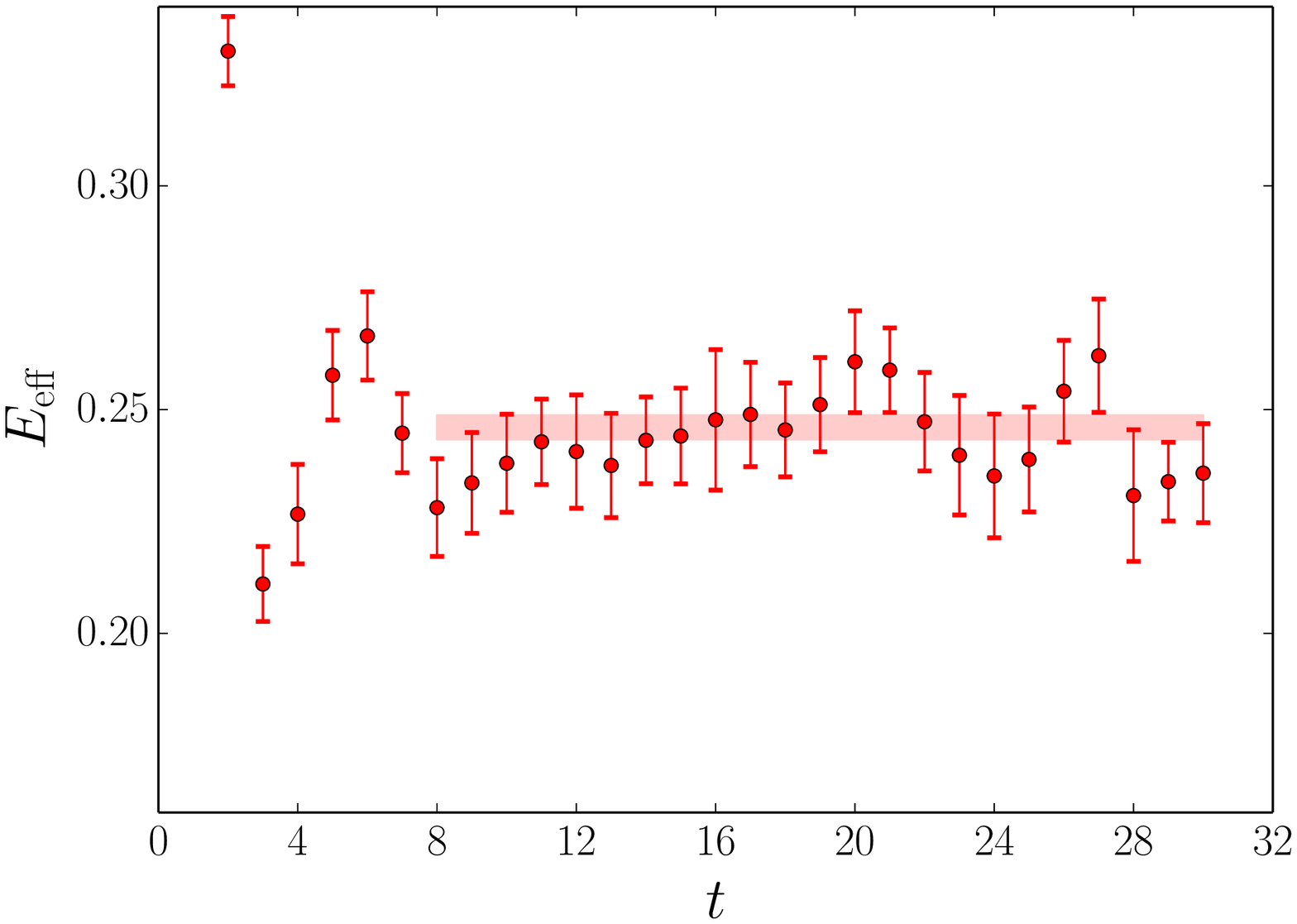}
\includegraphics[width=0.48\textwidth]{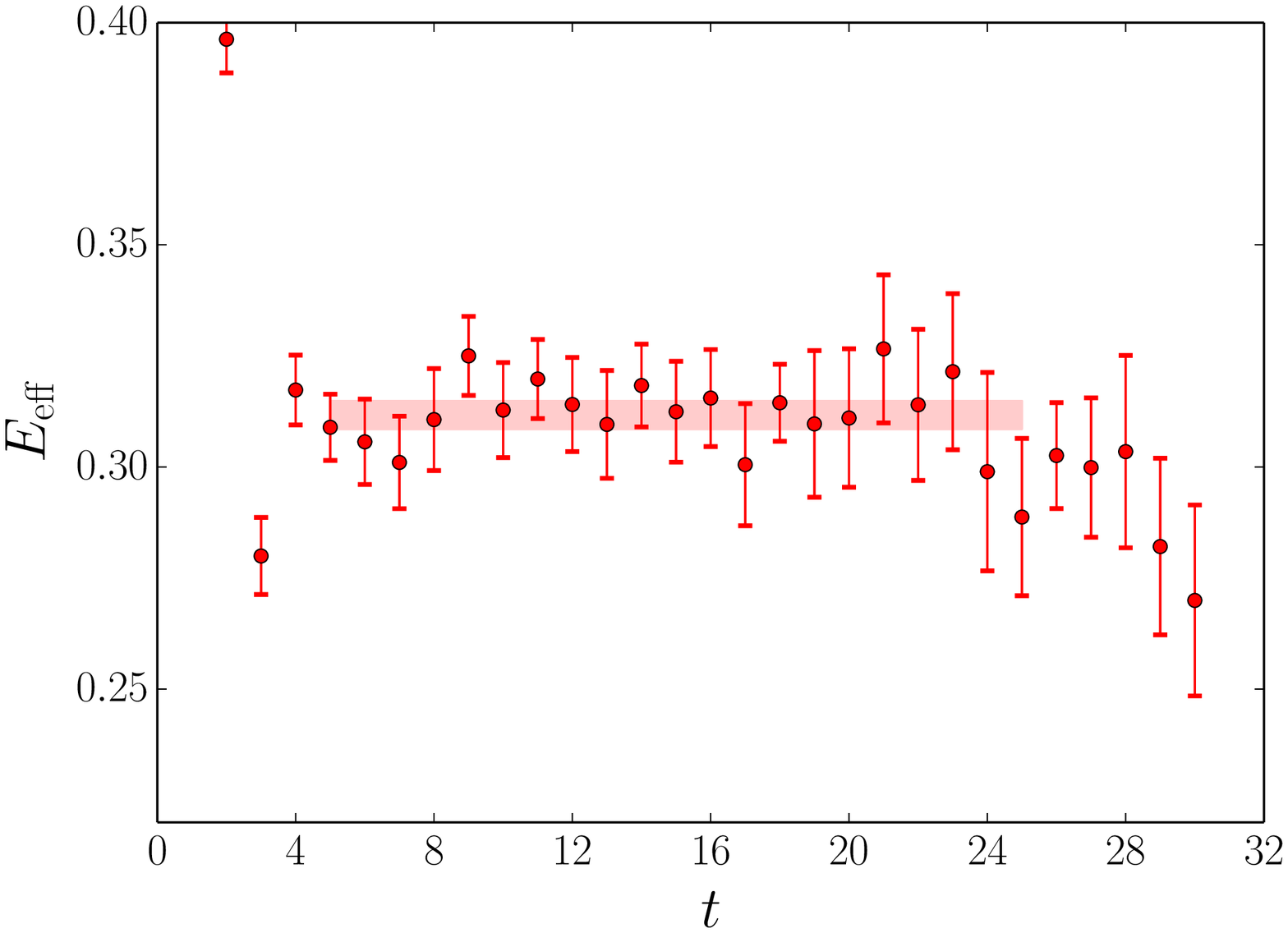}
\includegraphics[width=0.48\textwidth]{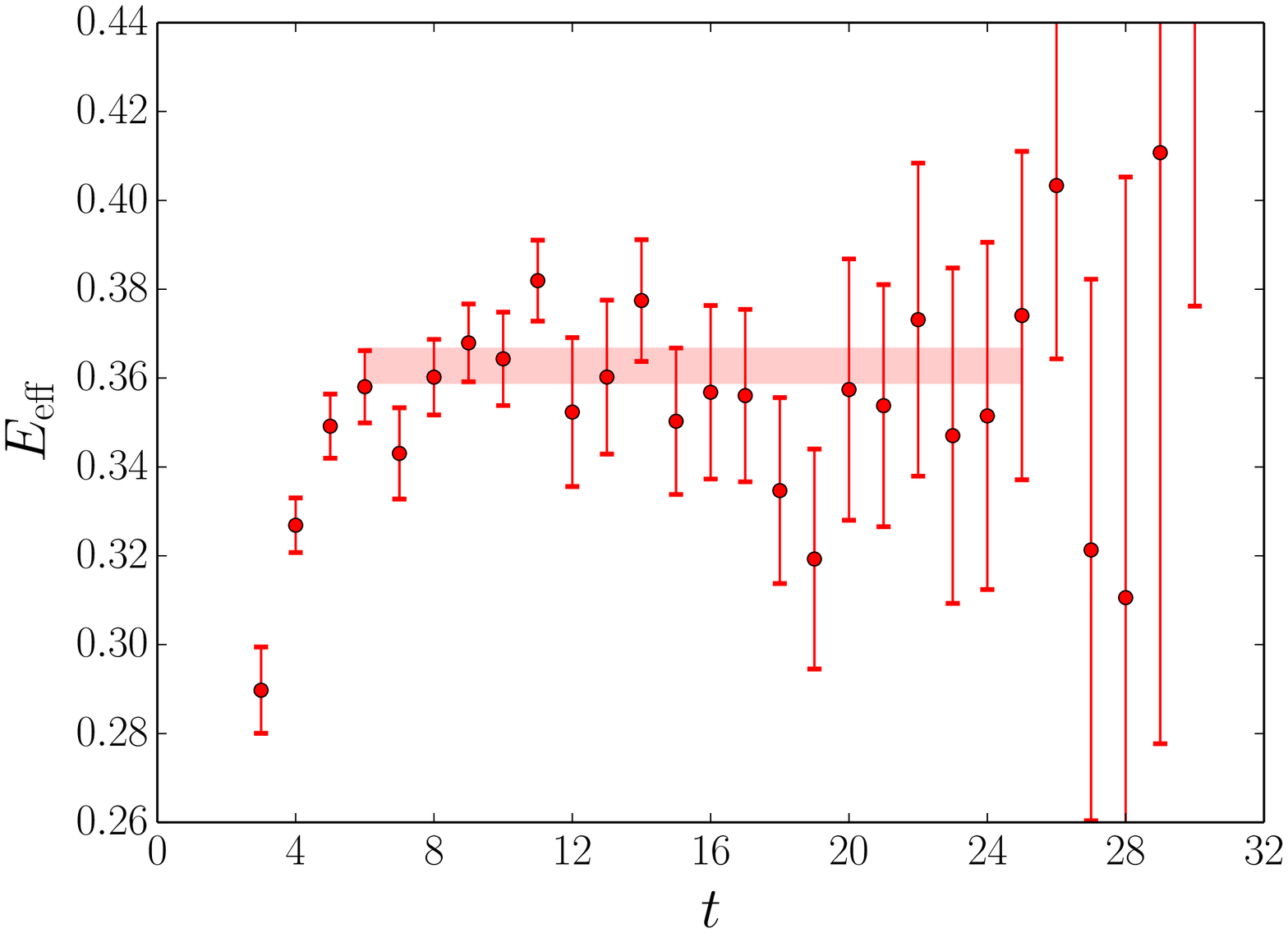}

\caption{The pion effective energy in the $PP$ channel overlaid by the fitted value on the GP0 (upper-left), GP1 (upper-right) and GP2 (lower) ensembles.\label{fig-pioneffmass} }
\end{figure}

The fitted masses and amplitudes that we obtain are given in Table~\ref{tab-fitmpi}, alongside the fit ranges chosen by eye based on effective mass plots (used uniformly for all correlators on a given ensemble) and the uncorrelated $\chi^2/{\rm dof}$. A far more precise measurement of $m_\pi$ on the GP0 ensemble was performed in Ref.~\cite{Liu:2012}, which we also include in this table. In Figure~\ref{fig-pioneffmass} we show effective mass plots for the $PP$ point-sink channel. 

In Table~\ref{tab-fitmpi} we also list the predicted values of the energy that one obtains by applying the continuum dispersion relation,
\begin{equation}
E_\pi = \sqrt{ m_\pi^2 + \frac{n\pi^2}{L^2} }
\end{equation}
where $n$ is the number of G-parity directions, $L=16$ is the spatial box size, and $\sqrt{n}\pi/L$ are the magnitudes of the expected pion momenta. Here $m_\pi$ is the fitted mass obtained from the GP0 data in Ref.~\cite{Liu:2012}. From the table we observe good agreement between the predicted and measured energies on the GP1 ensemble. The value on the GP2 ensemble is $1.8(1.0)\%$ smaller than the predicted value, although if we replace the continuum dispersion relation with the following lattice expression, $E_\pi = \sqrt{m_\pi^2 + n\sin^2(\pi/L)}$, the discrepancy drops to $1.4(1.0)\%$ which is statistically compatible with zero. On the other hand it is possible that this effect is related to the larger value for the chiral condensate observed on this ensemble that we discussed in Section~\ref{sec-ensproperties}.

\begin{table}[tp]
\begin{tabular}{l|c|c|c}
\hline\hline
Quantity  & GP0  & GP1  & GP2 \\
\hline
Fit range & 8--30 & 5--25 & 6--25 \\
$\chi^2/{\rm dof}$  & 0.140(86)  & 0.29(15)  & 0.48(35) \\
\hline
$E_\pi$  & $0.2460(27)$  & $0.3117(32)$  & $0.3628(38)$ \\
$E_\pi$~\cite{Liu:2012} & 0.24373(47) & - & - \\
$E_\pi^{\rm pred}$  & -  & 0.31298(37)  & 0.36947(31) \\
\hline
${\cal N}^{LW}_{PP}$  & $1.326(43)\times 10^{5}$  & $5.14(16)\times 10^{4}$  & $4.19(14)\times 10^{4}$ \\
${\cal N}^{LW}_{A_4A_4}$  & $7.01(28)\times 10^{3}$  & $5.14(15)\times 10^{3}$  & $5.98(22)\times 10^{3}$ \\
${\cal N}^{LW}_{A_4P}$  & $1.964(55)\times 10^{4}$  & $9.89(24)\times 10^{3}$  & $9.66(27)\times 10^{3}$ \\
${\cal N}^{LW}_{A_1P}$  & -  & $6.33(22)\times 10^{3}$  & $5.06(20)\times 10^{3}$ \\
${\cal N}^{LW}_{A_2P}$  & -  & -  & $5.14(23)\times 10^{3}$ \\
${\cal N}^{WW}_{PP}$  & $4.99(15)\times 10^{7}$  & $1.012(30)\times 10^{7}$  & $7.83(25)\times 10^{6}$ \\
\end{tabular}
\caption{The results of simultaneous fits to the $PP$, $A_4A_4$ and $A_\mu P$ correlation functions on each of the three ensembles ($\mu=4$ is the time direction). The superscripts $LW$ and $WW$ refer to wall-source-point-sink and wall-source-wall-sink correlators respectively. For the $A_1P$ and $A_2P$ correlators, the amplitude ${\cal N}$ is of the imaginary part of the correlator, and for the remainder it is of the real part. The value of $E_\pi$ on the fifth line is that measured in Ref.~\cite{Liu:2012}. $E_\pi^{\rm pred}$ is the predicted pion energy obtained by applying the continuum dispersion relation to this, more precise value. \label{tab-fitmpi}}
\end{table}

\subsection{Signal-to-noise ratio of pion two-point functions}

An interesting observation from Figure~\ref{fig-pioneffmass} is that the signal-to-noise ratio for the G-parity ensembles appears to reduce with time, and degenerates more quickly as we increase the number of G-parity directions. This will unfortunately hamper our ability to measure the $\pi\pi$ energy and $K\to\pi\pi$ amplitudes with G-parity boundary conditions. 

The origin of the falling signal-to-noise ratio can be understood via Lepage's argument~\cite{Lepage:1989hd} for the statistical error on a correlation function, which states that for a Green's function $\cal O$ computed on a gauge configuration $i$ in an ensemble of size $N$, the standard error on the mean, $\sigma$, of the measurement $\langle {\cal O} \rangle = \frac{1}{N}\sum_{i=1}^N {\cal O}[U_i]$ can be obtained as follows:
\begin{dmath}
\sigma^2 = \frac{1}{N}\left(\langle {\cal O} {\cal O}^\dagger \rangle - |\langle {\cal O} \rangle|^2\right)\,,
\end{dmath}
hence the error must grow as the square-root of the expectation value of the square of the Green's function. The signal to noise ratio ${\bf S}/{\bf N}$, where ${\bf S}=|\langle {\cal O} \rangle|$ and ${\bf N}=\sigma$ is
\begin{dmath}
{\bf S}/{\bf N} = \frac{1}{\frac{1}{\sqrt{N} }\sqrt{ \frac{\langle {\cal O} {\cal O}^\dagger \rangle}{|\langle  {\cal O}\rangle|^2} - 1} } \approx \frac{\sqrt{N}}{\sqrt{ \frac{\frac{1}{N}\sum_i ({\cal O}{\cal O}^\dagger)[U_i]}{ |\frac{1}{N}\sum_i {\cal O} [U_i] |^2 } } } \overset{t\rightarrow\infty}{\rightarrow} \frac{\sqrt{N}}{\sqrt{ \frac{ A_2\exp(-E_2 t) }{ A_1^2 \exp(-2E_1t) } } } \propto \sqrt{N} e^{-(E_1-\half E_2)t}\,.
\end{dmath}
Here $E_2$ and $A_2$ are the energy and amplitude of the lightest state in the squared correlation function $\langle {\cal O} {\cal O}^\dagger\rangle$, and $E_1$ and $A_1$ are those of the state that we are measuring (in this case the pion). We can see that the time dependence of the signal-to-noise ratio drops out when $E_2 = 2E_1$, and that the signal-to-noise ratio will decrease with time when $E_2 < 2E_1$.

For the two-point functions in the previous section, the Green's function ${\cal O}{\cal O}^\dagger$ describes the propagation of two quarks and two antiquarks. On the GP0 ensemble, the ground state of this system is just two pions at rest, hence the ${\bf S}/{\bf N}$ is constant. On the G-parity ensembles this state does not exist: instead the lightest two-pion state contains only moving pions, and we again have $E_2 = 2E_1$. We might therefore expect that ${\bf S}/{\bf N} $ would also remain constant for these ensembles. However, the four-quark Green's function also contains a contribution from two pseudoscalar SU(2) flavor singlet particles with quark content $\bar u u + \bar d d$. As only the connected components are present in the noise, this state has the same energy as the stationary pion on the GP0 ensemble, and the signal-to-noise will remain constant. However, on the G-parity ensembles this state is G-parity even and will therefore have the energy of the {\it stationary} pion rather than the moving pion, such that the states entering the noise are lighter than the moving pions in the signal. To confirm this we measured the energy of the flavor-singlet, whose operator is $\bar\psi \gamma^5 \psi$ in our formalism (cf. Table~\ref{tab-lllocop}), on the GP1 ensemble and obtained an energy of $E_{\rm singlet} = 0.2440(50)$ which is in excellent agreement with the stationary pion mass of $m_\pi = 0.2460(27)$ determined from the GP0 ensemble.

Note that if we replace one of the light quark propagators with that of the strange/strange-prime quark pair (which differ only in their input mass), the contractions for the flavor singlet are identical to those of the G-parity kaonic state, $\tilde K^0_+$ (see below). As a result the flavor singlet might also be considered a ``light kaon'' in this sense.

\begin{figure}[tp]
\includegraphics[width=0.48\textwidth]{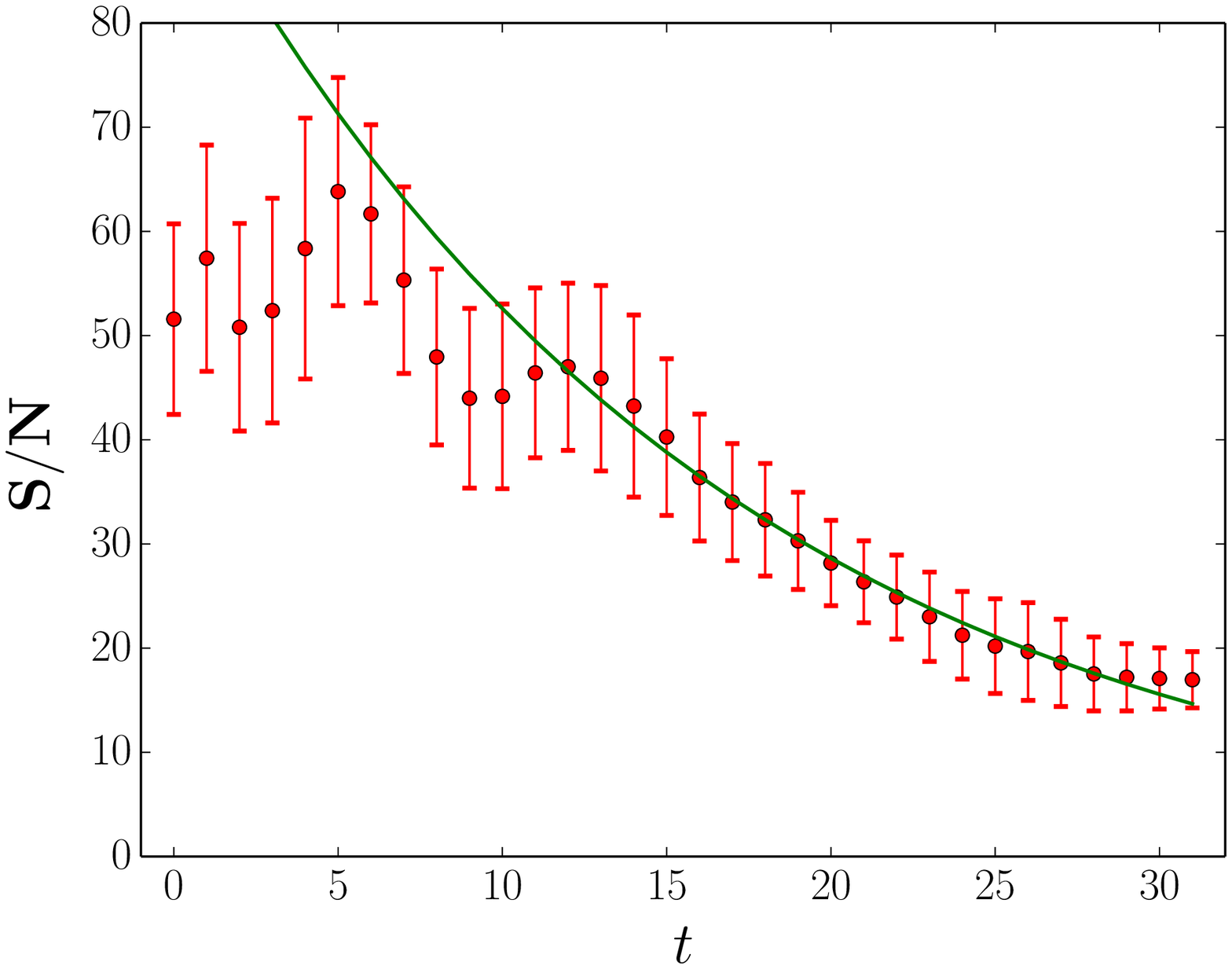}
\includegraphics[width=0.48\textwidth]{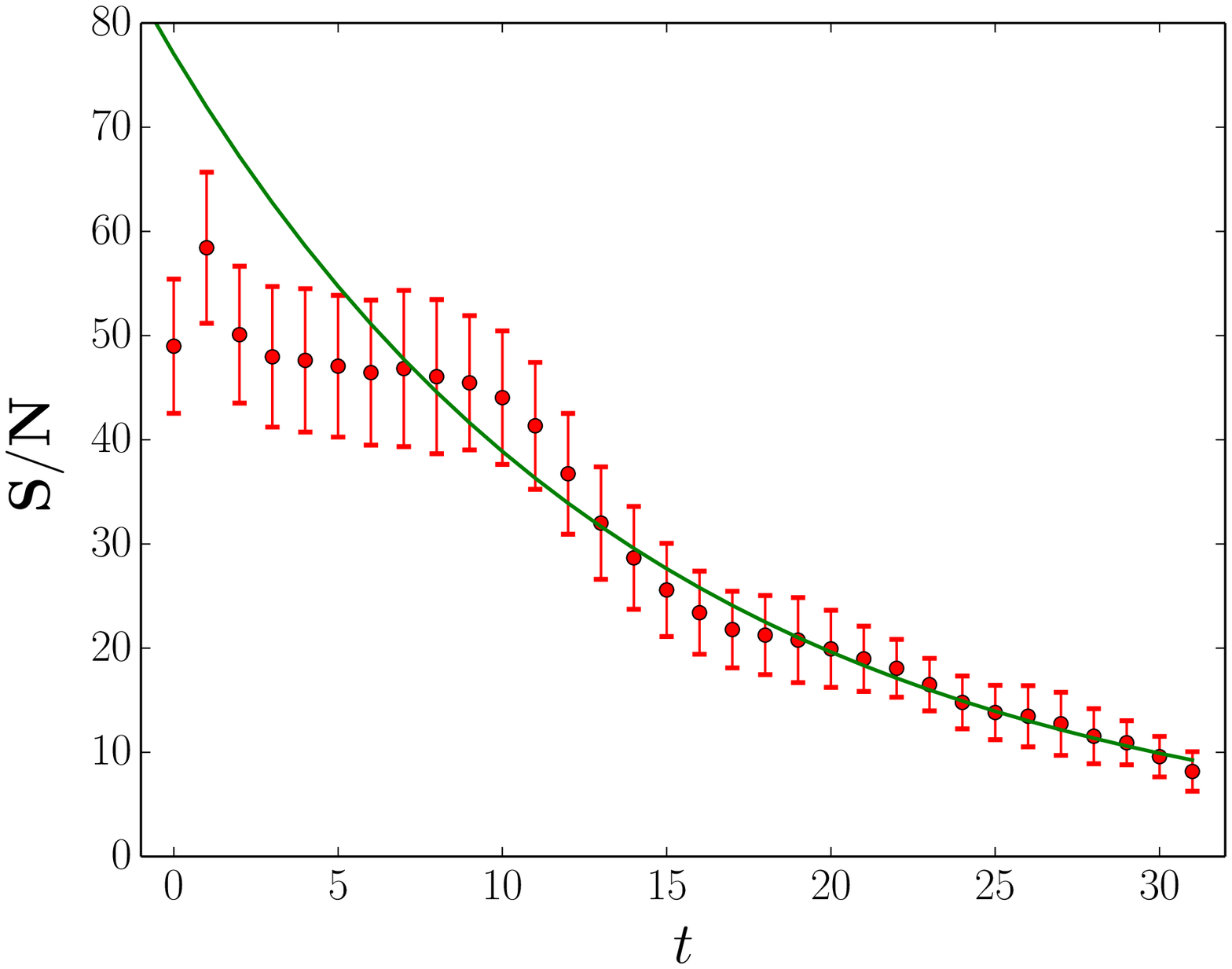}\\
\includegraphics[width=0.48\textwidth]{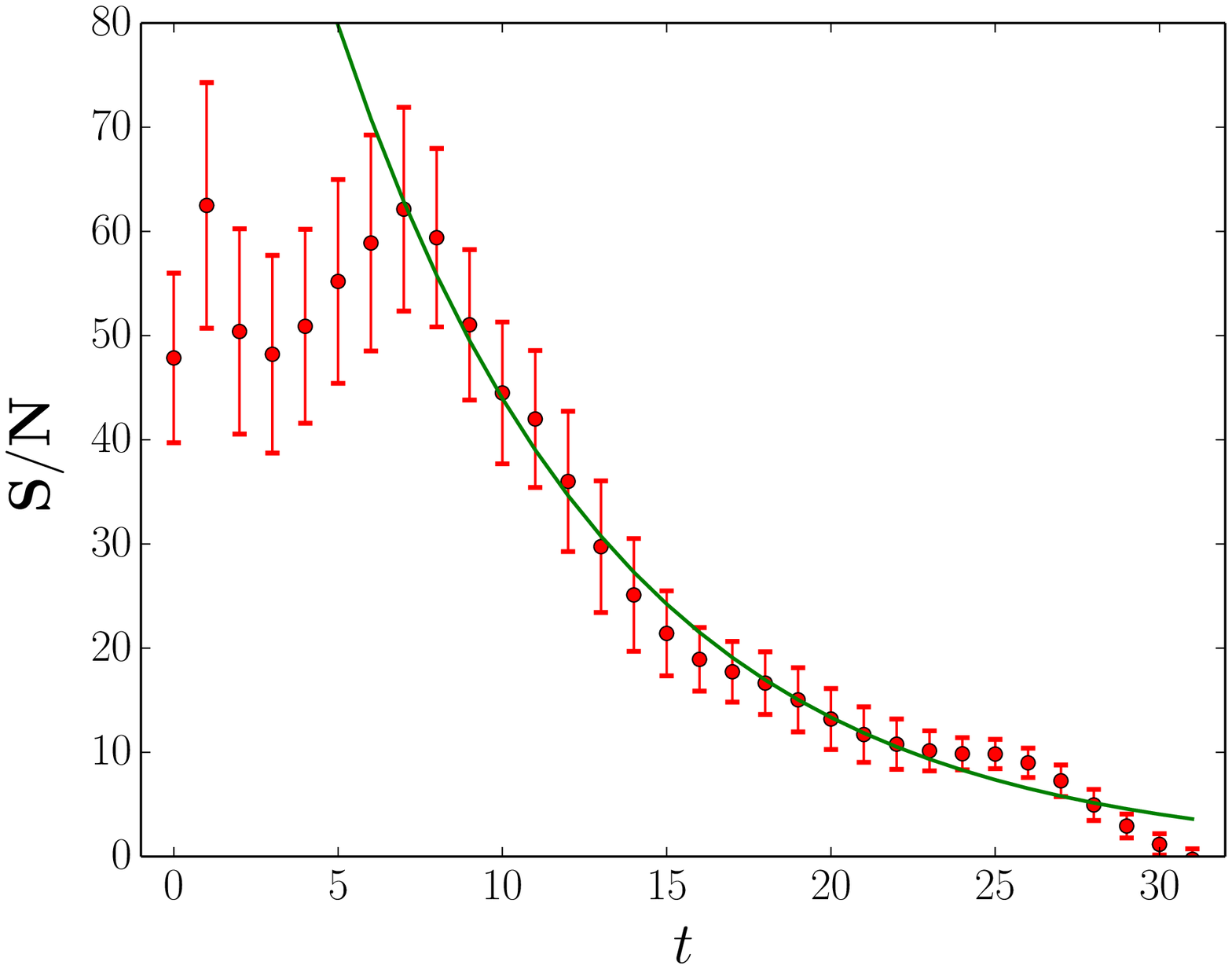}
\caption{ The signal-to-noise ratio of the $PP$ wall-source-point-sink correlation function computed using the jackknife error bars on the GP1 ensemble (left) and the GP2 ensemble (right), overlaid by the fitted exponential. \label{fig-mpisnr}}
\end{figure}

To demonstrate that it is indeed this flavor singlet state that dominates the noise we performed single-exponential fits to the signal-to-noise ratio of the $PP$ wall-source-point-sink correlation function on all three ensembles; the results are given in Table~\ref{tab-fitsnr} and the data in Figure~\ref{fig-mpisnr}. We observe excellent consistency between the predicted and measured exponential falloff of the signal-to-noise ratio on both the GP1 and GP2 ensembles. On the GP0 ensemble however the LePage argument predicts a constant behavior but instead we observe a significant exponential falloff in ${\bf S}/{\bf N} $ suggesting a state of energy $0.370(32)$ appears in the noise. This discrepancy has not been understood but may result from the presence of multiple heavier states contributing with different signs or the absence of disconnected diagrams.

\begin{table}[tp]
\begin{tabular}{l|c|c|c}
\hline\hline
Quantity  & GP0  & GP1  & GP2 \\
\hline
Fit range & 13--30 & 11--30 & 9--30 \\
$\chi^2/{\rm dof}$  & 0.042(54)  & 0.13(15)  & 1.01(44) \\
\hline
Measured  & $0.061(14)$  & $0.068(16)$  & $0.119(12)$ \\
Expected  & $0$          & $0.0657(42)$ & $0.1168(47)$ 
\end{tabular}
\caption{The results of fitting a single exponential to the signal-to-noise ratio in the $PP$ channel. The final line gives the values predicted by the LePage argument~\cite{Lepage:1989hd} \label{tab-fitsnr}}
\end{table}

\subsection{Effects of quark-level rotational symmetry breaking}
\label{sec-rotsymbreakmeas}

In Section~\ref{sec-rotsymact} we observed that on a lattice with GPBC, the quark's momentum components in orthogonal directions must differ only by integer multiples of $2\pi/L$. As a result the momentum $(1,1,0)\frac{\pi}{2L}$ for GPBC in two directions is allowed but $(1,-1,0)\frac{\pi}{2L}$ is not. Given that these two momenta are related by a cubic rotation, the above restriction constitutes a breaking of the cubic rotational symmetry at the quark level.

In this subsection we will describe a numerical investigation of these rotational symmetry breaking effects and a strategy for constructing interpolating operators which reduce the size of this cubic symmetry breaking to what we believe is an acceptable level. This investigation was performed on a separate $16^3\times 32\times 8$ ensemble of M\"obius domain wall fermions with $b+c=2$ and the Iwasaki+DSDR gauge action with $\beta=1.75$ that was generated as part of the preparation for our $K\to\pi\pi$ project~\cite{Bai:2015nea}. Aside from the lattice size, M\"obius parameter and a heavier pion mass of 440 MeV, this ensemble is otherwise identical to the $32^3\times 64\times 12$ ensemble used for the $K\to\pi\pi$ calculation, and has G-parity BCs in three directions. Instead of the Coulomb gauge fixed wall sources used elsewhere in this document, this investigation was performed using hydrogen wavefunction smeared sources and the correlation functions were computed using the all-to-all propagator technique~\cite{Foley:2005ac} used for the $K\to\pi\pi$ calculation. In this technique an approximation to the all-to-all propagator is generated by combining exact low eigenmodes (we used 100 eigenmodes here) computed using the Lanczos algorithm with a stochastic approximation to the high mode component. We perform full time, spin, color and flavor dilution such that the random numbers are required only to induce a delta function in the spatial coordinate. We use a single random hit per configuration such that the delta function is generated under the ensemble average.

For each of the four pion momenta $(-2,-2,-2)$  $(2,-2,-2)$ $(-2,2,-2)$ and $(-2,-2,2)$ (in units of $\frac{\pi}{2L}$) we examine the two-point correlation functions of the operators
\begin{dmath}
{\cal O}_\pi^\pm(\vec p,t) = -\frac{i}{\sqrt{2}}\sum_{\vec x, \vec y} e^{-i(\vec p_1\cdot \vec x + \vec p_2\cdot \vec y)}\Theta(|\vec x - \vec y|)\psibar_{\mp\ l}(\vec x,t) \gamma^5 \sigma_3 \psi_{\pm\ l}(\vec y,t)
\end{dmath}
where $\Theta(|\vec x - \vec y|) = \exp(-|\vec x - \vec y|/r)$, with radius $r=2$, is the smearing function, and $\vec p = \vec p_1 + \vec p_2$ is the pion momentum, with $\vec p_1$ and $\vec p_2$ chosen from the appropriate set of allowed momenta. For each operator and pion momentum we choose $\vec p_1$ and $\vec p_2$ as those that have the smallest magnitude subject to the constraints; the chosen momenta are given in Table~\ref{tab-rotstudymomchoices}. Here in each case the ${\cal O}_\pi^-$ form contains the lower magnitude quark momenta, and is therefore the natural primary choice of operator. (Note that for the parity partners of these four momenta, the ${\cal O}_\pi^+$ operator would be the natural choice.)

\begin{table}[t]
\centering
\begin{tabular}{c|cc|cc}
\hline\hline
$\vec p$ & Operator & $\vec p_1$ & $\vec p_2$\\
\hline
(-2,-2,-2) & ${\cal O}_\pi^-$ & (-1,-1,-1) & (-1,-1,-1) \\
           & ${\cal O}_\pi^+$ & (1,1,1) & (-3,-3,-3) \\
\hline
(2,-2,-2)  & ${\cal O}_\pi^-$ & (-1,-1,-1) & (3,-1,-1)\\
	   & ${\cal O}_\pi^+$ & (1,1,1)    & (1,-3,-3)\\
\hline
(-2,2,-2)  & ${\cal O}_\pi^-$ & (-1,-1,-1) & (-1,3,-1)\\
           & ${\cal O}_\pi^+$ & (1,1,1)    & (-3,1,-3)\\
\hline
(-2,-2,2)  & ${\cal O}_\pi^-$ & (-1,-1,-1) & (-1,-1,3)\\
           & ${\cal O}_\pi^+$ & (1,1,1)    & (-3,-3,1)\\
\end{tabular}
\caption{Choices of quark momenta for each operator and pion total momentum, in units of $\frac{\pi}{2L}$. \label{tab-rotstudymomchoices} }
\end{table}

We fit the correlation function 
\begin{dmath}
C^\pm(t) = \frac{1}{L_T} \sum_\tau \langle 0| \left[{\cal O}^\pm_\pi(\vec p,t+\tau)\right]^\dagger {\cal O}^\pm_\pi(\vec p,\tau) |0\rangle 
\end{dmath}
to the form
\begin{dmath}
C^\pm(t) =  {\cal N}_\pi^\pm\left( e^{-E^\pm_\pi t} + e^{-E^\pm_\pi (L_T-t)} \right)\,.
\end{dmath}
(Note that we do not employ the technique of combining propagators with different temporal BCs here.)

\begin{table}[t]
\centering
\begin{tabular}{c|cc|cc}
\hline\hline
$\vec p$ & $E^-_\pi$ & ${\cal N}_\pi^-$ & $E^+_\pi$ & ${\cal N}_\pi^+$\\
\hline
$(-2,-2,-2)$ & $ 0.4639(26)$ & $ 437.3(6.5)\times 10^{3}$ & $ 0.4634(24)$ & $ 1004.6(13.9)\times 10^{3}$\\
$(2,-2,-2)$ & $ 0.4632(22)$ & $ 380.8(5.5)\times 10^{3}$ & $ 0.4642(19)$ & $ 1055.3(12.0)\times 10^{3}$\\
$(-2,2,-2)$ & $ 0.4651(26)$ & $ 383.0(6.3)\times 10^{3}$ & $ 0.4660(26)$ & $ 1055.1(15.9)\times 10^{3}$\\
$(-2,-2,2)$ & $ 0.4640(23)$ & $ 380.6(5.6)\times 10^{3}$ & $ 0.4639(21)$ & $ 1041.3(14.0)\times 10^{3}$\\
\end{tabular}
\caption{Fit parameters for the $C^-$ and $C^+$ correlation function for each choice of pion momentum $\vec p$ (in units of $\frac{\pi}{2L}$). \label{tab-rotstudy-Opm-parms} }
\end{table}

We analyze 43 configurations separated by 8 MD time units. The results of fitting the folded two-point functions $C^\pm$ of the ${\cal O}_\pi^\pm$ operators over the time range $6-15$ are given in Table~\ref{tab-rotstudy-Opm-parms}. We observe that the pion energies are all consistent as expected, but for both operators the amplitudes fall into two distinct groups: those of the three momenta perpendicular to the $(1,1,1)$ diagonal are consistent, and that of the momentum parallel to the diagonal is ${\sim}16\%$ larger (about $9\sigma$) for the ${\cal O}_\pi^-$ operator and ${\sim}4.5\%$ smaller (about $3.5\sigma$) for the ${\cal O}_\pi^+$ operator, respectively. Thus we have clear and unambiguous evidence of the rotational symmetry breaking. The three momenta that have the same amplitude can be recognized as being related by the discrete group of 120$^\circ$ rotations around the diagonal of the spatial box, which can readily be seen as a residual symmetry of the action from the discussion in Section~\ref{sec-rotsymact}.

As part of this study we also measured the correlation function of the averaged operator 
\begin{dmath}
{\cal O}_\pi^{\rm avg}(\vec p) = \frac{1}{2}\left[ {\cal O}_\pi^{-}(\vec p) + {\cal O}_\pi^{+}(\vec p) \right]\,.
\end{dmath}
The amplitudes and energies of the corresponding fits are given in Table~\ref{tab-rotstudy-Oavg-parms}. We observe that in performing this average we significantly reduce the amount of rotational symmetry breaking, and in fact the amplitudes all now agree within statistics.

\begin{table}[t]
\centering
\begin{tabular}{c|cc}
\hline\hline
$\vec p$ & $E^{\rm avg}_\pi$ & ${\cal N}_\pi^{\rm avg}$\\
\hline
$(-2,-2,-2)$ & $ 0.4635(24)$ & $ 360.4(4.9)\times 10^{3}$\\
$(2,-2,-2)$ & $ 0.4639(20)$ & $ 358.9(4.2)\times 10^{3}$\\
$(-2,2,-2)$ & $ 0.4657(25)$ & $ 359.4(5.3)\times 10^{3}$\\
$(-2,-2,2)$ & $ 0.4639(21)$ & $ 355.4(4.7)\times 10^{3}$\\
\end{tabular}
\caption{Fit parameters for the $C^{\rm avg}$ correlation function of the averaged operator ${\cal O}^{\rm avg}$ for each choice of pion momentum $\vec p$ (in units of $\frac{\pi}{2L}$). \label{tab-rotstudy-Oavg-parms} }
\end{table}

\begin{table}[t]
\centering
\begin{tabular}{c|cc|cc}
\hline\hline
$\vec p$ & Operator & $\vec p_1$ & $\vec p_2$ \\
\hline
(-6,-2,-2) & ${\cal O}_\pi^-$ &  (-1,-1,-1) & (-5,-1,-1)\\
           & ${\cal O}_\pi^+$ &  (1,1,1) & (-7,-3,-3)\\
\hline
(6,-2,-2)  & ${\cal O}_\pi^-$ &  (-1,-1,-1) & (7,-1,-1) \\
	   & ${\cal O}_\pi^+$ &  (1,1,1)    & (5,-3,-3)\\
\hline
(-6,2,-2)  & ${\cal O}_\pi^-$ &  (-1,-1,-1) & (-5,3,-1) \\
           & ${\cal O}_\pi^+$ &  (1,1,1)    & (-7,1,-3) \\
\hline
(-6,-2,2)  & ${\cal O}_\pi^-$ &  (-1,-1,-1) & (-5,-1,3)  \\
           & ${\cal O}_\pi^+$ &  (1,1,1)    & (-7,-3,1) \\
\end{tabular}
\caption{Choices of quark momenta for each operator and pion total momentum, in units of $\frac{\pi}{2L}$, for the higher-momentum pions. The choices for the other cyclic permutations of momentum can be obtained by permuting the momenta in this table.\label{tab-rotstudymomchoices-311} }
\end{table}

In order to test this further we also generated pion two-point functions of higher momentum in the set $\{ (-6,-2,-2)$, $(6,-2,-2)$, $(-6,2,-2)$, $(-6,-2,2) \}$ plus cyclic permutations thereof. These measurements were performed on 86 configurations separated by 4 MD time units and binned over 3 successive configurations. The corresponding quark momentum choices are given in Table~\ref{tab-rotstudymomchoices-311}. To better discern the pattern of the symmetry breaking for this noisier data, we perform a simultaneous fit over all 12 correlation functions with a common energy. The fit is performed to time range 5-15. The results for the ${\cal O}_\pi^+$ and ${\cal O}_\pi^-$ operators alone are given in Table~\ref{tab-rotstudy-Opm-parms-311}. We observe that the amplitudes for each operator clearly fall into three discrete groups: two groups of three whose momenta are related by 120$^\circ$ rotations about the $(1,1,1)$ diagonal, and one group of six whose momenta are related by 60$^\circ$ rotations about the same axis. These groups are recognizable as the sets of points that lie on discrete planes whose normals are parallel to this diagonal. In Table~\ref{tab-rotstudy-Oavg-parms-311} we show the results for the averaged operator ${\cal O}_\pi^{\rm avg}$ where we again find that the symmetry breaking is heavily suppressed.

\begin{table}[t]
\centering
\begin{tabular}{cc|c||cc|c}
\hline\hline
Quantity & $\vec p$ & Value & Quantity & $\vec p$ & Value\\
\hline
$E_\pi^-$ &  & $ 0.721(3)$ & $E_\pi^+$ &  & $ 0.718(3)$\\
\hline
\multirow{4}{*}{${\cal N}^-_\pi$} & (-6,-2,-2) & $ 218(4)\times 10^{3}$ & \multirow{4}{*}{${\cal N}^+_\pi$} & (-6,-2,-2) & $ 462(8)\times 10^{3}$\\
 & (6,-2,-2) & $ 160(3)\times 10^{3}$ &  & (6,-2,-2) & $ 505(8)\times 10^{3}$\\
 & (-6,2,-2) & $ 192(3)\times 10^{3}$ &  & (-6,2,-2) & $ 471(8)\times 10^{3}$\\
 & (-6,-2,2) & $ 193(4)\times 10^{3}$ &  & (-6,-2,2) & $ 468(7)\times 10^{3}$\\
\hline
\multirow{4}{*}{${\cal N}^-_\pi$} & (-2,-6,-2) & $ 214(4)\times 10^{3}$ & \multirow{4}{*}{${\cal N}^+_\pi$} & (-2,-6,-2) & $ 461(6)\times 10^{3}$\\
 & (2,-6,-2) & $ 194(3)\times 10^{3}$ &  & (2,-6,-2) & $ 469(7)\times 10^{3}$\\
 & (-2,6,-2) & $ 157(3)\times 10^{3}$ &  & (-2,6,-2) & $ 498(8)\times 10^{3}$\\
 & (-2,-6,2) & $ 192(3)\times 10^{3}$ &  & (-2,-6,2) & $ 472(7)\times 10^{3}$\\
\hline
\multirow{4}{*}{${\cal N}^-_\pi$} & (-2,-2,-6) & $ 214(4)\times 10^{3}$ & \multirow{4}{*}{${\cal N}^+_\pi$} & (-2,-2,-6) & $ 461(7)\times 10^{3}$\\
 & (2,-2,-6) & $ 196(3)\times 10^{3}$ &  & (2,-2,-6) & $ 476(8)\times 10^{3}$\\
 & (-2,2,-6) & $ 190(4)\times 10^{3}$ &  & (-2,2,-6) & $ 465(7)\times 10^{3}$\\
 & (-2,-2,6) & $ 158(3)\times 10^{3}$ &  & (-2,-2,6) & $ 494(8)\times 10^{3}$\\
\end{tabular}
\caption{Fit parameters for the simultaneous fit to the $C^\pm$ correlation functions of the operator ${\cal O}^\pm$ over all 12 choices of pion momentum $\vec p$ (in units of $\frac{\pi}{2L}$), for the higher-momentum pions. Here the three separate groups of four correspond to the three cyclic permutations of the base set of momenta.\label{tab-rotstudy-Opm-parms-311} }
\end{table}

\begin{table}[t]
\centering
\begin{tabular}{cc|c}
\hline\hline
Quantity & $\vec p$ & Value\\
\hline
$E^{\rm avg}_\pi$ &  & $ 0.718(3)$\\
\hline
\multirow{4}{*}{${\cal N}^{\rm avg}_\pi$}  & (-6,-2,-2) & $ 170(3)\times 10^{3}$\\
                                 & (6,-2,-2)  & $ 166(3)\times 10^{3}$\\
                                 & (-6,2,-2)  & $ 166(3)\times 10^{3}$\\
                                 & (-6,-2,2)  & $ 165(3)\times 10^{3}$ \\
                                 \hline
 \multirow{4}{*}{${\cal N}^{\rm avg}_\pi$} & (-2,-6,-2) & $ 169(2)\times 10^{3}$\\
                                 & (2,-6,-2)  & $ 166(2)\times 10^{3}$\\
                                 & (-2,6,-2)  & $ 164(3)\times 10^{3}$\\
                                 & (-2,-6,2)  & $ 166(2)\times 10^{3}$\\
                                 \hline
 \multirow{4}{*}{${\cal N}^{\rm avg}_\pi$} & (-2,-2,-6) & $ 169(3)\times 10^{3}$\\
                                 & (2,-2,-6)  & $ 168(3)\times 10^{3}$\\
                                 & (-2,2,-6)  & $ 164(2)\times 10^{3}$\\
                                 & (-2,-2,6)  & $ 163(3)\times 10^{3}$
\end{tabular}
\caption{Fit parameters for the simultaneous fit to the $C^{\rm avg}$ correlation functions of the averaged operator ${\cal O}^{\rm avg}$ over all 12 choices of pion momentum $\vec p$ (in units of $\frac{\pi}{2L}$), for the higher-momentum pions. Here the three separate groups of four correspond to the three cyclic permutations of the base set of momenta.\label{tab-rotstudy-Oavg-parms-311} }
\end{table}

We make use of these averaged pion operators in Ref.~\cite{Bai:2015nea} when constructing $\pi\pi$ operators that are intended to transform under particular representations of the cubic symmetry group. Details of the operators and further tests will be presented in a forthcoming paper. 

\subsection{Residual mass}

The residual mass $m_{\rm res}$ is a measure of the explicit chiral symmetry breaking in the domain wall formalism due to the finite extent of the fifth dimension. It acts as an additive renormalization of the bare quark masses, $\tilde m = m + m_{\rm res}$, and it is the limit $\tilde m\to 0$ that defines the chiral limit. We make use of $m_{\rm res}$ below when computing the pseudoscalar decay constants, and when testing the Ward Takahashi identity, hence it is important to understand the effects of the G-parity boundary conditions on this quantity.

The residual mass is measured by fitting the ratio
\begin{equation}
m_{\rm res}'(t) = \frac{ \sum_{\vec x } \langle 0| J^a_{5q}(\vec x,t) | \pi^a(0) \rangle  }{ \sum_{\vec x } \langle 0| J^a_{5}(\vec x,t) | \pi^a(0) \rangle } \label{eq-mresdef}
\end{equation}
to a constant in time, and extrapolating the results to the massless limit. Here $J_5^a$ is the local pseudoscalar density
\begin{dmath}
J_5^a(x) = -\bar\Psi(x,L_s-1) P_L \tau^a \Psi(x,0) + \bar\Psi(x,0) P_L \tau^a \Psi(x,L_s-1)
      = \bar q(x)\tau^a \gamma^5 q(x)
\end{dmath}
where $\Psi$ are the 5d fields, $q$ the 4d surface fields and $\tau^a$ are the generators of the flavor symmetry (SU(2) in our case). $J_{5q}^a$ is also a local pseudoscalar density, but defined from the midpoint fields in the fifth direction:
\begin{dmath}
J_{5q}^a(x) = -\bar\Psi(x,L_s/2-1) P_L \tau^a \Psi(x,L_s/2) + \bar\Psi(x,L_s/2) P_L \tau^a \Psi(x,L_s/2-1)\,.
\end{dmath}

Up to a constant we can compute $m_{\rm res}'$ for a neutral pion ($\tau^a = \sigma_3/2$) on the GP0 ensemble via the following (wall source) correlation function
\begin{dmath}
C^{\rm std}_{J_5}(t) = \frac{-1}{2}\sum_{\vec x,\vec y\vec z} \langle 0| \left[\bar u(\vec x,t)\gamma^5 u(\vec x,t) -  \bar d(\vec x,t)\gamma^5 d(\vec x,t)\right]  \left[\bar u(\vec y,0)\gamma^5 u(\vec z,0) -  \bar d(\vec y,0)\gamma^5 d(\vec z,0)\right] | 0\rangle
= \sum_{\vec x,\vec y,\vec z} {\rm tr}\left\{  \prop_l^\dagger(\vec x,t;\vec z,0) \prop_l(\vec x,t;\vec y,0)     \right\}  \label{eq-Cj5corr}
\end{dmath}
and the equivalent for $J_{5q}$ in which the propagator sink is evaluated on the appropriate $s$-slices. Here we have used the degeneracy of the quarks to relate the up and down quark propagators, and we have not shown the Coulomb gauge fixing matrices on the source timeslice.

We would, of course, obtain the same expression as the above for any of the three pion states due to the exact isospin symmetry, but the neutral pion can be created by a particularly convenient operator in our G-parity formalism, giving the following expression for $C_{J_5}$ with GPBC:
\begin{dmath}
C^{\rm GPBC}_{J_5}(t) = \frac{-1}{2}\sum_{\vec x,\vec y,\vec z} \langle 0| \left[e^{2i\vec p\cdot \vec x}\bar\psi_l(\vec x,t)\gamma^5 \sigma_3 \psi_l(\vec x,t)\right]  \left[e^{-i\vec p\cdot (\vec y+\vec z)}\bar \psi_l(\vec y,0)\gamma^5 \sigma_3\psi_l(\vec z,0)\right] | 0\rangle
= \frac{1}{2}\sum_{\vec x,\vec y} e^{i\vec p\cdot(2\vec x-\vec y-\vec z)} {\rm tr}\left\{  \sigma_3\prop_l^\dagger(\vec x,t;\vec z,0) \sigma_3\prop_l(\vec x,t;\vec y,0)     \right\}\,, \label{eq-Cj5corrgpbc}
\end{dmath}
where $\psi$ are the usual two-flavor G-parity fields and we are using the same symbol $\mathcal{G}_l$ for the one-flavor and two-flavor light quark propagators in Eq.~\eqref{eq-Cj5corr} and~\eqref{eq-Cj5corrgpbc}. Note that we have not applied the $1+\sigma_2$ flavor projection operator at the source here; the result is that we potentially have a poorer projection onto the desired moving pion state. Nevertheless, the ratio $m_{\rm res}'$ is common to all pseudoscalar states, and therefore can be obtained just as well from a linear combination of pseudoscalar states as it can be from just the ground state.

\begin{figure}[tp]
\centering
\includegraphics[width=0.7\textwidth]{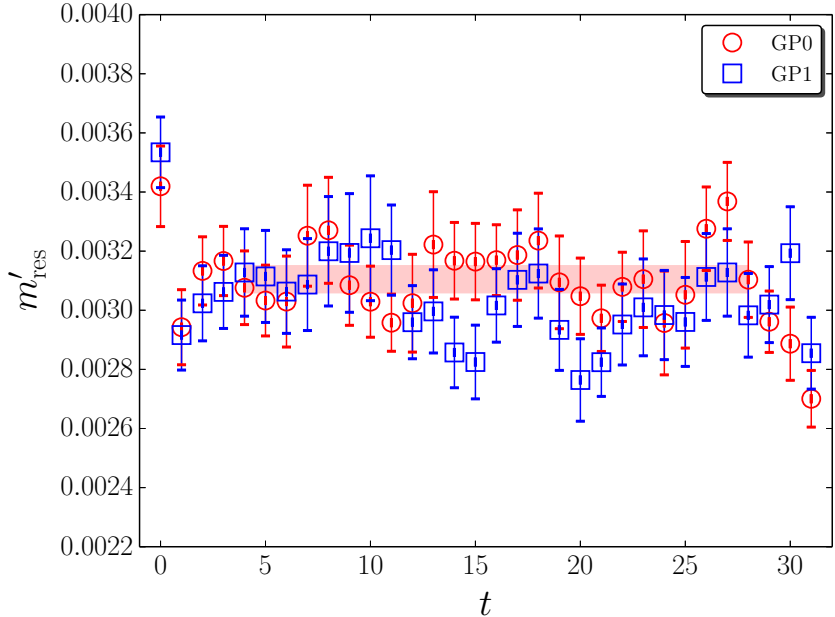}
\caption{$m_{\rm res}'$ on the GP0 and GP1 ensembles, overlaid by the fit to the GP0 data. \label{fig-mresgp0gp1} }
\end{figure}

We compute $m_{\rm res}'$ using just the light-quark ($m_l=0.01$) propagators with antiperiodic temporal boundary conditions. For simplicity we use a uniform fit range of $t=4$--$28$ and we perform uncorrelated fits. The values that we obtain are listed in Table~\ref{tab-16gp-mresmeas}. We observe very good agreement between the GP0 and GP2 ensembles, but the value on the GP1 ensemble is slightly lower by ${\sim}2\sigma$. Examining this in more detail, we plot $m_{\rm res}'$ as a function of time for the GP0 and GP1 ensembles in Figure~\ref{fig-mresgp0gp1}. Here we see no evidence of any systematic deviation between the ensembles, suggesting the discrepancy is merely due to statistical fluctuations. In principle we would not expect $m_{\rm res}$ to have any dependence on the boundary, it being simply a ratio of amplitudes separated in the fifth dimension, and our results are in line with this expectation. 

Given that we measure with only a single valence quark mass, we cannot extrapolate to the massless limit. However, the result measured on the GP0 ensemble agrees very well with the value $m_{\rm res}'=0.003102(25)$ quoted in Ref.~\cite{Allton:2007hx}. Recall these measurements were performed with a $25\%$ higher sea strange mass; this suggests the sea strange mass dependence is very weak, and therefore that we can use the value in the chiral limit of $m_{\rm res}=0.00308(4)$ given in that paper as our final value for the residual mass on all three ensembles.

\begin{table}[tp]
\centering
\begin{tabular}{c|cc}
\hline\hline
Approach (Source) & $m_{\rm res}'$ & $\chi^2/{\rm dof}$ \\ 
\hline
 16GP0      & 0.003105(45)  & 0.63(43) \\
 16GP1      & 0.003005(45)   & 0.77(61) \\
 16GP2      & 0.003106(65)   & 0.83(48)
\end{tabular}
\caption{The residual mass calculated on the 16GPx ensembles alongside their uncorrelated $\chi^2/{\rm dof}$.\label{tab-16gp-mresmeas}}
\end{table}

\subsection{Ward-Takahashi identity}

Pseudoscalar and axial correlators comprising $N_f$ degenerate flavors of domain wall fermion obey the Ward-Takahashi identity~\cite{Furman:1994ky}
\begin{equation}
\Delta_\mu \langle 0|{\cal A}_\mu^a(x) J_5^b(y)|0\rangle = 2m\langle 0|J_5^a(x) J_5^b(y)|0\rangle + 2\langle 0| J_{5q}^a(x) J_5^b(y) |0\rangle - \delta_{x,y}\langle 0|\bar q(y)\{\lambda^a,\lambda^b\}q(y)|0\rangle \label{eq-WTidentbasic}
\end{equation}
where $J_5$ and $J_{5q}$ were defined in the previous section. Here $\{\lambda^a\}$ are $N_f\times N_f$ flavor matrices, and $a$ and $b$ are flavor indices (of the conventional sort). $\Delta_\mu f(x) = f(x) - f(x-\hat\mu)$ is the discretized backwards derivative, with $\hat\mu$ a unit vector in the $\mu$-direction. The partially-conserved current ${\cal A}_\mu$ is composed of the 5d domain wall fields $\Psi$, and has the ``point-split'' form,
\begin{dmath}
{\cal A}_\mu^a(x) = -\frac{1}{2}\sum_{s=1}^{L_s} {\rm sign}(\frac{L_s}{2}-s+\half) \left[ \bar\Psi(x,s)(1+\gamma_\mu)U_\mu(x) \lambda^a \Psi(x+\hat\mu,s) - \bar\Psi(x+\hat\mu,s)(1-\gamma_\mu)U_\mu^\dagger(x)\lambda^a \Psi(x,s) \right]\,,
\end{dmath}
i.e. it is defined along the link {\it between} $x$ and $x+\hat\mu$ rather than on the site $x$. This current can be related to the {\it local} axial current $A^a_\mu = \bar q\gamma^\mu\gamma^5 \lambda^a q$, comprised of the domain wall surface fields $q$ as follows
\begin{equation}
{\cal A}^a_\mu(x) \approx \frac{Z_A}{Z_{\cal A}} A_\mu^a(x)\,,\label{eq-axialcurreln}
\end{equation}
where the relation becomes exact in the continuum limit. In the above, $Z_A$ is the renormalization factor relating the local current to the continuum-normalized Symanzik currents, and $Z_{\cal A}$ is the same for the partially-conserved current (for further information we refer the reader to Ref.~\cite{Aoki:2010dy}). 

Combining Eqs.~\eqref{eq-WTidentbasic} and~\eqref{eq-axialcurreln} we obtain
\begin{equation}
\frac{Z_A}{Z_{\cal A}} \partial_\mu \langle 0| A_\mu^a(\vec x,t) J_5^b(\vec y,0)|0\rangle = 2(m+m_{\rm res})\langle 0| J_5^a(\vec x,t) J_5^b(\vec y,0)|0\rangle \label{eq-WTcorr1}
\end{equation}
where we have used Eq.~\eqref{eq-mresdef} to remove the $J_{5q}$ terms. Inserting a complete set of states and taking the large time limit, the lhs of Eq.~\eqref{eq-WTcorr1} becomes
\begin{equation}\begin{array}{l}
\frac{Z_A}{Z_{\cal A}} \partial_\mu \langle 0|A_\mu^a(\vec 0,0)e^{-i\vec p\cdot x}e^{-E_{\vec p}t}|\pi(\vec p)\rangle \langle\pi(\vec p)|J_5^b(\vec y,0)|0\rangle \\
\hspace{1cm}= -i \frac{Z_A}{Z_{\cal A}}\sum_\mu p_\mu \langle 0|A_\mu^a(\vec x,t)|\pi(\vec p)\rangle \langle\pi(\vec p)|J_5^b(\vec y,0)|0\rangle \\
%
\end{array}\end{equation}
where we have allowed the ground state pion to have non-zero three-momentum $\vec p$ and corresponding Euclidean four-momentum $p_\mu = (-iE_{\vec p},\vec p)$. Inserting the above into Eq.~\eqref{eq-WTcorr1} and taking the spatial Fourier transform, we obtain
\begin{equation}
-i\frac{Z_A}{Z_{\cal A}} \sum_\mu p_\mu \sum_{\vec x,\vec y}e^{i\vec p\cdot(\vec x-\vec y)}\langle 0| A_\mu^a(\vec x,t) J_5^b(\vec y,0)|0\rangle = 2(m+m_{\rm res})\sum_{\vec x,\vec y}e^{i\vec p\cdot(\vec x-\vec y)}\langle 0| J_5^a(\vec x,t) J_5^b(\vec y,0)|0\rangle 
\end{equation}
Noticing that, in the large time limit, the source matrix element $\langle \pi(\vec p)|J_5^b|0\rangle$ appears on both sides of the equation, and that the time-dependence $e^{-E_{\vec p}t}$ is identical, we can transform the above into an expression that acts upon the ground-state amplitudes of general point-sink two-point functions with a pseudoscalar source:
\begin{equation}
-i\frac{Z_A}{Z_{\cal A}} \sum_\mu p_\mu {\cal N}^{LS}_{A_\mu^a P^b} = 2(m+m_{\rm res}){\cal N}^{LS}_{J_5^a P^b} \label{eq-WTcorramps}
\end{equation}
where $S$ represents a general source smearing. We can therefore utilize the wall-point amplitudes determined in Section~\ref{sec-measpion} and $m_{\rm res}$ determined above to obtain the relative normalization of the local and partially-conserved domain wall axial currents, $Z_A/Z_{\cal A}$.

Notice that the rhs of Eq.~\eqref{eq-WTcorramps} is pure-real, whereas the lhs contains a factor of $i$ for the spatial components of the sum. This implies that the spatial amplitudes must be pure-imaginary, which we previously found to be the case. The expression therefore reduces to
\begin{equation}
\frac{Z_A}{Z_{\cal A}}\left( -E_{\vec p}{\rm Re}({\cal N}^{LS}_{A_0^a P^b}) + \sum_i p_i {\rm Im}({\cal N}^{LS}_{A_i^a P^b})\right) = 2(m+m_{\rm res}){\rm Re}({\cal N}^{LS}_{J_5^a P^b}) \label{eq-WTcorramps2}
\end{equation}

We combine the amplitudes in Table~\ref{tab-fitmpi} with $m_{\rm res}$ obtained in the previous section to obtain $\frac{Z_A}{Z_{\cal A}}$ from the above expression. We use $a=b=3$, i.e. the third isospin component of the pion triplet. Note that ${\cal N}^{LS}_{J_5 P} = -{\cal N}^{LS}_{P P}$ where the minus sign occurs because in the latter we compute
$$\langle [\bar\psi(x)\gamma^5\sigma_3\psi(x)]^\dagger \bar\psi(y)\gamma^5\sigma_3\psi(y)\rangle = \langle [J^3_5(x)]^\dagger \bar\psi(y)\gamma^5\sigma_3\psi(y)\rangle = -\langle J^3_5(x) \bar\psi(y)\gamma^5\sigma_3\psi(y)\rangle\,.$$
The values that we obtain are listed in Table~\ref{tab-za}. The value measured on the GP0 ensemble agrees very well with $Z_A/Z_{\cal A}(m_l=0.01)=0.71807(14)$ given in Ref.~\cite{Allton:2007hx}, but those obtained from the G-parity ensembles are ${\sim}2\%$ higher. This discrepancy may arise due to the boundary conditions but also due to differing discretization errors as the G-parity Green's functions are evaluated at non-zero momentum. Nevertheless the effect is small. For use below, we take use $Z_A/Z_{\cal A} = 0.7162(2)$ obtained in the massless limit in Ref.~\cite{Allton:2007hx} for all ensembles.

\begin{table}[tp]
\begin{tabular}{cc}
\hline\hline
Ensemble & $Z_A/Z_{\cal A}(m_l=0.01)$ \\
\hline
GP0 & 0.7182(25) \\
GP1 & 0.7312(38) \\
GP2 & 0.7291(92)
\end{tabular}
\caption{Measured values of $Z_A/Z_{\cal A}$.\label{tab-za}}
\end{table}

\subsection{Pion decay constant}

The decay constants for a pseudoscalar meson $P$ are defined on the lattice as~\cite{Blum:2000kn,Antonio:2006px}
\begin{equation}
f_P = \frac{i}{q_\mu} \langle 0 | {\cal A}_\mu (0) |P(\vec q)\rangle\,,
\end{equation}
where the states are normalized as $\langle P(\vec q)|P(\vec q') \rangle = (2\pi)^3 2 E_{\vec q}\ \delta^{(3)}(\vec q - \vec q')$ and ${\cal A}_\mu = Z_A A_\mu$. 

$f_P$ can be obtained from the following ratio of the pseudoscalar correlator amplitudes computed in Section~\ref{sec-measpion}:
\begin{dmath}
\frac{\left|{\cal N}^{LS}_{A_\mu P}\right|^2}{ {\cal N}^{SS}_{P P} } = \frac{V}{2E_{\vec q}}\left|\langle 0|A_\mu(\vec 0,0)|P(\vec q)\rangle\right|^2
= f_P^2 \frac{q_\mu^2 V}{2Z_A^2 E_{\vec q}}.
\end{dmath}
Here $Z_A$ is the renormalization factor relating the local domain wall current to the continuum rather than the ratio $Z_A/Z_{\cal A}$ that we obtained in the previous section. However, as $Z_{\cal A} = 1 + {\cal O}(m_{\rm res})$ we can obtain a good approximation to $f_P$ using this ratio in place of $Z_A$ in the above.

For the pion we label the decay constant $f_{ll}$. In Table~\ref{tab-fpi} we list our measured values for this quantity, computed on all ensembles using the amplitudes and energies given in Table~\ref{tab-fitmpi}. We measure using the temporal ($\mu=4$) component of the axial-vector operator at the sink, and also using the non-zero-momentum spatial components on the G-parity ensembles (for which $q_\mu = \pi/L$). As expected we observe excellent agreement between the results on all three ensembles and between spatial and temporal determinations.

\begin{table}[tp]
\centering
\begin{tabular}{c|ccc}
\hline\hline
$\mu$  & GP0  & GP1  & GP2 \\
\hline
$4$  & 0.0887(11)  & 0.0882(12)  & 0.0908(11) \\
$1$  & -  & 0.0896(25)  & 0.0878(28) \\
$2$  & -  & -  & 0.0891(35) \\
\end{tabular}
\caption{$f_{ll}$ computed from the temporal and spatial components of the axial-vector operator. \label{tab-fpi} }
\end{table}

\subsection{Impact of the flavor projection}

It is interesting to observe the effect of the the $1\pm \sigma_2$ flavor projection for non-local sources. To do so we consider the pseudoscalar point-sink correlation functions
\begin{dmath}
C^{LW}(t)
= \sum_{\vec x,\vec y, \vec z} \left\langle \left[e^{\pm 2i\vec p\cdot \vec x}\frac{-i}{\sqrt{2}}\bar\psi_l(\vec x,t)\sigma_3\gamma^5\psi_l(\vec x,t)\right]\left[ e^{-i\vec p\cdot (\vec y+\vec z)}\frac{-i}{\sqrt{2}}\bar\psi_l(\vec y,0)\sigma_3\half(1\pm\sigma_2)\gamma^5\psi_l(\vec z,0) \right]\right\rangle\,, \label{eq-flavprojtestcorr1}
\end{dmath}
where we vary both the sign in the source projection operator and the sign of the sink momentum projection. We fit each independently to obtain the amplitude and ground-state energy. For simplicity we consider only the GP1 ensemble (with $\vec p = (\pi/2L,0,0)$) and use a common fit range of 5--25.

\begin{table}[tp]
\centering
\begin{tabular}{c|c||c|c}
\hline\hline
Source proj. sign  & Sink mom. sign  & ${\cal N}$  & $E$ \\
\hline
+  & +  & $5.22(17)\times 10^{4}$  & 0.3132(34) \\
+  & -  & $-1.1(1.1)\times 10^{3}$  & 0.271(64) \\
-  & +  & $3.31(13)\times 10^{4}$  & 0.3126(38) \\
-  & -  & $-1.566(93)\times 10^{4}$  & 0.3127(64) \\
\end{tabular}
\caption{Amplitudes (${\cal N}$) and energies ($E$) determined for different source flavor projections $(1\pm\sigma_2)$ and sink momentum directions $\pm 2\vec p$ on the GP1 ensemble. The first column gives the sign of the source flavor projection, and the second the sign of the sink momentum.\label{tab-flavproj1} }
\end{table}

We list the fit results in Table~\ref{tab-flavproj1}. The first line of the table gives the values with the `correct' flavor and sink momentum projections, i.e. for which the sink momentum projection equals the total source momentum, and we use the correct $1+\sigma_2$ projector to make the source translationally covariant. The amplitude and energy for this correlation function agree within statistics with those obtained in our simultaneous fit and listed in Table~\ref{tab-fitmpi}. If we attempt to project onto the opposite sink momentum we find an amplitude (second line of the table) statistically consistent with zero as expected due to momentum conservation.

For the `wrong' source flavor projection we observe from the third line of Table~\ref{tab-flavproj1} that this source also couples strongly to the forwards-moving pion state with momentum $\pi/L$. However we find (fourth line) that it also projects onto the {\it backwards-moving} pion with momentum $-\pi/L$, despite our explicit source momentum projection onto the forwards-moving state, seemingly violating momentum conservation. To see how this comes about, recall that the $1-\sigma_2$ operator projects out the $\psi_-$ field, for which the allowed momenta are $q \in \frac{\pi}{2L}\{\ldots, -9, -5, -1, 3, 7,\ldots\} = -\pi(1+4n)/2L$ in each G-parity direction, where $n$ is an integer. 

In the following discussion we suppress the coordinates in the $y$ and $z$ and $t$ directions. The phases $\exp(i\pi[1+4n]x/2L)$ associated with the allowed momenta for $\psi_-$, form an orthonormal basis:
\begin{dmath}
\sum_{x=0}^{L-1} \exp\left(i\pi[1+4n]x/2L\right)\times\exp\left(-i\pi[1+4m]x/2L\right) \\
= \frac{ 1-\exp\left(2\pi i[n-m]\right) }{ 1-\exp\left(2\pi i[n-m]/L\right) }
= L\delta_{n,m}
\end{dmath}
We can therefore define a Fourier transform and its inverse into a momentum space indexed by $n$:
\begin{dgroup}
\begin{dmath}
\hat\psi_-(n) = \sum_x \exp\left(i\pi[1+4n]x/2L\right)\psi_-(x)
\end{dmath}
\begin{dmath}
\psi_-(x) = \frac{1}{L}\sum_n \exp\left(-i\pi[1+4n]x/2L\right)\hat\psi_-(n)
\end{dmath}
\end{dgroup}

The Fourier transform of the quark field $\psi_-(x)$ with the non-allowed momentum $\pi/2L$ can then be written as
\begin{dmath}
\sum_{x} e^{-i \pi x/2L}  \psi_-(x)
= \frac{1}{L}\sum_{n} \hat\psi_-(n)\sum_{x} e^{-i \pi x/2L}e^{-i\pi[1+4n]x/2L} = \frac{1}{L}\sum_{n} R(n)\hat\psi_-(n)\,, \label{eq-flavprojfourier}
\end{dmath}
where
\begin{dmath}
R(n) = \sum_{x} e^{-i \pi x/2L}e^{-i\pi[1+4n]x/2L}
= \frac{2}{ 1 - e^{-i\pi[1+2n]/L} } \label{eq-ftpsiminusp1}
\end{dmath}
we observe that the coefficient is generally complex and is non-zero for {\it all} $n$. The same will be true for the $\overline\psi_+$ field operator, which also has the same set of allowed momenta. 

The source operator can then be written as
\begin{dmath}
\sum_{x, x'}  e^{-i\pi(x+x')/2L}\overline\psi_+(x)\sigma_3\gamma^5\psi_-(x') \\
= \frac{1}{L^2}\sum_{n,n'}  R(n)R(n')\hat{\overline\psi}_+(n)\sigma_3\gamma^5\hat\psi_-(n')
= \frac{1}{L^2}\sum_{n,n'}  R(n)R(n')\sum_{x,x'} e^{i\pi[2+4n+4n']x/2L}\psi_-(x)\sigma_3\gamma^5\psi_-(x')
\end{dmath}
which has non-zero projection onto all states of momentum $-[2+4n + 4n']\pi/2L$. This includes the pion state with momentum $-\pi/L$, for which $n=-n'$. For example, with $n=n'=0$ we have
\begin{dmath}
R(n)R(n') = \frac{4}{ (1 - e^{-i\pi/L})^2 } 
\end{dmath}
which has a non-zero imaginary component. We therefore predict that the correlation function with source projection $1-\sigma_2$ and sink momentum sign $-1$ should also have an imaginary component. Performing the fit to our data we indeed find a statistically-resolvable imaginary component with amplitude $-4.1(1.0)\times 10^{3}$ and an exponent $E=0.331(26)$ consistent with the pion energy.

Combinations with $n' = -1 - n$ instead project onto the forwards-moving pion with momentum $\pi/L$. For example with $n=0$ and $n'=-1$, we have
\begin{dmath}
R(n)R(n') = \frac{4}{(1-e^{-\pi i/L})(1-e^{\pi i/L})}
\end{dmath}
which is pure-real. We therefore predict that we will {\it not} observe an imaginary component for the correlation function with source projection $1-\sigma_2$ and sink momentum sign $+1$. In practise we found that we could not fit to a cosh-like form to the imaginary component (the fitter did not converge) unless we performed a simultaneous fit including the real-part with a shared energy; there we found an amplitude for the imaginary component of $1.7(3.9)\times 10^2$, consistent with zero as expected. We can fit the imaginary part alone to a constant, for which we obtain a value $0.68(85)$, again consistent with zero.

We conclude that it is vital to perform the flavor projection if one is concerned with the specific direction of the resulting pion's momentum. This is important, for example, when constructing a two-pion state with zero total momentum.

\subsection{Kaon mass and decay constant}

We obtain the mass and correlator amplitudes for the neutral kaon on the GP0 ensemble and those of the mixed state $|\tilde K^0_+\rangle = \frac{1}{\sqrt{2}}\left(|K^0\rangle + |K^{\prime\,0}\rangle\right)$, discussed in 
Section~\ref{sec-hlbilinears} above, on the G-parity ensembles. 

On the GP0 ensemble we compute wall-point correlators of the form
\begin{dmath}
C^{LW}_{PP}(t) = \sum_{\vec x,\vec y,\vec z}\langle 0 | \left[ i\bar s(\vec x,t)\gamma^5 d(\vec x,t) \right]\left[ i\bar d(\vec y,0)\gamma^5 s(\vec z,0) \right] |0\rangle
\end{dmath}
using our Coulomb gauge fixed wall source propagators. Likewise, on the G-parity ensembles we compute
\begin{dmath}
C^{LW}_{PP}(t) = \sum_{\vec x,\vec y,\vec z}\langle 0 | \left[ \frac{i}{\sqrt{2}}\bar \psi_h(\vec x,t)\gamma^5 \psi_l(\vec x,t) \right]\left[ e^{-i\vec p\cdot(\vec y-\vec z)}\frac{i}{\sqrt{2}}\bar \psi_l(\vec y,0)\gamma^5 \half(1-\sigma_2)\psi_h(\vec z,0) \right] |0\rangle\,.
\end{dmath}
Recall that the mixed combination $|\tilde{K}^0\rangle$ was introduced in Section~\ref{sec-hlbilinears} as the eigenstate of the QCD Hamiltonian which is an energy and momentum eigenstate with the energy of the kaon and zero spatial momentum. In order to obtain a zero-momentum state under the constraint that the quark momenta are odd-integer multiples of $\pi/2L$, we assign momentum $+\vec p$ to the anti-light quark and $-\vec p$ to the strange quark as indicated above, where $p_i=\frac{\pi}{2L}$ for each G-parity direction and zero otherwise. Here we must be careful with the momentum projection required to create a translationally covariant source operator: The momentum $-\vec p$ ($n_{\vec p}=-1$) of the strange quark requires the projection $\psi_h \to \half(1-\sigma_2)\psi_h$, and the momentum $+\vec p$ of the anti-light quark requires $\bar\psi_l \to \bar\psi_l\half(1-\sigma_2)$ (cf. Eq.~\eqref{eqn-transconvbarfieldndir}). If we swap the momentum assignment we must also swap the sign of the projection.

For the kaon wall-sink two-point function we measure the following:
\begin{dmath}
C^{WW}_{PP}(t) = \sum_{\vec r,\vec s,\vec y,\vec z}\langle 0 | \left[e^{-i\vec p\cdot(\vec r-\vec s)} \frac{i}{\sqrt{2}}\bar \psi_h(\vec r,t)\gamma^5 \half(1-\sigma_2)\psi_l(\vec s,t) \right]\left[ e^{-i\vec p\cdot(\vec y-\vec z)}\frac{i}{\sqrt{2}}\bar \psi_l(\vec y,0)\gamma^5 \half(1-\sigma_2)\psi_h(\vec z,0) \right] |0\rangle\,.
\end{dmath}

Computing the decay constant for the physical kaon requires a little more thought. The continuum temporal axial-vector operator that annihilates the $K^0$ is $A_4 = -i\bar s \gamma^4\gamma^5 d$, where we have chosen a phase convention such that $\langle 0|A_4|K^0\rangle$ is real and positive for the operators specified in Section~\ref{sec-hlbilinears}. From the G-parity fields $\psi_h$ and $\psi_l$ we can construct such a bilinear operator (here for a generic spin-matrix $\Gamma$) as
\begin{dmath}
\bar s \Gamma d = \psi_h F_{11} \Gamma F_{11} \psi_l \label{eq-sbardbilgp}
\end{dmath}
where $F_{11}=\half(1+\sigma_3)$, such that $A_4 = -i\bar \psi_h F_{11}\gamma^4\gamma^5 \psi_l$. The axial-sink pseudoscalar source correlator is therefore
\begin{dmath}
C^{LW}_{A_4P}(t) = \sqrt{2}\sum_{\vec x,\vec y,\vec z}\langle 0 | \left[ -i\bar \psi_h(\vec x,t)F_{11}\gamma^4\gamma^5 \psi_l(\vec x,t) \right]\left[ e^{-i\vec p\cdot(\vec y-\vec z)}\frac{i}{\sqrt{2}}\bar \psi_l(\vec y,0)\gamma^5 \half(1-\sigma_2)\psi_h(\vec z,0) \right] |0\rangle\,,\label{eq-A4P-sonly}
\end{dmath}
where the coefficient of $\sqrt{2}$ compensates for the fact that only the physical component of the incoming state couples to the operator (up to exponential corrections) as explained in Section~\ref{sec-opactphyskaon}. (Note, as discussed in that section we have limited the observable operator appearing in Eq.~\ref{eq-A4P-sonly} to one containing only the $s$-quark.) We can also compute the matrix element with a similar axial operator that connects to the {\it unphysical} $K^{0\,'}$ state, $A'_4 = -i\bar \psi_h F_{22}\gamma^4\gamma^5 \psi_l$, where $F_{22}=\half(1-\sigma_3)$. The operators $A_0$ and $A'_0$ interchange under the G-parity operation, hence the G-parity symmetry of the action implies $\langle A_4 | \tilde K^0\rangle = \langle A'_4 | \tilde K^0\rangle$. This is easily verified numerically. In order to improve statistics we can therefore take the average of the two.

\begin{figure}[tp]
\includegraphics[width=0.48\textwidth]{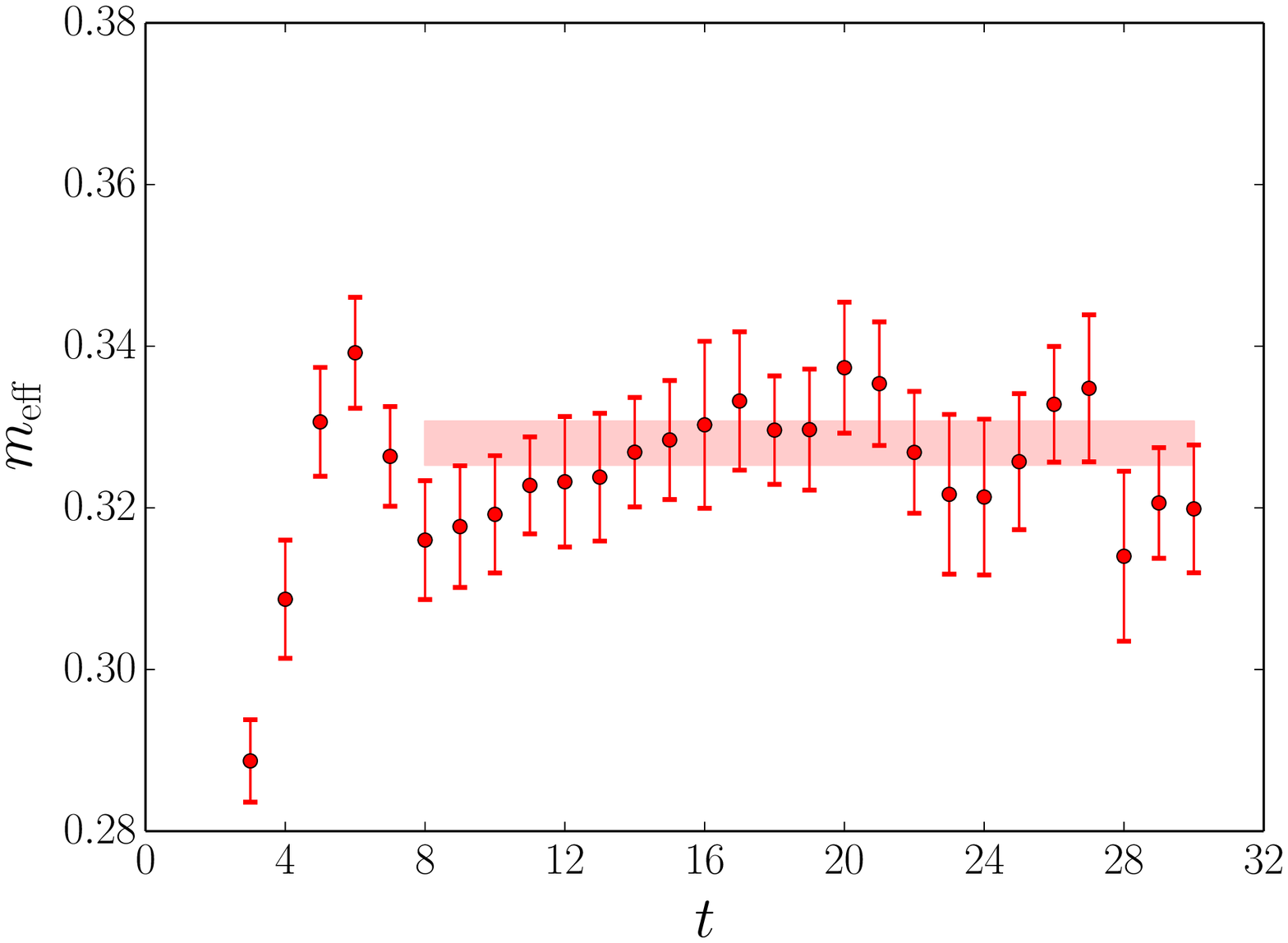}
\includegraphics[width=0.48\textwidth]{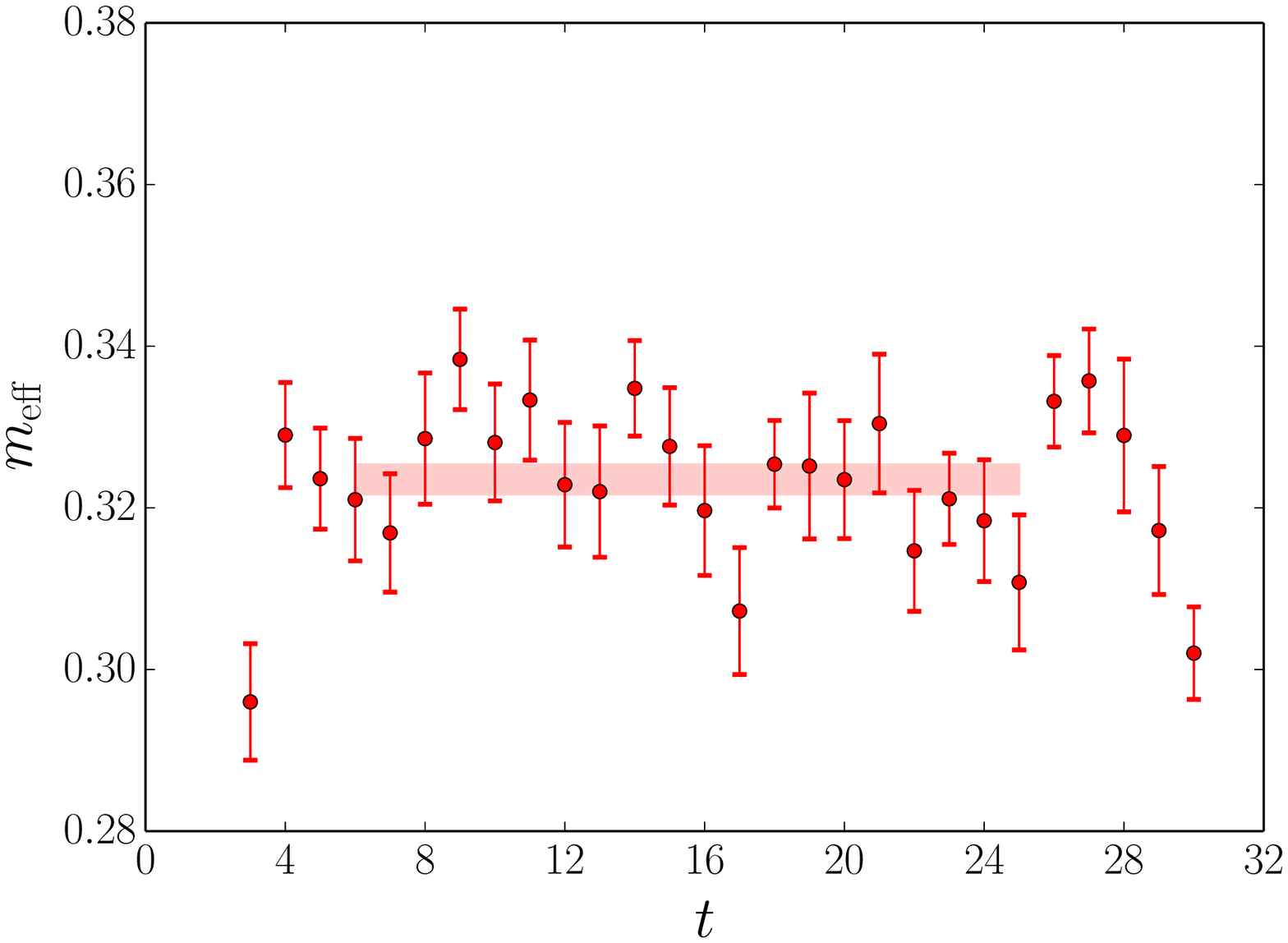}
\includegraphics[width=0.48\textwidth]{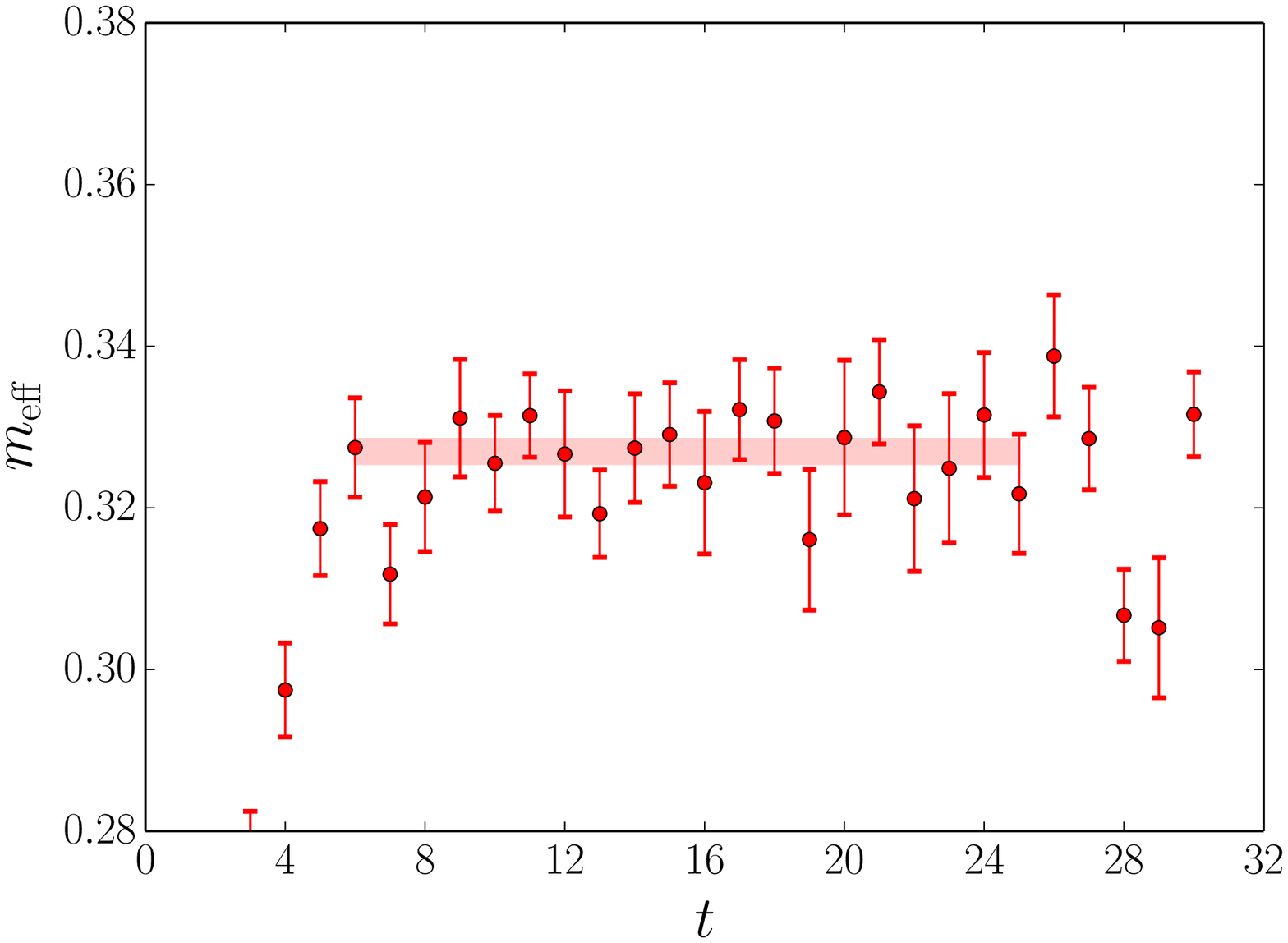}

\caption{The kaon effective mass in the $PP$ channel overlaid by the fitted value on the GP0 (upper-left), GP1 (upper-right) and GP2 (lower) ensembles.\label{fig-kaoneffmass} }
\end{figure}

The data are fit to the functional form given in Eq.~\ref{eq-twoptfitform}. The fitted masses and amplitudes as well as the corresponding values of the decay constant $f_K$ are given in Table~\ref{tab-fitmkfk}, alongside the fit ranges chosen by eye based on effective mass plots (used uniformly for all correlators on a given ensemble) and the uncorrelated $\chi^2/{\rm dof}$. In Figure~\ref{fig-kaoneffmass} we show effective mass plots for the $PP$ point-sink channel. The errors shown in the table are only statistical but appear to be sufficient to explain any discrepancies between the results found for $m_K$ and $f_K$ among the three ensembles.

We observe good agreement between the values of both the masses and decay constants of all three ensembles, suggesting that masses of the degenerate $K$ and $K^'$ particles are not significantly altered when they are allowed to mix by the boundary and that we are able to extract the decay constants of the physical kaon despite this mixing.

\begin{table}[tp]
\begin{tabular}{l|c|c|c}
\hline\hline
Quantity  & GP0  & GP1  & GP2 \\
\hline
Fit range  & 8--30  & 6--25  & 6--25 \\
$\chi^2/{\rm d.o.f.}$  & 0.14(8)  & 0.37(24)  & 0.21(10) \\
$m_K$  & 0.3280(27)  & 0.3235(18)  & 0.3270(15) \\
$f_K$  & 0.0969(11)  & 0.0956(10)  & 0.0992(8) \\
${\cal N}^{LW}_{PP}$  & $1.103(34)\times 10^{5}$  & $5.27(12)\times 10^{4}$  & $5.29(11)\times 10^{4}$ \\
${\cal N}^{LW}_{A_4A_4}$  & $1.039(33)\times 10^{4}$  & $5.13(14)\times 10^{3}$  & $4.85(13)\times 10^{3}$ \\
${\cal N}^{LW}_{A_4P}$  & $2.233(58)\times 10^{4}$  & $1.074(22)\times 10^{4}$  & $1.078(20)\times 10^{4}$ \\
${\cal N}^{WW}_{PP}$  & $4.05(11)\times 10^{7}$  & $9.78(23)\times 10^{6}$  & $9.04(19)\times 10^{6}$ \\
\end{tabular}
\caption{The results of simultaneous fits to the $PP$, $A_4A_4$ and $A_\mu P$ heavy-light (kaonic) correlation functions on each of the three ensembles ($\mu=4$ is the time direction). The superscripts $LW$ and $WW$ refer to wall-source-point-sink and wall-source-wall-sink correlators respectively. Here ${\cal N}$ are the fitted amplitudes.\label{tab-fitmkfk}}
\end{table}

\subsection{Neutral kaon mixing parameter}

The mixing between neutral kaons occurs via a second-order weak process that can be realized at low energies by the $\Delta S=2$ effective four-quark operator $\mathcal{O}_{\rm VV+AA}$,
\begin{dmath}
{\cal O}_{\rm VV+AA} = (\bar s\gamma^\mu d)(\bar s\gamma^\mu d) + (\bar s\gamma^5\gamma^\mu d)(\bar s\gamma^5\gamma^\mu d)
= (\bar \psi_h F_{11}\gamma^\mu \psi_l)(\bar \psi_h F_{11}\gamma^\mu \psi_l) + (\bar \psi_h F_{11}\gamma^5\gamma^\mu \psi_l)(\bar \psi_h F_{11}\gamma^5\gamma^\mu \psi_l)\,.
\end{dmath}
The desired matrix element is 
\begin{equation}
\langle K^0(t_2)|\mathcal{O}_{\rm VV+AA}(t)|\bar K^0(t_1)\rangle \eqsim (\sqrt{2})^2 \langle \tilde K^0(t_2)|\mathcal{O}_{\rm VV+AA}(t)|\bar {\tilde K}^0(t_1)\rangle
\end{equation}
where the factors of $\sqrt{2}$ on the right side again enter because the physical operator only couples to the neutral kaon component of the incoming and outgoing states. Including these coefficients, the above Wick contracts to $F(\gamma^\mu) + F(\gamma^\mu\gamma^5)$, where
\begin{dmath}
F(\Gamma) =  2{\rm tr}\left\{ \Gamma \mathcal{G}_l(\vec y,t;\vec x_2,t_2) \mathcal{G}^{\dagger}_h(\vec y,t;\vec x_2,t_2) F_{11} \right\}     {\rm tr}\left\{ \Gamma \mathcal{G}_l(\vec y,t;\vec x_1,t_1) \mathcal{G}^{\dagger}_h(\vec y,t;\vec x_1,t_1) F_{11} \right\}    
  -2{\rm tr}\left\{ \Gamma \mathcal{G}_l(\vec y,t;\vec x_2,t_2) \mathcal{G}^{\dagger}_h(\vec y,t;\vec x_2,t_2)\Gamma F_{11} \mathcal{G}_l(\vec y,t;\vec x_1,t_1) \mathcal{G}^{\dagger}_h(\vec y,t;\vec x_1,t_1) F_{11} \right\}\,.
\label{eqn-BK_G}
\end{dmath}
Here we have made use of Eqs.~\eqref{eqn-propconjreln} and~\eqref{eq-g5herm} to simplify the result. Without G-parity the corresponding Wick contractions are
\begin{dmath}
F(\Gamma) =    2   {\rm tr}\left\{ \Gamma\mathcal{G}_d(\vec y,t;\vec x_2,t_2) \mathcal{G}^\dagger_s(\vec y,t;\vec x_2,t_2)  \right\}     {\rm tr}\left\{ \Gamma \mathcal{G}_d(\vec y,t;\vec x_1,t_1)\mathcal{G}^\dagger_s(\vec y,t;\vec x_1,t_1)  \right\}    -2   {\rm tr}\left\{ \Gamma\mathcal{G}_d(\vec y,t;\vec x_2,t_2) \mathcal{G}^\dagger_s(\vec y,t;\vec x_2,t_2) \Gamma \mathcal{G}_d(\vec y,t;\vec x_1,t_1) \mathcal{G}^\dagger_s(\vec y,t;\vec x_1,t_1)  \right\}     
\label{eqn-BK_Conv}
\end{dmath}
which are identical in form up to the presence of the additional flavor matrix $F_{11}$, further demonstrating the utility of our notation.

The expressions given in Eqs.~\eqref{eqn-BK_G} and \eqref{eqn-BK_Conv} assume that the kaon state is created by a point source at $(x_1,t_1)$ and absorbed by a point sink located at $(x_2,t_2)$.  In a modern calculation of neutral kaon mixing, improved source and sink wave functions are used which better project onto the kaon state and which fix the momentum of the kaon to be zero in order to reduce systematic errors arising from the contributions of excited or moving kaon states.   Specifically we might use Coulomb gauge fixed wall sources and sinks for the light and strange quarks introducing explicit position-dependent phase factors and the associated projection operator $(1\pm\sigma_2)$ to give these quarks the momenta required by the boundary conditions.  If we do not show the Coulomb gauge fixing matrices, a $\bar{\tilde K}^0$ interpolating operator at the time $t$ would be written as
\begin{dmath}
\bar{\tilde K}^0(t) = \sum_{\vec x,\vec y}e^{-i\vec p\cdot(\vec x-\vec y)} \bar\psi_l(\vec x,t)\gamma^5 \half(1-\sigma_2) \psi_h(\vec y,t)
\label{eqn-kaon_source}
\end{dmath}
where as before the momentum $\vec p$ has components $+\pi/2L$ for directions in which G-parity boundary conditions are imposed and zero otherwise.

The bag parameter $B_K$ is constructed from the Green's functions containing the ${\cal O}_{{\rm VV+AA}}$ operator as follows:
\begin{equation}
B_K^{\rm lat}(t) = \frac{\langle K^0(t_2) \mathcal{O}_{\rm VV+AA}(t) \bar K^0(t_1)\rangle}{\frac{8}{3}\langle K^0(t_2) A_4(t)\rangle\langle A_4(t) \bar K^0(t_1)\rangle}\,,\label{eqn-bkoperator}  
\end{equation}
where the denominator serves to divide out the normalizations of the source and sink operators. On the G-parity ensembles the $K^0$ operators in the above are replaced by $\tilde K^0$ operators and a factor of 2 is required in the numerator, as described above.

Following Ref.~\cite{Aoki:2010pe} we set $t_2=T$ and $t_1=0$ and utilize the forwards ($F$) propagators (obtained by using quark propagators which are the sum of propagators obeying periodic and anti-periodic boundary conditions in the time) to form the kaon between $t_1$ and $t$ such that it falls off exponentially as $\exp(-m_K[t-t_1])$. Similarly we form the kaon between $t$ and $t_2$ from the backwards ($B$) propagators (obtained by using quark propagators which are the difference of propagators obeying periodic and anti-periodic boundary conditions in the time) such that it falls off exponentially in the negative-$t$ direction, $\exp(-m_K[t_2-t])$. Similarly the $\langle A_4(t) \bar K^0(t_1)\rangle$ term in the denominator of Eq.~\eqref{eqn-bkoperator} is constructed with the $F$ propagators and the $\langle A_4(t) K^0(t_2)\rangle$ from the $B$ propagators.  The exponential time dependence then cancels exactly between the numerator and denominator such that $B_K^{\rm lat}(t)$ is constant up to excited state contamination. 

We will next present the results from such a calculation of $B_K$.  We begin by explaining that this calculation was carried at the start of our study of G-parity boundary conditions and the flavor projection matrix $\half(1-\sigma_2)$ shown in Eq.~\eqref{eqn-kaon_source} was not included.  This omission might lead to the presence of additional  kaon states with non-zero momenta in both the numerator and denominator of Eq.~\eqref{eqn-bkoperator}, resulting in systematic errors.  However, for our particular operators and kinematics, we can argue that such effects are at or below the 1\% level. 

First we observe that the operator $\bar{\tilde K}^0(t)$ defined in Eq.~\eqref{eqn-kaon_source} and its conjugate $\tilde K^0(t)$ have positive G-parity.  Therefore, they obey periodic boundary conditions in each spatial direction and can only create states whose momentum components are integral multiples of $2\pi/L$.  Second, both the weak mixing operator $O_{\rm VV+AA}$ and the axial current operators which appear in the numerator and denominator of Eq.~\eqref{eqn-bkoperator} are summed over the full spatial volume.  Since these operators do not contain a translationally invariant combination of our four quark flavors, this sum over space is insufficient to ensure that these operators cannot create or destroy momentum.  However, for the case of momentum components that are multiples of $2\pi/L$, ({\it i.e.} $\vec p = 2\pi \vec n/L$ where $\vec n$ is a vector of integers) these spatial sums will give zero unless $\vec n = \vec 0$.  Thus, for the factors in the denominator, only zero-momentum kaons can contribute.  Similarly the spatial sum in the numerator guarantees that the momentum of the initial and final kaon must agree. Third, such an unwanted kaon with non-zero momentum traveling from the source to the sink will be suppressed relative to the kaon with zero momentum by the factor $\exp\{-(\sqrt{M_K^2+(2\pi/L)^2} - M_K)T\} = 0.0027$ for our $T=32$, supporting our assertion that the resulting errors are at or below 1\%.

\begin{table}[tp]
\begin{tabular}{l|c|c|c}
\hline\hline
Ens.  & Fit range  & $\chi^2/{\rm dof}$  & $B_K$ \\
\hline
GP0  & 6--26  & 1.32(68)  & 0.5997(63) \\
GP1  & 6--26  & 0.79(63)  & 0.5945(59) \\
GP2  & 6--26  & 0.42(40)  & 0.5956(50) \\
\end{tabular}
\caption{The values of $B_K$ and the uncorrelated $\chi^2/{\rm dof}$ obtained by fitting to the specified range on each ensemble. \label{tab-fitbk} }
\end{table}

\begin{figure}[tp]
\includegraphics[width=0.48\textwidth]{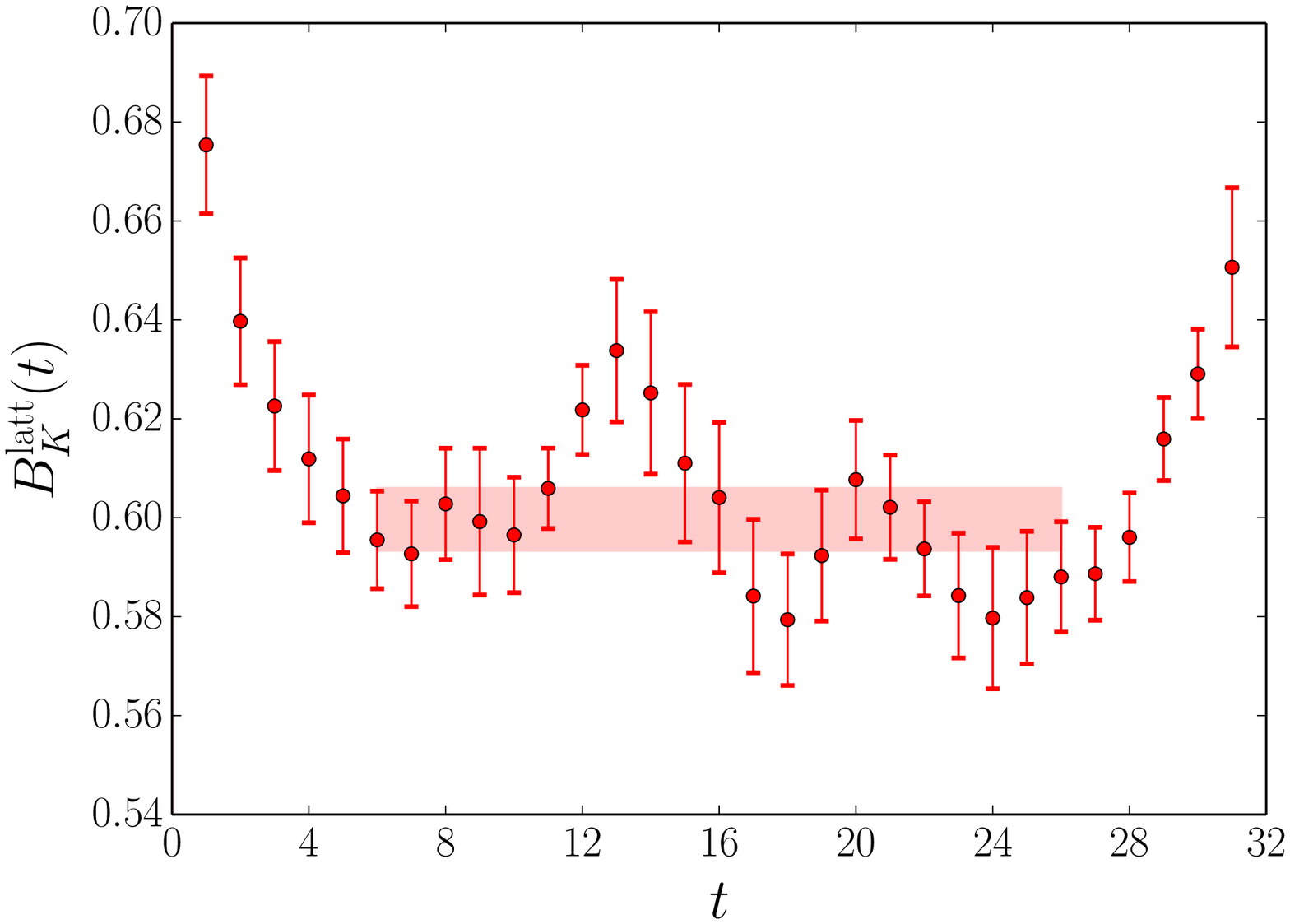}
\includegraphics[width=0.48\textwidth]{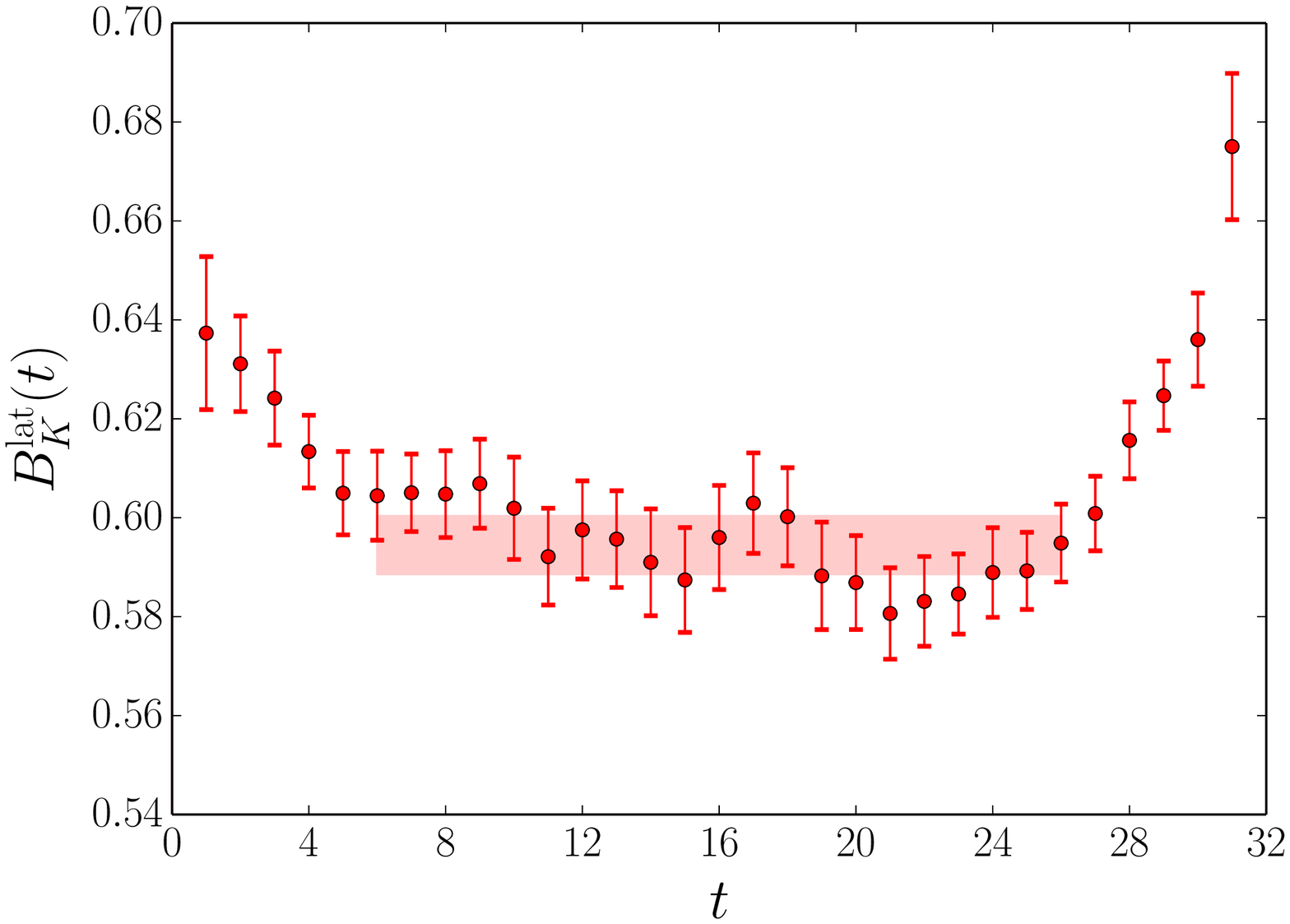}
\includegraphics[width=0.48\textwidth]{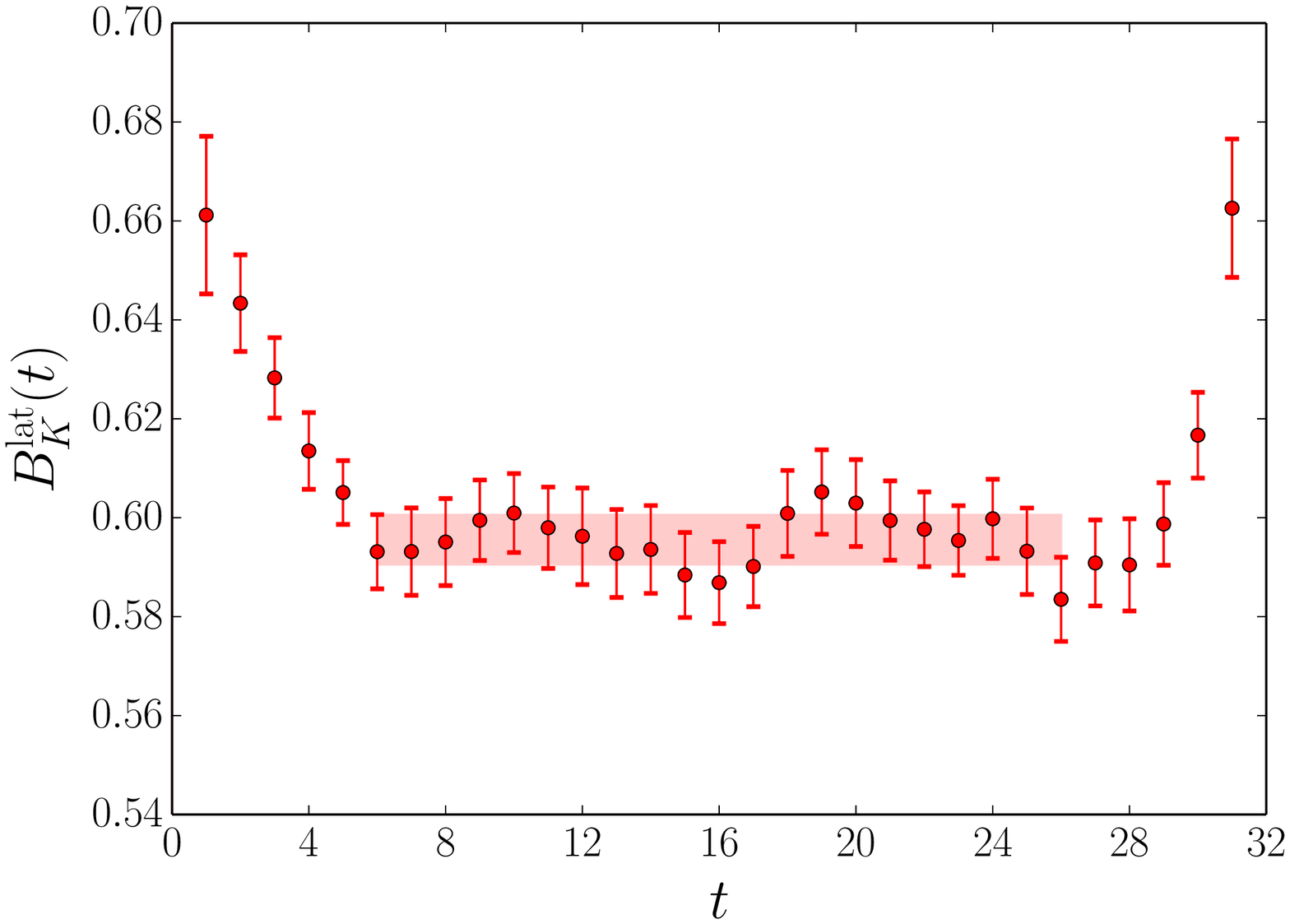}

\caption{Results for $B_K^{\rm lat}(t)$ from Eq.~\eqref{eqn-bkoperator} overlaid by the fitted value on the GP0 (upper-left), GP1 (upper-right) and GP2 (lower) ensembles.\label{fig-bkplots} }
\end{figure}

In Table~\ref{tab-fitbk} we list the hand-chosen fit ranges, the fitted values of $B_K$ and the associated $\chi^2/{\rm dof}$. Plots of $B_K^{\rm lat}(t)$ are given in Figure~\ref{fig-bkplots}. We observe excellent consistency between all three ensembles, demonstrating our ability to compute physical kaon matrix elements using the G-parity mixed kaon states.

\FloatBarrier
\section{Conclusions}
\label{sec:Conclusions}

This document contains a thorough investigation of the use of G-parity spatial boundary conditions (GPBC)~\cite{Wiese:1991ku,Kim:2003xt,Kim:2009fe} in zero temperature lattice simulations. The G-parity operation is a combination of charge conjugation and an isospin rotation by $\pi$ radians around the y-axis under which the charged and neutral pions are all odd eigenstates. GPBC are therefore equivalent to antiperiodic boundary conditions for the pions and the allowed discretized pion momenta become odd-integer multiples of $\pi/L$, where $L$ is the lattice spatial extent. This results in the removal of the stationary pion state from the spectrum, making this technique ideal for studying interactions of moving pions as it is no longer necessary to battle to isolate the weaker excited-state contribution involving moving pions from the dominant ground-state contribution. This is similar to the technique of applying antiperiodic boundary conditions to just the down or up quark~\cite{Kim:2003xt}, which also results in antiperiodic charged pions and has been used successfully to compute the $\Delta I=3/2$ $K\to\pi\pi$ decay amplitude~\cite{Blum:2012uk,Blum:2015ywa}. GPBC are superior however as they also give rise to moving neutral pions, and preserve the isospin symmetry which is explicitly broken if different boundary conditions are applied to the up and down quarks.

We demonstrated this technique numerically in Section~\ref{sec:Results} by performing measurements of a variety of quantities including the pion and kaon masses and decay constants, the bag parameter $B_K$, as well as the domain wall residual mass and axial current renormalization, using two custom-generated $16^3\times 32\times 16$, 2+1 flavor domain wall ensembles with the Iwasaki gauge action at $\beta=2.13$ ($a^{-1}=1.73(3)$) and G-parity boundary conditions in one and two spatial directions. These were compared with the same quantities computed on an identical lattice but with periodic boundary conditions in all spatial directions. The generation of custom ensembles is necessary here as gauge invariance requires the gauge fields to obey charge conjugate (complex conjugate) boundary conditions (cf. Section~\ref{sec:DiscAct}). Using these ensembles we show, for example, that the pion ground-state energies indeed obey the continuum dispersion relation as the number of G-parity directions increases, and that we can extract decay constants from both the spatial and temporal axial current operators that agree in all cases.

The implementation of GPBC is complicated by the flavor mixing under the isospin rotation at the boundary. In Section~\ref{sec-1fequiv} we demonstrated that GPBC in a single direction is equivalent to antiperiodic boundary conditions on a lattice of doubled spatial extent in the G-parity direction where we force the gauge links on the second half to be complex conjugates of those on the first. While this provides a convenient implementation, extending this technique to multiple G-parity directions requires either further (unnecessary) doubling, increasing the computational cost, or applying unusual non-local boundary conditions in the directions perpendicular to the first doubling. We concluded that a general implementation of the lattice action with an explicitly two-flavor fermion field that mixes at the lattice boundary is more practical, and it is this approach that we use for our later measurements and in Ref.~\cite{Bai:2015nea}.

In Section~\ref{sec:Symmetries} we investigated the symmetries of the lattice action. The mixing of quark flavor at the boundary explicitly breaks baryon number conservation as, for example, it transforms a proton ($uud$) into an anti-neutron ($\bar d\bar d \bar u$). There is also a more subtle symmetry breaking of the flavor non-singlet axial current, which was originally recognized by Wiese~\cite{Wiese:1991ku}. This potentially gives rise to a change in the value of the chiral condensate and a corresponding shift in the energy spectrum of the pion states, but in Section~\ref{sec:Results} we argue that this is exponentially suppressed in $m_\pi L$ and indeed found the effect to be small on our $16^3\times 32$ ensembles, for which $m_\pi L \sim 4$. 

We also observed that, in contrast to anti-periodic boundary conditions, the imposition of G-parity boundary conditions in all three direction for a cubic box does not result in a system which is symmetric under cubic rotations.  For example the quark free-field eigenstates of the system are not symmetric under cubic rotations but instead, for our conventions, must have the three components of their momentum equal modulo $2\pi/L$. This implies, for example, that the momentum $(\frac{\pi}{2L},\frac{\pi}{2L},0)$ (for GPBC in two directions) is allowed, but $(\frac{\pi}{2L},-\frac{\pi}{2L},0)$ is not. This symmetry breaking does not extend to the pions, which obey standard antiperiodic boundary conditions in the G-parity directions as we demonstrated explicitly in Section~\ref{sec:Results} by showing that the energies of pions moving in orthogonal directions are in good agreement. However, the fact that we are restricted in our allowed choices of quark momentum components forces us to use operators that are inequivalent under cubic rotations in order to create pions moving in different orthogonal directions. This results in values of the two-point function {\it amplitude} that depend somewhat strongly on the direction of the pion momentum. We found that by choosing operators that are averages over multiple choices of quark momenta with fixed total momentum we drastically reduce the variation in the amplitude. This ultimately enabled us to to construct an interpolating operator that was rotationally symmetric within our measured accuracy to absorb the $\pi\pi$ state in our $K\to\pi\pi$ calculation~\cite{Bai:2015nea}.

A further difficulty of the G-parity approach involves the treatment of the strange quark. In Section~\ref{sec:StrangeQuark} we describe how charge-conjugation boundary conditions, while consistent with the gauge field boundary conditions, are not suitable for constructing the stationary kaon state required for a $K\to\pi\pi$ calculation. Such a state can be constructed if we instead impose G-parity boundary conditions between the strange quark and a fictional degenerate partner, $s'$, and comprises an admixture of the continuum kaon and an unphysical, degenerate partner state. We discuss how operators that interact only with the physical kaon retain their continuum values up to a known numerical factor and additional exponentially-suppressed finite-volume corrections that can be neglected. The effects of the unphysical strange sea quark introduced here are removed by using the square root of the resulting 2-quark fermion determinant in order to revert to a three-flavor simulation.  This can be accomplished, for example by using the RHMC algorithm.  We show that the non-local effects of applying this square root are exponentially suppressed in the spatial lattice size. In Section~\ref{sec:Results} we study the strange quark states numerically and observe that both $B_K$ and the kaon masses and decay constants remain in excellent agreement as we impose GPBC in a successively larger number of directions.

We conclude that G-parity boundary conditions provide a novel and useful means with which to study mesonic interactions and decays involving pions carrying above-threshold momenta. The RBC \& UKQCD collaborations have already made use of these boundary conditions in order to compute the $I=0$ $K\to\pi\pi$ decay amplitude with physical kinematics~\cite{Bai:2015nea}, which requires moving pions. Here the statistical noise is very large due to disconnected contributions, which makes isolating an excited state contribution in such a setup difficult. Further investigations of the momentum dependence of the $\pi\pi$ scattering phase shift are among the future plans for this technique.

\section*{Acknowledgments}

We thank Tianle Wang and our other collaborators in the RBC \& UKQCD collaborations for their insights and comments. N.H.C and D.Z. were partially supported by U.S. DOE grant \#DE-SC0011941 and C.K. by the Intel corporation. The software used includes the CPS QCD library (https://usqcd-software.github.io/CPS.html), supported in part by the USDOE SciDAC program and the Bfm/BAGEL (https://github.com/paboyle/BFM) library~\cite{Boyle:2009vp} for many of the high-performance optimized kernels.   Measurements and ensemble generation were performed using the IBM Bluegene/Q computers~\cite{Boyle:2012iy} formerly at Brookhaven National Laboratory (BNL).  We thank the RIKEN BNL Research Center and the Brookhaven National Laboratory for the use of these machines.

\appendix
\section{Domain wall fermions with G-parity boundary conditions}
\label{appendix-dwf}

In Section~\ref{sec:DiscAct} we formulated the lattice action for G-parity boundary conditions in terms of two four-dimensional fields $\psi_1(x)=d(x)$ and $\psi_2(x)=C\bar u^T(x)$ that transform as $\hat T\psi_1(L-1)\hat T^{-1} = \psi_2(0)$ and $\hat T\psi_2(L-1)\hat T^{-1} = -\psi_1(0)$ under translations across the boundary in a G-parity direction (suppressing orthogonal coordinates). In this Appendix we show how to extend this formulation to the five-dimensional quark fields in the domain wall fermion framework.

For domain wall fermions the fundamental fields $\Phi(x,s)$ are functions of the fifth dimensional coordinate $s$, and the four-dimensional fields are constructed as surface fields as follows:
\begin{equation}\begin{array}{ll}
\phi(x) &= P_R \Phi(x,0) + P_L \Phi(x,L_s-1)\\
{\bar \phi}(x) &= \bar\Phi(x,0)P_L + \bar \Phi(x,L_s-1)P_R  \label{eqn-DWF4d5d}
\end{array}\end{equation}
where $\phi$ and $\Phi$ here are generic 4d and 5d fields respectively, and we have used $\phi/\Phi$ rather than the conventional $\psi/\Psi$ to avoid confusion with our G-parity fields. Here and below we capitalize the field variables to indicate that they are five dimensional, and we explicitly display the coordinate in only one of the four space-time directions, which is assumed to have GPBC.

Starting with the known action of the charge conjugation operation on the 4d field, we can induce its action on those in five dimensions.  This can be accomplished by expressing the charge-conjugated 4d field, $\hat C \phi(x) \hat C^{-1}$ in two ways:
\begin{eqnarray}
   \hat C\phi(x)\hat C^{-1} &=& P_R \hat C\Phi(x,0)\hat C^{-1} + P_L \hat C\Phi(x,L_s-1)\hat C^{-1} \\
&& \quad \mbox{and} \nonumber  \\
 \hat C\phi(x)\hat C^{-1} &=& C\bar\phi^T(x) \nonumber \\
 &=&  P_L C \bar \Phi^T(x,0) + P_R C\bar \Phi^T(x,L_s-1)
\end{eqnarray}
and then equating the $P_L$ and $P_R$ terms found in these two equations.  We conclude that charge conjugation acting on the 5d domain wall fields involves a reflection in the fifth dimension:
\begin{eqnarray}
\hat C \Phi(x,s) \hat C^{-1}       &=& C \bar\Phi^T(x,L_s-1-s) \\
\hat C \bar\Phi(x,s) \hat C^{-1} &=& -\Phi^T(x,L_s-1-s) C^{-1},
\end{eqnarray}
where the second equation can be obtained by applying the same analysis to the conjugate fields $\bar\phi$ and $\bar\Phi$.  

Next we generalize the single-flavor 5d field $\Phi(x,s)$ to a doublet of two fields $U(x,s)$ and $D(x,s)$ and define the action of the G-parity operator $\hat G$ on this doublet as the combination $\hat C e^{-i\hat I_y}$ given in Eq.~(1):
\begin{eqnarray}
\hat G\colvectwo{U(x,s)}{D(x,s)}\hat G^{-1} &=& \colvectwo{-C \bar D^T(x,L_s-1-s)}{C \bar U^T(x,L_s-1-s)}, \label{eq-GP5d} \\
\hat G\rowvectwo{\bar D(x,s)}{- \bar U(x,s)}\hat G^{-1} &=& \rowvectwo{U^T(x,L_s-1-s) C^{-1}}{-D^T(x,L_s-1-s)C^{-1}}, \label{eq-GP5d-bar} 
\end{eqnarray}
in analogy to Eq.~\ref{eqn-GPonud}.

Following Section~\ref{sec:notation} we next define two 5d two-component fields $\Psi$ and $\bar\Psi$ which are composed of $U$, $D$, $\bar U$ and $\bar D$ and defined so that each transforms into itself under G-parity:
\begin{eqnarray}
\Psi(x,s)       &=& \colvectwo{D(x,s)}{C\bar U^T(x,L_s-1-s)} \\
\bar\Psi(x,s) &=& \rowvectwo{\bar D(x,s)}{U^T(x,L_s-1-s) C}.
\end{eqnarray}
With these definitions we have extended our four-dimensional formalism to five dimensions:
\begin{eqnarray}
\hat G \Psi(x,s) \hat G^{-1}       &=& i\sigma_2\Psi(x,s) \label{eq-G-5d}\\
\hat G \bar\Psi(x,s) \hat G^{-1} &=& -\bar\Psi i\sigma_2 \label{eq-G-5d-bar}
\end{eqnarray}
while maintaining the usual relation between the 4d and 5d fields $\psi$, $\bar\psi$, $\Psi$, and $\bar\Psi$:
\begin{eqnarray}
\psi(x)       &=& P_R\Psi(x,0) + P_L\Psi(x,L_s-1) \\
\bar\psi(x) &=& \bar\Psi(x,0) P_L + \bar\Psi(x,L_s-1) P_R\,.
\end{eqnarray}

Equations~\eqref{eq-G-5d} and \eqref{eq-G-5d-bar} ensure that $\Psi(x,s)$ and $\bar\Psi(x,s)$ transform under translations in the same way as the fields $\psi(x)$ and $\bar\psi(x)$:
\begin{eqnarray}
\hat T_\mu \Psi(x_\mu,s) \hat T_\mu^{-1} &=& \left\{ \begin{array}{l|r}
\Psi(x_\mu+1)   &  0 \leq x_\mu < L-1 \\
(i\sigma_2)\Psi(0,s)  & x_\mu = L-1 \end{array} \right. \\
\hat T_\mu \bar\Psi(x_\mu,s) \hat T_\mu^{-1} &=& \left\{ \begin{array}{l|r}
\Psi(x_\mu+1)   &  0 \leq x_\mu < L-1  \\
          \bar\Psi(0,s)(-i\sigma_2)  & x_\mu = L-1 \end{array} \right.
\label{eq-T-5d}
\end{eqnarray}
in analogy with Eq. (16).

\FloatBarrier
\bibliography{paper}

\end{document}